\documentclass{jfm}
\usepackage{graphicx}
\usepackage{newtxtext}
\usepackage{newtxmath}
\usepackage{natbib}
\usepackage{hyperref}
\usepackage{comment}
\usepackage{scalerel}
\usepackage{amsfonts}
\usepackage{amssymb}
\usepackage{amsmath}
\usepackage{mathtools}
\usepackage{mathrsfs}
\usepackage{xcolor}
\definecolor{tpgray}{gray}{0.90}

\usepackage[capitalise]{cleveref}
\usepackage[percent]{overpic}
\usepackage{tikz}
\usepackage{tikz-3dplot}
\usetikzlibrary{arrows,shapes,trees,backgrounds,calc}
\newtheorem{theorem}{Theorem}

\numberwithin{theorem}{section}
\numberwithin{lemma}{section}
\numberwithin{corollary}{section}
\numberwithin{proposition}{section}
\numberwithin{definition}{section}
\numberwithin{example}{section}

\hypersetup{
    colorlinks = true,
    urlcolor   = blue,
    citecolor  = black,
}

\newcommand{\RomanNumeralCaps}[1]
\linenumbers

\usepackage{algorithm}
\usepackage{hyperref}
\usepackage{algpseudocode}
\makeatletter 
\newcounter{algorithmicH}
\let\oldalgorithmic\algorithmic
\renewcommand{\algorithmic}{
  \stepcounter{algorithmicH}
  \oldalgorithmic}
\renewcommand{\theHALG@line}{ALG@line.\thealgorithmicH.\arabic{ALG@line}}
\makeatother

\title{Residual Dynamic Mode Decomposition:\\
Robust and verified Koopmanism}
\author{Matthew J. Colbrook\aff{1,}\aff{2}
  \corresp{\email{m.colbrook@damtp.cam.ac.uk}},
  Lorna J. Ayton\aff{2}
 \and Máté Sz{\H{o}}ke\aff{3}}

\affiliation{\aff{1}Centre Sciences des Données, École Normale Supérieure, Paris, France
\aff{2}Department of Applied Mathematics and Theoretical Physics, University of Cambridge, UK
\aff{3}Kevin T. Crofton Department of Aerospace and Ocean Engineering, Virginia Tech, VA, US}

\begin{document}
\maketitle

\begin{abstract}
Dynamic Mode Decomposition (DMD) describes complex dynamic processes through a hierarchy of simpler coherent features. DMD is regularly used to understand the fundamental characteristics of turbulence and is closely related to Koopman operators. However, verifying the decomposition, equivalently the computed spectral features of Koopman operators, remains a major challenge due to the infinite-dimensional nature of Koopman operators. Challenges include spurious (unphysical) modes, and dealing with continuous spectra, both of which occur regularly in turbulent flows. Residual Dynamic Mode Decomposition (ResDMD), introduced by \citep{colbrook2021rigorous},  overcomes some of these challenges through the data-driven computation of residuals associated with the full infinite-dimensional Koopman operator. ResDMD computes spectra and pseudospectra of general Koopman operators with error control, and computes smoothed approximations of spectral measures (including continuous spectra) with explicit high-order convergence theorems. ResDMD thus provides robust and verified Koopmanism. We implement ResDMD and demonstrate its application in a variety of fluid dynamic situations, at varying Reynolds numbers, arising from both numerical and experimental data. Examples include: vortex shedding behind a cylinder; hot-wire data acquired in a turbulent boundary layer; particle image velocimetry data focusing on a wall-jet flow; and acoustic pressure signals of laser-induced plasma. We present some advantages of ResDMD, namely, the ability to verifiably resolve non-linear, transient modes, and spectral calculation with reduced broadening effects.  We also discuss how a new modal ordering based on residuals enables greater accuracy with a smaller dictionary than the traditional modulus ordering. This paves the way for greater dynamic compression of large datasets without sacrificing accuracy.
\end{abstract}



\section{Introduction}

Koopman operator theory, first introduced by Koopman and von Neumann~\citep{koopman1931hamiltonian,koopman1932dynamical}, provides a ``linearisation'' of non-linear dynamical systems \eqref{eq:DynamicalSystem}~\citep{mezicAMS}. Sparked by \citet{mezic2005spectral,mezic2004comparison}, recent attention has shifted to Koopman operators in data-driven methods for studying dynamical systems from trajectory data~\citep{brunton2021modern}. The ensuing explosion in popularity, dubbed ``Koopmanism''~\citep{budivsic2012applied}, includes 1000s of articles over the last decade. Prominent applications are: fluid dynamics~\citep{schmid2010dynamic,rowley2009spectral,mezic2013analysis,giannakis2018koopman}, the focus of the current paper; epidemiology~\citep{proctor2015discovering}; neuroscience~\citep{brunton2016extracting}; finance~\citep{mann2016dynamic}; robotics~\citep{berger2015estimation,bruder2019modeling}; power grids~\citep{susuki2011coherent,susuki2011nonlinear}; and molecular dynamics~\citep{nuske2014variational,klus2018data,schwantes2015modeling,schwantes2013improvements}. Whilst dynamical systems theory has a rich history of using data for generating models and forecasts, there is currently a renaissance of Koopmanism in the age of big data.

Herein we specifically consider dynamical systems whose state $\pmb{x}$ evolves over a state-space $\Omega\subseteq\mathbb{R}^d$ in discrete time-steps according to a function $F:\Omega \rightarrow \Omega$. That is,
\begin{equation} 
\pmb{x}_{n+1} = F(\pmb{x}_n), \qquad n\geq 0, 
\label{eq:DynamicalSystem} 
\end{equation}
for an initial state $\pmb{x}_0$. Such a dynamical system forms a trajectory of iterates $\pmb{x}_0$, $\pmb{x}_1$, $\pmb{x}_2,\ldots$ in $\Omega$. We consider the discrete-time setting since we are interested in analysing data collected from experiments and numerical simulations, where a continuum of data cannot practically be obtained. The Koopman operator for such dynamical systems acts on functions $g:\Omega\rightarrow\mathbb{C}$, also known as `observables' because they indirectly measure the state of the dynamical system. For example, any measurement taken experimentally is an example of such an observable. The Koopman operator $\mathcal{K}$ is defined by the composition formula
\begin{equation} 
[\mathcal{K}g](\pmb{x}) = (g\circ F)(\pmb{x}), \qquad \pmb{x}\in\Omega, \qquad g\in \mathcal{D}(\mathcal{K}), 
\label{eq:KoopmanOperator} 
\end{equation}
where $\mathcal{D}(\mathcal{K})$ is a suitable domain of observables. For a fixed $g$, $[\mathcal{K}g](\pmb{x})$ measures the state of the dynamical system after one time-step. On the other hand, for a fixed $\pmb{x}$,~\eqref{eq:KoopmanOperator} is a linear composition operator with $F$. $\mathcal{K}$ is therefore a \textit{linear} operator, regardless of whether the dynamics are linear or non-linear. The Koopman operator transforms the non-linear dynamics in the state variable $\pmb{x}$ into equivalent linear dynamics in the observables $g$. However, there is price to pay for this linearisation -- $\mathcal{K}$ acts on an \textit{infinite-dimensional} space. The behavior of the dynamical system \eqref{eq:DynamicalSystem} is determined by the spectral information of $\mathcal{K}$. In the data-driven setting of this paper, we do not assume knowledge of the function $F$ in \eqref{eq:DynamicalSystem}. Instead, the name of the game is to recover spectral properties of $\mathcal{K}$ from measured trajectory data.

The leading numerical algorithm used to approximate Koopman operators is dynamic mode decomposition (DMD)~\citep{tu2014dynamic,kutz2016dynamic}. First introduced by~\citet{schmid2010dynamic,schmid2009dynamic} in the fluids community and connected to the Koopman operator in~\citet{rowley2009spectral}, there are now numerous extensions and variants of DMD~\citep{chen2012variants,wynn2013optimal,jovanovic2014sparsity,williams2015data,proctor2016dynamic,kutz2016multiresolution,noack2016recursive,loiseau2018constrained,korda2018convergence,klus2020data,deem2020adaptive,herrmann2021data,wang2021geometrically,baddoo2021physics,baddoo2022kernel}. The reader is encouraged to consult~\citet{schmid2022dynamic} for a recent overview of DMD-type methods, and~\citet{taira2017modal,taira2020modal,towne2018spectral} for an overview of modal-decomposition techniques in fluid flows. Koopman modes are projections of the physical field onto eigenfunctions of the Koopman operator. For fluid flows, Koopman modes provide collective motions of the fluid that are occurring at the same spatial frequency, growth, or decay rate, according to an (approximate) eigenvalue of the Koopman operator. DMD breaks apart a high-dimensional spatiotemporal signal into a triplet of Koopman modes, scalar amplitudes, and purely temporal signals. This decomposition allows the user to describe complex flow patterns by a hierarchy of simpler processes. When linearly superimposed, these simpler processes approximately recover the full flow. 

DMD is undoubtedly a powerful data-driven algorithm for the approximation of Koopman operators that has led to rapid progress over the past decade~\citep{grilli2012analysis,mezic2013analysis,motheau2014mixed,sarmast2014mutual,subbareddy2014direct,sayadi2014reduced,chai2015numerical,brunton2016koopman,hua2016dynamic,priebe2016low,pasquariello2017unsteady,brunton2019data,page2018koopman,page2019koopman,page2020searching,korda2020data}. However, the infinite-dimensional nature of $\mathcal{K}$ leads to several fundamental challenges. The spectral information of infinite-dimensional operators can be far more complicated, and also far more challenging to compute, than that of a finite matrix~\citep{colbrook2020foundations,SCI_big}. For example, $\mathcal{K}$ can have both discrete and continuous spectra~\citep{mezic2005spectral}. Recently,~\citet{colbrook2021rigorous} introduced \textit{residual DMD} (ResDMD) aimed at tackling some of the challenges associated with infinite dimensions. More specifically, ResDMD addresses the challenges of:

\vspace{2mm}

\begin{itemize}
\item \textbf{Spectral pollution (spurious modes).} A well-known difficulty of computing spectra of infinite-dimensional operators is spectral pollution, where discretisations to a finite matrix cause spurious eigenvalues to appear that have nothing to do with the operator~\citep{lewin2010spectral,Colbrook2019}. DMD-type methods typically suffer from spectral pollution, and can produce modes with no physical relevance to the system being investigated~\citep{budivsic2012applied,williams2015data}. Heuristics, such as comparing different discretization sizes, are common to reduce a user's concern. In some cases, it is possible to approximate the spectral information of a Koopman operator without spectral pollution. For example, when working in a finite-dimensional invariant subspace of known dimension, the Hankel-DMD algorithm computes the corresponding eigenvalues for ergodic systems~\citep{arbabi2017ergodic}. Instead, it is highly desirable to have a principled way of detecting spectral pollution with as few assumptions as possible. By computing residuals associated with the spectrum with error control, ResDMD allows the computation of spectra of general Koopman operators with rigorous convergence guarantees and without spectral pollution. Moreover, ResDMD provides a verification of computations by computing pseudospectra with error control. ResDMD thus allows data-driven study of dynamical systems with error control.
	\vspace{1mm}
	\item \textbf{Invariant subspaces.} A finite-dimensional invariant subspace of $\mathcal{K}$ is a space of observables $\mathcal{G} = {\rm span}\!\left\{g_1,\ldots,g_k\right\}$ such that $\mathcal{K}g\in \mathcal{G}$ for all $g\in\mathcal{G}$. A common, but often incorrect (e.g., when the system is mixing), assumption in the literature is that $\mathcal{K}$ has a non-trivial finite-dimensional invariant space. Even if there is such a subspace, it can be challenging to compute, or may not capture all of the dynamics of interest. Often, one must settle for \textit{approximate invariant subspaces}, and DMD-type methods can result in closure issues~\citep{brunton2016koopman}. By computing upper bounds on residuals, ResDMD provides a way of measuring how invariant a candidate subspace is. It is important to stress that ResDMD computes residuals associated with the underlying infinite-dimensional operator $\mathcal{K}$, in contrast to earlier work that computes residuals associated with finite DMD discretisation matrices (which can never be used as error bounds for spectral information of $\mathcal{K}$ and suffer from issues such as spectral pollution)~\citep{drmac2018data}.	Residuals computed by ResDMD also allow a natural ranking or ordering of approximated eigenvalues and Koopman modes, which can be exploited for efficient compression of the system (see~\cref{sec:verif_method_example_4}).
	\vspace{1mm}
	\item \textbf{Continuous spectra and chaos.} $\mathcal{K}$ can have a continuous spectrum, which is a generic feature of chaotic or turbulent flow~\citep{basley2011experimental,arbabi2017study}. A formidable challenge is how to deal with the continuous spectrum~\citep{colbrook2019computing,colbrook2020,mezic2013analysis}, since discretising to a finite-dimensional operator destroys its presence. Most existing nonparametric approaches for computing continuous spectra of $\mathcal{K}$ are restricted to ergodic systems~\citep{giannakis2019data,arbabi2017study,korda2020data,das2021reproducing}, as this allows relevant integrals to be computed using long-time averages. Many of these methods are related to techniques in signal processing, but can be challenging to apply in the presence of noise or if a pair of eigenvalues are close together, and often rely on heuristic parameter choices or cleanup procedures. ResDMD \citep{colbrook2021rigorous} provides a general computational framework to deal with continuous spectra through smoothed approximations of spectral measures, leading to explicit and rigorous high-order convergence. The methods deal jointly with both discrete and continuous spectra, do not assume ergodicity, and can be applied to data from either short or long trajectories.
	\vspace{1mm}
	\item \textbf{Non-linearity and high-dimensional state-space.} For many fluid flows, e.g., turbulent phenomena, the corresponding dynamical system is strongly non-linear and has a very large state-space dimension. ResDMD can be combined with learned dictionaries to tackle this issue. A key advantage compared to traditional learning methods is that one still has convergence theory and can perform a-posteriori verification that the learned dictionary is suitable. In this paper, we demonstrate this for two choices of dictionary: kernelized extended dynamic mode decomposition (kEDMD)~\citep{williams2015kernel} and (rank-reduced) DMD.
\end{itemize}

\vspace{2mm}


In this paper, we take the framework of ResDMD and demonstrate its use in a wide range of fluid dynamic situations, at varying Reynolds numbers, arising from both numerical and experimental data. We discuss how new ResDMD methods can be reliably used for turbulent flows such as aerodynamic boundary layers and we include, for the first time, a rigorous link between the spectral measure and the turbulent power spectra. We illustrate this link explicitly for experimental measurements of boundary layer turbulence. We also discuss how a new modal ordering based on the residual permits greater accuracy with a smaller DMD dictionary than when using traditional modal modulus ordering. This paves the way for greater dynamic compression of large data sets without sacrificing accuracy.

\begin{figure}
 \centering
 \begin{minipage}[b]{0.95\textwidth}
  \begin{overpic}[width=\textwidth,trim={0mm 0mm 0mm 0mm},clip]{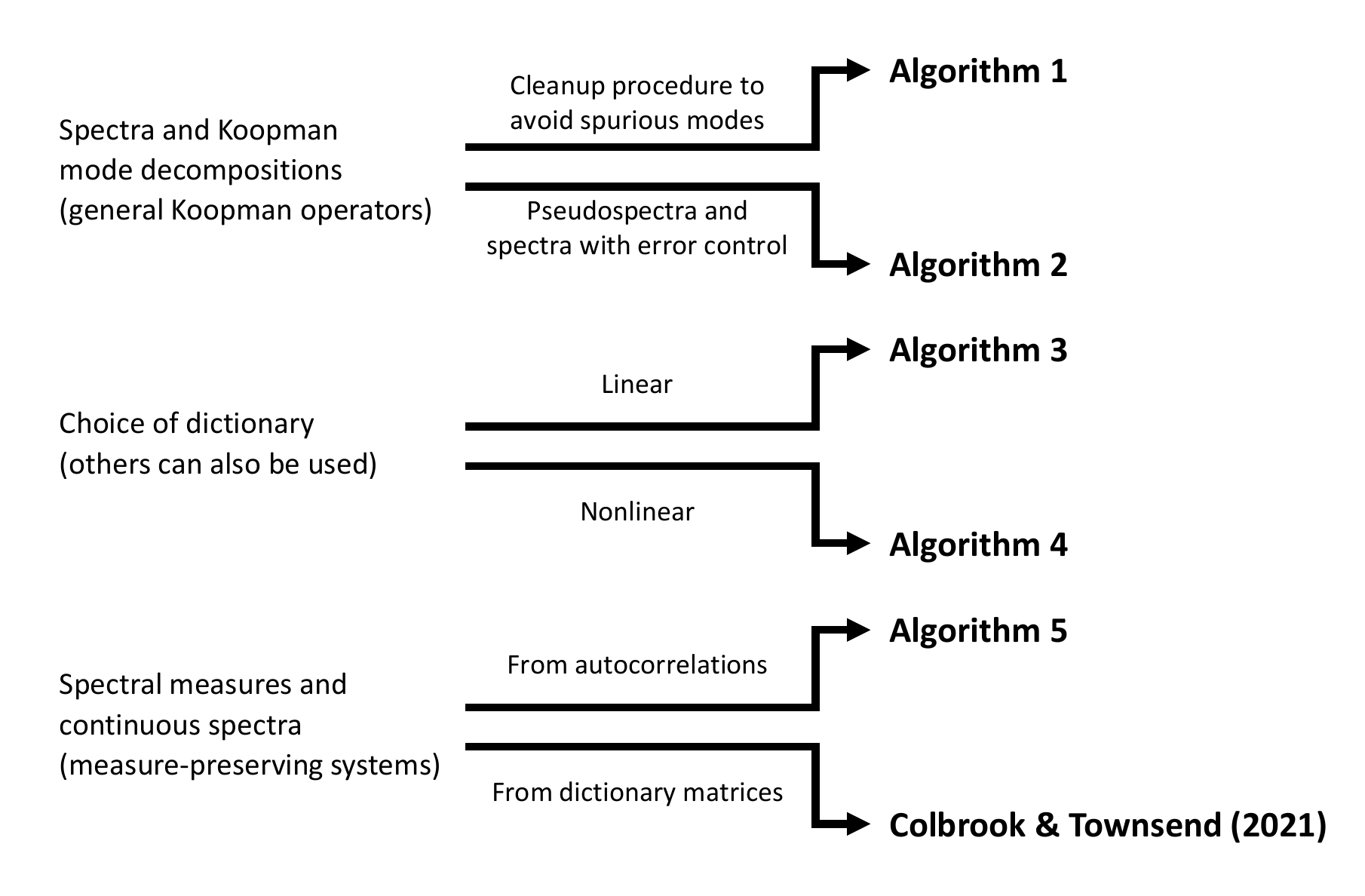}
   \end{overpic}
 \end{minipage}
\caption{A diagrammatic chart for the algorithms used in this paper. The computational problem is shown on the left, and the relevant algorithms on the right.}
\label{fig:flowchart}
\end{figure}

The paper is outlined as follows. In~\cref{sec:DMD_def_recap}, we recap extended dynamic mode decomposition (EDMD), for which DMD is a special case, and interpret the algorithm as a Galerkin method. ResDMD is then introduced in~\cref{sec:introduce_resDMD}, where we recall and expand upon the results of~\citet{colbrook2021rigorous}. We stress that ResDMD does not make any assumptions on the Koopman operator or dynamical system. In~\cref{sec:spectral_measures_new_and_old}, we recall one of the algorithms of~\citet{colbrook2021rigorous} for computing spectral measures (dealing with continuous spectra) under the assumption that the dynamical system is measure-preserving, and make a new link between an algorithm for spectral measures of Koopman operators and the power spectra of signals. We then validate and apply our methods to four different flow cases. We treat flow past a cylinder (numerical data) in \cref{sec:verif_method_example_1}, turbulent boundary layer flow (hot-wire experimental data) in \cref{sec:verif_method_example_2}, wall-jet boundary layer flow (PIV experimental data) in \cref{sec:verif_method_example_3}, and laser-induced plasma (experimental data collected with a microphone) in \cref{sec:verif_method_example_4}. In each case, only the flow or pressure fields are used to extract relevant dynamical information. We end with a discussion and conclusions in~\cref{sec:last_section_conc}. Code for our methods can be found at \url{https://github.com/MColbrook/Residual-Dynamic-Mode-Decomposition} and we provide a diagrammatic chart for implementation in \cref{fig:flowchart}.

\section{Recalling dynamic mode decomposition}
\label{sec:DMD_def_recap}

We recall the definition of EDMD from~\citet{williams2015data} and its interpretation as a Galerkin method. As a special case, DMD can be interpreted as producing a Galerkin approximation of the Koopman operator using a dictionary of linear functions. We assume that the dynamics are governed by a general dynamical system~\eqref{eq:DynamicalSystem} and that we have access to discrete time snapshots of this system, i.e., a finite set of $M$ pairs of measurements
\begin{equation}
\label{snapshot_data}
\{\pmb{x}^{(m)},\pmb{y}^{(m)}\}_{m=1}^M\quad \text{such that}\quad \pmb{y}^{(m)}=F(\pmb{x}^{(m)}),\quad m=1,...,M,
\end{equation}
where the operator $F$ evolves the system along one discrete time unit. For example, these snapshots could be measurements of unsteady velocities across $M$ discrete spatial grid points taken via Particle Image Velocimetry (PIV). Suitable data could be collected from one long time trajectory, corresponding to $\pmb{x}^{(m)}=F^{m-1}(\pmb{x}_0)$, or from multiple shorter trajectories.

We work in the Hilbert space $L^2(\Omega,\omega)$ of observables for a positive measure $\omega$ on $\Omega$. We do not assume that this measure is invariant, and the most common choice of $\omega$ is the standard Lebesgue measure. This choice is natural for Hamiltonian systems for which the Koopman operator is unitary on $L^2(\Omega,\omega)$. For other systems, we can select $\omega$ according to the region where we wish to study the dynamics, such as a Gaussian measure. To be precise, we consider $\mathcal{K} : \mathcal{D}(\mathcal{K})\rightarrow L^2(\Omega,\omega)$, where $\mathcal{D}(\mathcal{K}) \subseteq L^2(\Omega,\omega)$ is a suitable domain of observables. In this section and \cref{sec:introduce_resDMD}, we do not make any assumptions on $\mathcal{K}$.

\subsection{Extended dynamic mode decomposition}
\label{sec:recalling_EDMD_1}

Given a dictionary $\{\psi_1,\ldots,\psi_{N}\}\subset\mathcal{D}(\mathcal{K})\subseteq L^2(\Omega,\omega)$ of observables, EDMD constructs a matrix $\mathbb{K}\in\mathbb{C}^{N\times N}$ from the snapshot data \eqref{snapshot_data} that approximates $\mathcal{K}$ on the finite-dimensional subspace ${V}_{{N}}=\mathrm{span}\{\psi_1,\ldots,\psi_{N}\}$. The choice of dictionary is up to the user, with some common hand-crafted choices given in~\citet[Table 1]{williams2015data}. When the state-space dimension $d$ is large, as in this paper, it is beneficial to use a data-driven choice of dictionary. We discuss this in \cref{sec:how_to_chose_functions}, where we present DMD~\citep{kutz2016dynamic} and kEDMD~\citep{williams2015kernel} in a unified framework.

Assuming we have chosen our dictionary of observables, we define the vector-valued observable or ``quasimatrix''
$$
\Psi(\pmb{x})=\begin{bmatrix}\psi_1(\pmb{x}) & \cdots& \psi_{{N}}(\pmb{x}) \end{bmatrix}\in\mathbb{C}^{1\times {N}}.
$$
Any new observable $g\in V_{N}$ can then be written as $g(\pmb{x})=\sum_{j=1}^{N}\psi_j(\pmb{x})g_j=\Psi(\pmb{x})\,\pmb{g}$ for some vector of constant coefficients $\pmb{g}\in\mathbb{C}^{N}$. It follows from \eqref{eq:KoopmanOperator} that
\begin{equation}
\label{resid_def}
[\mathcal{K}g](\pmb{x})=\Psi(F(\pmb{x}))\,\pmb{g}=\Psi(\pmb{x})(\mathbb{K}\,\pmb{g})+R(\pmb{g},\pmb{x}),
\end{equation}
where
$$
R(\pmb{g},\pmb{x})=\left(\sum_{j=1}^{N}\psi_j(F(\pmb{x}))g_j-\Psi(\pmb{x})(\mathbb{K}\,\pmb{g})\right).
$$
Typically, our subspace $V_{N}$ generated by the dictionary is not an invariant subspace of $\mathcal{K}$, and hence there is no choice of $\mathbb{K}$ that makes the error $R(\pmb{g},\pmb{x})$ zero for all choices of $g\in V_N$ and $\pmb{x}\in\Omega$. Instead, it is natural to select $\mathbb{K}$ as a solution of
\begin{equation} 
\underset{B\in\mathbb{C}^{N\times N}}{\mathrm{argmin}} \left\{\int_\Omega \max_{\|\pmb{g}\|=1}|R(\pmb{g},\pmb{x})|^2\,d\omega(\pmb{x})=\int_\Omega \left\|\Psi(F(\pmb{x})) - \Psi(\pmb{x})B\right\|^2\,d\omega(\pmb{x})\right\}.
\label{eq:ContinuousLeastSquaresProblem}
\end{equation} 
Here, $\|\cdot\|$ denotes the standard Euclidean norm of a vector. Given a finite amount of snapshot data, we cannot directly evaluate the integral in~\eqref{eq:ContinuousLeastSquaresProblem}. Instead, we approximate it via a quadrature rule by treating the data points $\{\pmb{x}^{(m)}\}_{m=1}^{M}$ as quadrature nodes with weights $\{w_m\}_{m=1}^{M}$. Note that in the original definition of EDMD, $\omega$ is a probability measure and the quadrature weights are $w_m=1/M$. We consider the more general case. The discretised version of~\eqref{eq:ContinuousLeastSquaresProblem} is the following weighted least-squares problem:
\begin{equation}
\label{EDMD_opt_prob2}
\mathbb{K}\in\underset{B\in\mathbb{C}^{N\times N}}{\mathrm{argmin}}\sum_{m=1}^{M} w_m\left\|\Psi(\pmb{y}^{(m)})-\Psi(\pmb{x}^{(m)})B\right\|^2.
\end{equation}
Define the following two matrices
\begin{equation}
\begin{split}
\Psi_X=\begin{pmatrix}
\Psi(\pmb{x}^{(1)})\\
\vdots\\
\Psi(\pmb{x}^{(M)})
\end{pmatrix}\in\mathbb{C}^{M\times N},\quad
\Psi_Y=\begin{pmatrix}
\Psi(\pmb{y}^{(1)})\\
\vdots \\
\Psi(\pmb{y}^{(M)})
\end{pmatrix}\in\mathbb{C}^{M\times N},
\label{psidef}
\end{split}
\end{equation}
and let $W=\mathrm{diag}(w_1,\ldots,w_M)$ be the diagonal weight matrix of the quadrature rule. A solution to~\eqref{EDMD_opt_prob2} can then be written down explicitly as
$$
\mathbb{K}=(\Psi_X^*W\Psi_X)^{\dagger}(\Psi_X^*W\Psi_Y)=(\sqrt{W}\Psi_X)^\dagger\sqrt{W}\Psi_Y,
$$
where `$\dagger$' denotes the pseudoinverse. In some applications, the matrix $\Psi_X^*W\Psi_X$ may be ill-conditioned, so it is common to consider truncated singular value decompositions or other forms of regularisation.

\subsection{Quadrature and Galerkin methods}

We can view EDMD as a Galerkin method. Note that
\begin{align*}
\Psi_X^*W\Psi_X= \sum_{m=1}^{M} w_m \Psi(\pmb{x}^{(m)})^*\Psi(\pmb{x}^{(m)}),\quad \Psi_X^*W\Psi_Y= \sum_{m=1}^{M} w_m \Psi(\pmb{x}^{(m)})^*\Psi(\pmb{y}^{(m)}).
\end{align*}
If the quadrature converges, it follows that
\begin{equation}
\label{quad_convergence}
\lim_{M\rightarrow\infty}[\Psi_X^*W\Psi_X]_{jk} = \langle \psi_k,\psi_j \rangle\quad \text{ and }\quad \lim_{M\rightarrow\infty}[\Psi_X^*W\Psi_Y]_{jk} = \langle \mathcal{K}\psi_k,\psi_j \rangle,
\end{equation}
where $\langle \cdot,\cdot \rangle$ is the inner product associated with the Hilbert space $L^2(\Omega,\omega)$. Let $\mathcal{P}_{V_{N}}$ denote the orthogonal projection onto $V_{N}$. As $M\rightarrow \infty$, the above convergence means that $\mathbb{K}$ approaches a matrix representation of $\mathcal{P}_{V_{N}}\mathcal{K}\mathcal{P}_{V_{N}}^*$. Thus, EDMD can be viewed as a Galerkin method in the large data limit $M\rightarrow \infty$. It also follows that the EDMD eigenvalues approach the spectrum of $\mathcal{P}_{V_{N}}\mathcal{K}\mathcal{P}_{V_{N}}^*$ as $M\rightarrow\infty$. Thus, approximating the spectrum of $\mathcal{K}$, $\sigma(\mathcal{K})$, by the eigenvalues of $\mathbb{K}$ is closely related to the so-called finite section method~\citep{bottcher1983finite}. Since the finite section method can suffer from spectral pollution (spurious modes), spectral pollution is also a concern for EDMD~\citep{williams2015data}. It is important to have an independent way to measure the accuracy of the candidate eigenvalue-eigenvector pairs, which is what we propose in our ResDMD method presented in \cref{sec:introduce_resDMD}.

\subsubsection{Convergence of the quadrature rule}
\label{sec:three_quad_conv_scen}

There are typically three scenarios for which the convergence in \eqref{quad_convergence} holds:

\vspace{2mm}

\begin{itemize}
	\item[(i)] {\textbf{Random sampling:}} In the initial definition of EDMD, $\omega$ is a probability measure and $\{\pmb{x}^{(m)}\}_{m=1}^M$ are drawn independently according to $\omega$ with the quadrature weights $w_m=1/M$. The strong law of large numbers shows that \eqref{quad_convergence} holds with probability one~\citep[Section 3.4]{2158-2491_2016_1_51}, provided that $\omega$ is not supported on a zero level set that is a linear combination of the dictionary~\citep[Section 4]{korda2018convergence}. Convergence is typically at a Monte Carlo rate of $\mathcal{O}(M^{-1/2})$~\citep{caflisch1998monte}.\vspace{1mm}
	\item[(ii)] {\textbf{High-order quadrature:}} If the dictionary and $F$ are sufficiently regular and we are free to choose the $\{\pmb{x}^{(m)}\}_{m=1}^{M}$, then it is beneficial to select $\{\pmb{x}^{(m)}\}_{m=1}^{M}$ as an $M$-point quadrature rule with weights $\{w_m\}_{m=1}^{M}$. This can lead to much faster convergence rates in~\eqref{quad_convergence}~\citep{colbrook2021rigorous}, but can be difficult if $d$ is large.\vspace{1mm}
	\item[(iii)] {\textbf{Ergodic sampling:}} For a single fixed initial condition $\pmb{x}_0$ and $\pmb{x}^{(m)}=F^{m-1}(\pmb{x}_0)$ (i.e., data collected along one trajectory), if the dynamical system is ergodic, then one can use Birkhoff's Ergodic Theorem to show \eqref{quad_convergence}~\citep{korda2018convergence}. One chooses $w_m=1/M$ but the convergence rate is problem dependent~\citep{kachurovskii1996rate}.
\end{itemize}

\vspace{2mm}

From an experimental point of view, an example of random sampling could be $\{\pmb{x}^{(m)}\}$ observed with a sampling rate that is lower than the characteristic time period of the system of interest. An example of ergodic sampling could be a time-resolved particle image velocimetry dataset of a flow field over a long time period. The examples in this paper use (i) random sampling, and (iii) ergodic sampling, which are typical for experimental data collection in practice as they arise from long time trajectory measurements.

\subsubsection{Relationship with DMD}
\label{DMD_is_EDMD}
\citet{williams2015data} observed that if we choose a dictionary of observables of the form $\psi_j(\pmb{x})=e_j^*\pmb{x}$ for $j=1,...,d$, the matrix $\mathbb{K}=(\sqrt{W}\Psi_X)^\dagger\sqrt{W}\Psi_Y$ with $w_m=1/M$ is the transpose of the usual DMD matrix
$$
\mathbb{K}_{\mathrm{DMD}}=\Psi_Y^\top\Psi_X^{\top\dagger}=\Psi_Y^\top\sqrt{W}(\Psi_X^\top\sqrt{W})^\dagger=((\sqrt{W}\Psi_X)^\dagger\sqrt{W}\Psi_Y)^\top=\mathbb{K}^\top.
$$
Thus, DMD can be interpreted as producing a Galerkin approximation of the Koopman operator using the set of linear monomials as basis functions. EDMD can be thought of as an extension that allows non-linear functions in the dictionary.

\subsection{Koopman mode decomposition}

Given an observable $h:\Omega\rightarrow\mathbb{C}$, we approximate the orthogonal projection $\mathcal{P}_{V_N}h$ via
$$
[\mathcal{P}_{V_N}h](\pmb{x})\approx \Psi(\pmb{x})(\Psi_X^*W\Psi_X)^\dagger\Psi_X^* W\begin{pmatrix}
h(\pmb{x}^{(1)})&\cdots&
h(\pmb{x}^{(M)})
\end{pmatrix}^\top.
$$
Assuming convergence of the quadrature rule, the right-hand side converges to the projection $\mathcal{P}_{V_N}h$ in $L^2(\Omega,\omega)$ as $M\rightarrow\infty$. Assuming that $\mathbb{K}V=V\Lambda$ for a diagonal matrix $\Lambda$ of eigenvalues and a matrix $V$ of eigenvectors, we obtain
\begin{equation}
\label{gen_kp_m_decm}
[\mathcal{P}_{V_N}h](\pmb{x})\approx \Psi(\pmb{x})V\left[V^{-1}(\sqrt{W}\Psi_X)^\dagger \sqrt{W}\begin{pmatrix}
h(\pmb{x}^{(1)})&\cdots&
h(\pmb{x}^{(M)})
\end{pmatrix}^\top\right].
\end{equation}
As a special case, and vectorising, we have the Koopman mode decomposition
\begin{equation}
\label{koop_mode_estimate}
\pmb{x}\approx \underbrace{\Psi(\pmb{x})V}_{\text{Koopman e-functions}} \underbrace{\left[V^{-1}(\sqrt{W}\Psi_X)^\dagger \sqrt{W}\begin{pmatrix}
\pmb{x}^{(1)}&\cdots&
\pmb{x}^{(M)}
\end{pmatrix}^\top\right]}_{N\times d\text{ matrix of Koopman modes}}.
\end{equation}
The $j$th Koopman mode corresponds to the $j$th row of the matrix in square brackets, and $\Psi V$ is a quasimatrix of approximate Koopman eigenfunctions.

\section{Residual dynamic mode decomposition (ResDMD)}
\label{sec:introduce_resDMD}

We now present ResDMD for computing spectral properties of Koopman operators, which in turn allows us to analyse fluid flow structures such as turbulence. ResDMD, first introduced in~\citet{colbrook2021rigorous}, uses an additional matrix constructed from the measured data, $\{\pmb{x}^{(m)},\pmb{y}^{(m)}\}_{m=1}^M$. The key difference between ResDMD and other DMD-type algorithms is that we construct Galerkin approximations for not only $\mathcal{K}$, but also $\mathcal{K}^*\mathcal{K}$. This difference allows us to have rigorous convergence guarantees for ResDMD and obtain error guarantees on the approximation. In other words, we can tell a-posteriori which parts of the computed spectra and Koopman modes are reliable, thus rectifying issues such as spectral pollution that arise in previous DMD-type methods.

\begin{figure}
 \centering
 \begin{minipage}[b]{1\textwidth}
  \begin{overpic}[width=\textwidth,trim={0mm 50mm 0mm 30mm},clip]{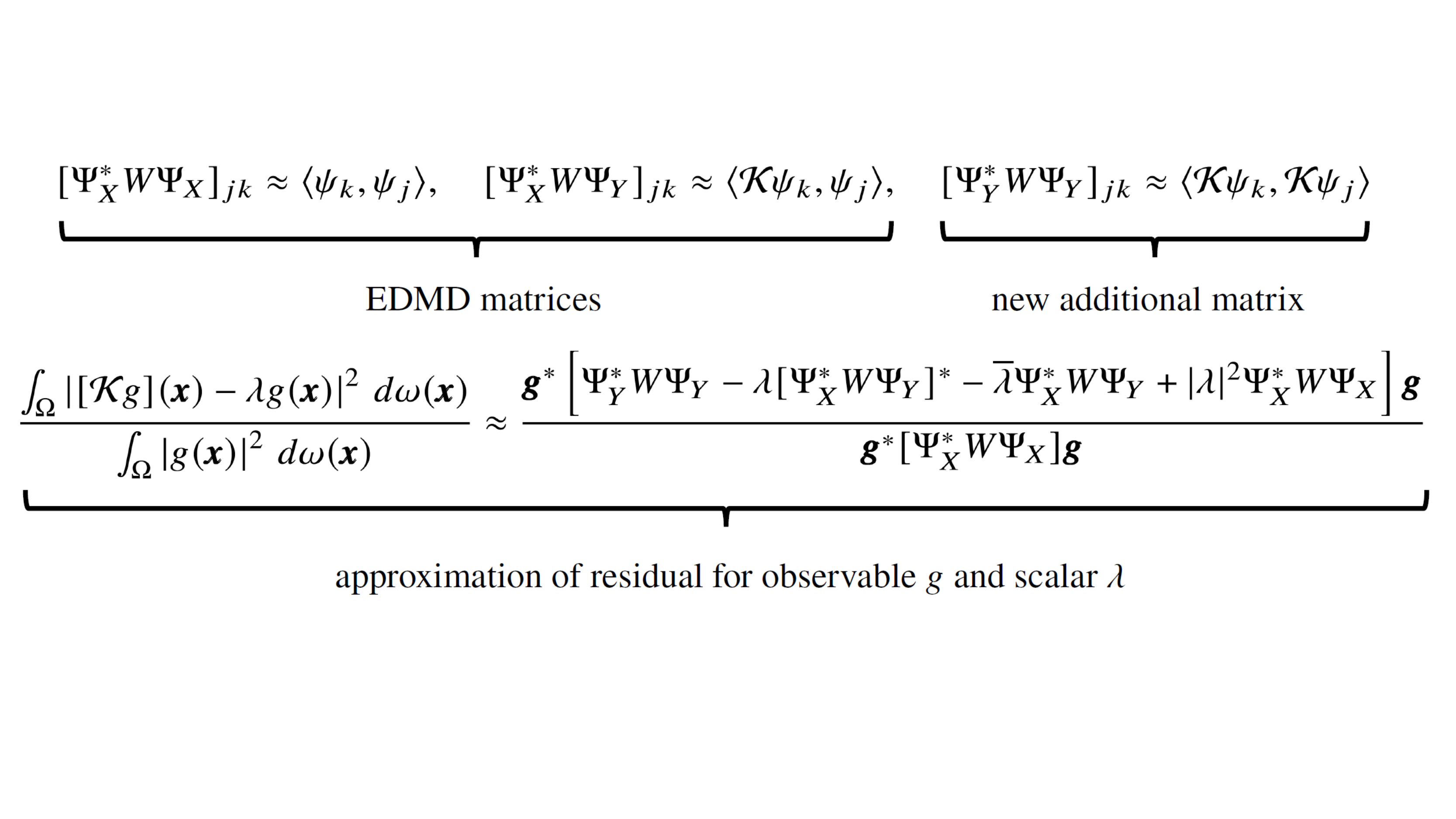}
   \end{overpic}
 \end{minipage}
\caption{The basic idea of ResDMD. The matrices $\Psi_X$ and $\Psi_Y$ are defined in \eqref{psidef} and correspond to the dictionary evaluated at the snapshot data. The matrix $W=\mathrm{diag}(w_1,\ldots,w_M)$ is a diagonal weight matrix.}
\label{fig:idea}
\end{figure}

\subsection{ResDMD and a new data-matrix}
Whilst EDMD obtains eigenvalue-eigenvector pairs for an approximation of the Koopman operator, it cannot verify the accuracy of the computed pairs. By rigorously rejecting inaccurate predictions (spurious modes) we can confidently identify true physical turbulent flow structures. This rigorous measure of accuracy is the linchpin of our new ResDMD method and shown pictorially in \cref{fig:idea}.

To measure the accuracy of a candidate eigenvalue-eigenvector pair $(\lambda,g)$, we approximate residuals. For $\lambda\in\mathbb{C}$ and $g=\Psi\,\pmb{g}\in V_{N}$, the squared relative residual is
\begin{align}
\label{residual_form1}
&\frac{\int_{\Omega}\left|[\mathcal{K}g](\pmb{x})-\lambda g(\pmb{x})\right|^2\, d\omega(\pmb{x})}{\int_{\Omega}\left|g(\pmb{x})\right|^2\, d\omega(\pmb{x})}\\
&\quad\quad=\frac{\sum_{j,k=1}^{N}\overline{{g}_j}{g}_k\left[\langle \mathcal{K}\psi_k,\mathcal{K}\psi_j\rangle -\lambda\langle \psi_k,\mathcal{K}\psi_j\rangle -\overline{\lambda}\langle \mathcal{K}\psi_k,\psi_j\rangle+|\lambda|^2\langle \psi_k,\psi_j\rangle\right]}{\sum_{j,k=1}^{N}\overline{{g}_j}{g}_k\langle \psi_k,\psi_j\rangle}.\notag
\end{align}
We approximate this residual using the same quadrature rule in \cref{sec:recalling_EDMD_1},
\begin{equation}
\mathrm{res}(\lambda,g)^2 = \frac{\pmb{g}^*\left[\Psi_Y^*W\Psi_Y- \lambda[\Psi_X^*W\Psi_Y]^* - \overline{\lambda}\Psi_X^*W\Psi_Y + |\lambda|^2\Psi_X^*W\Psi_X\right]\pmb{g}}{\pmb{g}^*[\Psi_X^*W\Psi_X]\pmb{g}}. 
\label{eq:abs_res}
\end{equation} 
As well as the matrices $\Psi_X^*W\Psi_X$ and $\Psi_X^*W\Psi_Y$ found in EDMD, \eqref{eq:abs_res} has the \textit{additional matrix} $\Psi_Y^*W\Psi_Y$. Since $\Psi_Y^*W\Psi_Y= \sum_{m=1}^{M} w_m \Psi(\pmb{y}^{(m)})^*\Psi(\pmb{y}^{(m)})$, if the quadrature converges,
$$
\lim_{M\rightarrow\infty}[\Psi_Y^*W\Psi_Y]_{jk} = \langle \mathcal{K}\psi_k,\mathcal{K}\psi_j \rangle.
$$
Hence, $\Psi_Y^*W\Psi_Y$ formally corresponds to an approximation of $\mathcal{K}^*\mathcal{K}$. We also have
\begin{equation}
\label{residual_convergence}
\lim_{M\rightarrow\infty} \mathrm{res}(\lambda,g)^2=\frac{\int_{\Omega}\left|[\mathcal{K}g](\pmb{x})-\lambda g(\pmb{x})\right|^2\, d\omega(\pmb{x})}{\int_{\Omega}\left|g(\pmb{x})\right|^2\, d\omega(\pmb{x})}.
\end{equation}
In~\citet{colbrook2021rigorous}, it was shown that the quantity $\mathrm{res}(\lambda,g)$ can be used to rigorously compute spectra and pseudospectra of $\mathcal{K}$. Next, we summarise some of the algorithms and results of~\citet{colbrook2021rigorous}.

\subsection{ResDMD for computing spectra and pseudospectra}
\label{sec:RES_spec_comp1}

\subsubsection{Avoiding spurious eigenvalues}

\Cref{alg:mod_EDMD} uses the residual defined in \eqref{eq:abs_res} to avoid spectral pollution (spurious modes). As is usually done, we assume that $\mathbb{K}$ is diagonalisable. We first find the eigenvalues and eigenvectors of $\mathbb{K}$, i.e., we solve $(\Psi_X^*W\Psi_X)^\dagger(\Psi_X^*W\Psi_Y){\pmb g} = \lambda {\pmb g}$. One can solve this eigenproblem directly, but it is often numerically more stable to solve the generalised eigenproblem $(\Psi_X^*W\Psi_Y){\pmb g} = \lambda (\Psi_X^*W\Psi_X){\pmb g}$. Afterward, to avoid spectral pollution, we discard eigenpairs with a relative residual larger than a specified accuracy goal $\epsilon>0$.

The procedure is a simple modification of EDMD, as the only difference is a clean-up step where spurious eigenpairs are discarded based on their residual. This clean-up avoids spectral pollution and also removes eigenpairs that are inaccurate because of numerical errors associated with non-normal operators, up to the relative tolerance $\epsilon$. The following result~\citep[Theorem 4.1]{colbrook2021rigorous} makes this precise.

\begin{algorithm}[t]\linespread{1.1}\selectfont{}
\textbf{Input:} Snapshot data $\{\pmb{x}^{(m)},\pmb{y}^{(m)}=F(\pmb{x}^{(m)})\}_{m=1}^{M}$, quadrature weights $\{w_m\}_{m=1}^{M}$, a dictionary of observables $\{\psi_j\}_{j=1}^{N}$ and an accuracy goal $\epsilon>0$.\\
\vspace{-4mm}
\begin{algorithmic}[1]
\State Compute $\Psi_X^*W\Psi_X$, $\Psi_X^*W\Psi_Y$, and $\Psi_Y^*W\Psi_Y$, where $\Psi_X$, $\Psi_Y$ are given in~\eqref{psidef}. 
\State Solve $(\Psi_X^*W\Psi_Y)\pmb{g}=\lambda (\Psi_X^*W\Psi_X)\pmb{g}$ for eigenpairs $\{(\lambda_j,g_{(j)}=\Psi\pmb{g}_j)\}$.
\State Compute $\mathrm{res}(\lambda_j,g_{(j)})$ for all $j$ (see~\eqref{eq:abs_res}) and discard if $\mathrm{res}(\lambda_j,g_{(j)})>\epsilon$.
\end{algorithmic} \textbf{Output:} A collection of accurate eigenpairs $\{(\lambda_j,\pmb{g}_j):\mathrm{res}(\lambda_j,g_{(j)})\leq\epsilon\}$.
\caption{: \textbf{ResDMD for computing eigenpairs without spectral pollution. The corresponding Koopman mode decomposition is given in \eqref{koop_mode_estimate2}.}}\label{alg:mod_EDMD}\linespread{1}\selectfont{}
\end{algorithm}

\begin{theorem}
\label{triv_prop}
Let $\mathcal{K}$ be the associated Koopman operator of~\eqref{eq:DynamicalSystem} from which snapshot data is collected. Let $\Lambda_{M}$ denote the eigenvalues in the output of~\cref{alg:mod_EDMD}. Then, assuming convergence of the quadrature rule in \cref{sec:recalling_EDMD_1},
\[
\limsup_{M\rightarrow\infty} \max_{\lambda\in\Lambda_{M}}\|(\mathcal{K}-\lambda)^{-1}\|^{-1}\leq \epsilon.
\]
\end{theorem}

We can also use \cref{alg:mod_EDMD} to cleanup the Koopman mode decomposition in \eqref{koop_mode_estimate}. To do this, we simply let $V^{(\epsilon)}$ denote the matrix whose columns are the eigenvectors $\pmb{g}_j$ with $\mathrm{res}(\lambda_j,g_{(j)})\leq\epsilon$ and compute the Koopman mode decomposition with respect to $\Psi_X^{(\epsilon)}=\Psi_X V^{(\epsilon)}$ and $\Psi_Y^{(\epsilon)}=\Psi_Y V^{(\epsilon)}$. Since $(\sqrt{W}\Psi_X V^{(\epsilon)})^\dagger \sqrt{W}\Psi_Y V^{(\epsilon)}$ is diagonal, the Koopman mode decomposition now becomes

\begin{equation}
\label{koop_mode_estimate2}
\pmb{x}\approx \underbrace{\Psi(\pmb{x})V^{(\epsilon)}}_{\text{Koopman e-functions}} \underbrace{\left[(\sqrt{W}\Psi_XV^{(\epsilon)})^\dagger  \sqrt{W}\begin{pmatrix}
\pmb{x}^{(1)}&\cdots&
\pmb{x}^{(M)}
\end{pmatrix}^\top\right]}_{N\times d\text{ matrix of Koopman modes}}.
\end{equation}

\subsubsection{Computing the full spectrum and pseudospectra}

In the large data limit,~\cref{triv_prop} tells us that \cref{alg:mod_EDMD} computes eigenvalues inside the $\epsilon$-pseudospectrum of $\mathcal{K}$ and hence, avoids spectral pollution and returns reasonable eigenvalues. Recall that the $\epsilon$-pseudospectrum of $\mathcal{K}$ is~\citep{roch1996c,trefethen2005spectra}
$$
\mathrm{\sigma}_{\epsilon}(\mathcal{K}):=\mathrm{cl}\left(\{\lambda\in\mathbb{C}:\|(\mathcal{K}-\lambda)^{-1}\| > 1/\epsilon\}\right)=\mathrm{cl}\left(\cup_{\|\mathcal{B}\|< \epsilon}\mathrm{\sigma}(\mathcal{K}+\mathcal{B})\right),
$$
where ${\rm cl}$ is the closure of the set, and that $\lim_{\epsilon\downarrow0}\mathrm{\sigma}_{\epsilon}(\mathcal{K})=\mathrm{\sigma}(\mathcal{K})$. Despite~\cref{triv_prop}, \cref{alg:mod_EDMD} may not approximate the whole $\epsilon$-pseudospectrum of $\mathcal{K}$, even as $M\rightarrow \infty$ and $N\rightarrow\infty$. This is because the eigenvalues of $\mathcal{P}_{V_{N}}\mathcal{K}\mathcal{P}_{V_{N}}^*$ may not approximate the whole spectrum of $\mathcal{K}$ as $N\rightarrow\infty$~\citep{colbrook2021rigorous,mezic2020numerical}, even if $\cup_{N}V_N$ is dense in $L^2(\Omega,\omega)$.

To overcome this issue, we further include \cref{alg:res_EDMD} which computes practical approximations of $\epsilon$-pseudospectra with rigorous convergence guarantees. Assuming convergence of the quadrature rule in \cref{sec:recalling_EDMD_1}, in the limit $M\rightarrow\infty$, the key quantity
\begin{equation}
\label{tau_factor}
\tau_N(\lambda) = \min_{\pmb{g}\in\mathbb{C}^{N}} \mathrm{res}(\lambda,\Psi\pmb{g})
\end{equation}
is an upper bound for $\|(\mathcal{K}-\lambda)^{-1}\|^{-1}$ and the output of \cref{alg:res_EDMD} is guaranteed to be inside the $\epsilon$-pseudospectrum of $\mathcal{K}$. As $N\rightarrow\infty$ and the grid of points is refined, \cref{alg:res_EDMD} converges to the pseudospectrum uniformly on compact subsets of $\mathbb{C}$ \citep[Theorem B.1, B.2]{colbrook2021rigorous}. Strictly speaking, we converge to the approximate point pseudospectrum, a more complicated algorithm leads to computation of the full pseudospectrum - see \cite[Appendix B]{colbrook2021rigorous}. For brevity, we have not included a statement of the results. We can then compute the spectrum by taking $\epsilon\downarrow 0$. \cref{alg:res_EDMD} also computes observables $g$ with $\mathrm{res}(\lambda,g)< \epsilon$, known as $\epsilon$-approximate eigenfunctions.

\begin{algorithm}[t]\linespread{1.1}\selectfont{}
\textbf{Input:} Snapshot data $\{\pmb{x}^{(m)},\pmb{y}^{(m)}=F(\pmb{x}^{(m)})\}_{m=1}^{M}$, quadrature weights $\{w_m\}_{m=1}^{M}$, a dictionary of observables $\{\psi_j\}_{j=1}^{N}$, an accuracy goal $\epsilon>0$, and a grid $z_1,\ldots,z_k\in\mathbb{C}$.\\
\vspace{-4mm}
\begin{algorithmic}[1]
\State Compute $\Psi_X^*W\Psi_X$, $\Psi_X^*W\Psi_Y$, and $\Psi_Y^*W\Psi_Y$, where $\Psi_X$, $\Psi_Y$ are given in~\eqref{psidef}. 
\State For each $z_j$, compute $\tau_N(z_j) = \min_{\pmb{g}\in\mathbb{C}^{N}} \mathrm{res}(z_j,\Psi\pmb{g})$ (see~\eqref{eq:abs_res}), which is a generalised SVD problem, and the corresponding singular vectors $\pmb{g}_j$.
\end{algorithmic} \textbf{Output:} Estimate of the $\epsilon$-pseudospectrum $\{z_j: \tau_N(z_j)<\epsilon\}$ and $\epsilon$-approximate eigenfunctions $\{\pmb{g}_j: \tau_N(z_j)<\epsilon\}$.
\caption{: \textbf{ResDMD for estimating $\epsilon$-pseudospectra.}}\label{alg:res_EDMD}\linespread{1}\selectfont{}
\end{algorithm}

\subsection{Choice of dictionary}
\label{sec:how_to_chose_functions}

When $d$ is large, it can be impractical to store or form the matrix $\mathbb{K}$, since the initial value of $N$ is very large. We consider two common methods to overcome this issue:

\vspace{2mm}
\begin{itemize}
\item [(i)] \textbf{DMD:} In this case, the dictionary consists of all monomials over $\Omega$ (see~\cref{DMD_is_EDMD}). It is standard to form a low-rank approximation of $\sqrt{W}\Psi_X$ via a truncated singular value decomposition (SVD) as
\begin{equation}
\label{lw_rk_SVD}
\sqrt{W}\Psi_X\approx U_r\Sigma_rV_r^*.
\end{equation}
Here, $\Sigma_r\in\mathbb{C}^{r\times r}$ is diagonal with strictly positive diagonal entries, and $V_r\in\mathbb{C}^{N\times r}$ and $U_r\in\mathbb{C}^{M\times r}$ have $V_r^*V_r=U_r^*U_r=I_r$. We then form the matrix
\begin{equation}
\label{low_rank_koopman}
\tilde{\mathbb{K}}=(\sqrt{W}\Psi_XV_r)^\dagger\sqrt{W}\Psi_YV_r=\Sigma_r^{-1}U_r^*\sqrt{W}\Psi_YV_r=V_r^*\mathbb{K}V_r\in\mathbb{C}^{r\times r}.
\end{equation}
Note that to fit into our Galerkin framework, this matrix is the transpose of the DMD matrix that is commonly computed in the literature.\vspace{1mm}
\item [(ii)] \textbf{Kernelized EDMD (kEDMD):} kEDMD~\citep{williams2015kernel} aims to make EDMD practical for large $d$. Supposing that $\Psi_X$ is of full rank, kEDMD constructs a matrix with an identical formula to \eqref{low_rank_koopman} with $r=M$, for which we have the equivalent form
\begin{equation}
\label{low_rank_EDMD_koopman}
\tilde{\mathbb{K}}=(\Sigma_M^\dagger U_M^*)(\sqrt{W}\Psi_Y\Psi_X^*\sqrt{W})(U_M\Sigma_M^\dagger).
\end{equation}
Suitable matrices $U_M$ and $\Sigma_M$ can be recovered from the eigenvalue decomposition $\sqrt{W}\Psi_X\Psi_X^*\sqrt{W}=U_M\Sigma_M^2U_M^*$. Moreover, both matrices $\sqrt{W}\Psi_X\Psi_X^*\sqrt{W}$ and $\sqrt{W}\Psi_Y\Psi_X^*\sqrt{W}$ can be computed using inner products. kEDMD applies the kernel trick to compute the inner products in an implicitly defined reproducing Hilbert space $\mathcal{H}$ with inner product $\langle\cdot,\cdot\rangle_{\mathcal{H}}$~\citep{scholkopf2001kernel}. A positive-definite kernel function $\mathcal{S}:\Omega\times \Omega\rightarrow\mathbb{R}$ induces a feature map $\varphi:\mathbb{R}^d\rightarrow\mathcal{H}$ so that $\langle\varphi(\pmb{x}),\varphi(\pmb{y})\rangle_{\mathcal{H}}=\mathcal{S}(\pmb{x},\pmb{y})$. This leads to a choice of (typically non-linear) dictionary $\Psi(\pmb{x})$ so that $\Psi(\pmb{x})\Psi(\pmb{y})^*=\langle\varphi(\pmb{x}),\varphi(\pmb{y})\rangle_{\mathcal{H}}=\mathcal{S}(\pmb{x},\pmb{y})$. Often $\mathcal{S}$ can be evaluated in $\mathcal{O}(d)$ operations, meaning that $\tilde{\mathbb{K}}$ is constructed in $\mathcal{O}(dM^2)$ operations.
\end{itemize}

\vspace{2mm}
In either of these two cases, the approximation of $\mathcal{K}$ is equivalent to using a new dictionary with feature map $\Psi(\pmb{x})V_r\in\mathbb{C}^{1\times r}$. In the case of DMD, we found it beneficial to use the mathematically equivalent choice $\Psi(\pmb{x})V_r\Sigma_r^{-1}$, which is numerically better conditioned. To see why, note that $\sqrt{W}\Psi_XV_r\Sigma_r^{-1}\approx U_r$ and $U_r$ has orthonormal columns.

\subsubsection{The problem of vanishing residual estimates}
\label{sec:problem_vanishing_resids}

 Suppose that $\sqrt{W}\Psi_XV_r$ has full row rank, so that $r=M$, and that $\pmb{v}\in\mathbb{C}^{M}$ is an eigenvector of $\tilde{\mathbb{K}}$ with eigenvalue $\lambda$. This means that $(\sqrt{W}\Psi_XV_M)^\dagger\sqrt{W}\Psi_YV_M \pmb{v}=\lambda\pmb{v}$. Since $\sqrt{W}\Psi_XV_M$ has independent rows, $\sqrt{W}\Psi_XV_M(\sqrt{W}\Psi_XV_M)^\dagger=I_M$ and hence $\sqrt{W}\Psi_YV_M\pmb{v}=\lambda\sqrt{W}\Psi_XV_M\pmb{v}$. The corresponding observable is $g(\pmb{x})=\Psi(\pmb{x})V_M\pmb{v}$ and the numerator of $\mathrm{res}(\lambda,g)^2$ in \eqref{eq:abs_res} is equal to $\|\sqrt{W}\Psi_YV_M\pmb{v}-\lambda\sqrt{W}\Psi_XV_M\pmb{v}\|^2.$ It follows that $\mathrm{res}(\lambda,g)=0$. Similarly, if $r$ is too large, $\mathrm{res}(\lambda,g)$ will be a bad approximation of the true residual.

In other words, the regime $r\sim M$ prevents the large data convergence $(M\rightarrow\infty)$ of the quadrature rule and \eqref{residual_convergence}, which holds for a \textit{fixed basis} and hence a fixed basis size. In turn, this prevents us from being able to apply the results of~\cref{sec:RES_spec_comp1}. We next discuss how to overcome this issue by using two sets of snapshot data; these could arise from two independent tests of the same system, or by partitioning the measured data into two groups.

\subsubsection{Using two subsets of the snapshot data}
\label{sec:prob_sol_2_np_data}

A simple remedy to avoid the problem in \cref{sec:problem_vanishing_resids} is to consider \textit{two sets of snapshot data}. We consider an initial set $\{\tilde{\pmb{x}}^{(m)},\tilde{\pmb{y}}^{(m)}\}_{m=1}^{M'}$, which we use to form our dictionary. We then apply ResDMD to the computed dictionary with a second set of snapshot data $\{\hat{\pmb{x}}^{(m)},\hat{\pmb{y}}^{(m)}\}_{m=1}^{M''}$, allowing us to prove convergence as $M''\rightarrow\infty$.

Exactly how to acquire a second set of snapshot data depends on the problem and method of data collection. Given snapshot data with random and independent $\{\pmb{x}^{(m)}\}$, one can simply split up the snapshot data into two parts. For initial conditions that are distributed according to a high-order quadrature rule, if one already has access to $M'$ snapshots then one must typically go back to the original dynamical system and request $M''$ further snapshots. For ergodic sampling along a trajectory, we can let $\{\tilde{\pmb{x}}^{(m)},\tilde{\pmb{y}}^{(m)}\}_{m=1}^{M'}$ correspond to the initial $M'+1$ points of the trajectory ($\tilde{\pmb{x}}^{(m)}=F^{m-1}(\pmb{x}_0)$ for $m=1,...,M'$) and let $\{\hat{\pmb{x}}^{(m)},\hat{\pmb{y}}^{(m)}\}_{m=1}^{M''}$ correspond to the initial $M''+1$ points of the trajectory ($\hat{\pmb{x}}^{(m)}=F^{m-1}(\pmb{x}_0)$ for $m=1,...,M''$).

\begin{algorithm}[t]\linespread{1.1}\selectfont{}
\textbf{Input:} Snapshot data $\{\tilde{\pmb{x}}^{(m)},\tilde{\pmb{y}}^{(m)}\}_{m=1}^{M'}$ and $\{\hat{\pmb{x}}^{(m)},\hat{\pmb{y}}^{(m)}\}_{m=1}^{M''}$, positive integer $N\leq M'$.\\
\vspace{-4mm}
\begin{algorithmic}[1]
\State Set $\Psi_{\mathrm{DMD}}(\pmb{x})=\begin{bmatrix}e_1^*\pmb{x} & \cdots& e_d^*\pmb{x} \end{bmatrix}$.
\State Compute a truncated SVD 
$$
\frac{1}{\sqrt{M'}}\begin{pmatrix}
\Psi_{\mathrm{DMD}}(\tilde{\pmb{x}}^{(1)})^\top&\cdots&\Psi_{\mathrm{DMD}}(\tilde{\pmb{x}}^{(M')})^\top
\end{pmatrix}^\top\approx U_N\Sigma_NV_N^*.
$$ 
\State Apply \cref{alg:res_EDMD,alg:mod_EDMD} with the matrices
\begin{equation}
\Psi_X=\begin{pmatrix}
\Psi_{\mathrm{DMD}}(\hat{\pmb{x}}^{(1)})\\
\vdots \\
\Psi_{\mathrm{DMD}}(\hat{\pmb{x}}^{(M'')})
\end{pmatrix}V_N\Sigma_N^\dagger,\quad
\Psi_Y=\begin{pmatrix}
\Psi_{\mathrm{DMD}}(\hat{\pmb{y}}^{(1)})\\
\vdots \\
\Psi_{\mathrm{DMD}}(\hat{\pmb{y}}^{(M'')})
\end{pmatrix}V_N\Sigma_N^\dagger.
\label{psidef_2DMD}
\end{equation}
\end{algorithmic} \textbf{Output:} Spectral properties of Koopman operator according to \cref{alg:res_EDMD,alg:mod_EDMD}.
\caption{: \textbf{ResDMD with DMD selected observables.}}\label{alg:DMD2}\linespread{1}\selectfont{}
\end{algorithm}

\begin{algorithm}[t]\linespread{1.1}\selectfont{}
\textbf{Input:} Snapshot data $\{\tilde{\pmb{x}}^{(m)},\tilde{\pmb{y}}^{(m)}\}_{m=1}^{M'}$ and $\{\hat{\pmb{x}}^{(m)},\hat{\pmb{y}}^{(m)}\}_{m=1}^{M''}$, positive-definite kernel function $\mathcal{S}:\Omega\times \Omega\rightarrow\mathbb{R}$, and positive integer $N\leq M'$.\\
\vspace{-4mm}
\begin{algorithmic}[1]
\State Apply kEDMD to $\{\tilde{\pmb{x}}^{(m)},\tilde{\pmb{y}}^{(m)}\}_{m=1}^{M'}$ with kernel $\mathcal{S}$ to compute the matrices $\tilde{\mathbb{K}}$, $U_{M'}$ and $\Sigma_{M'}$ using the kernel trick.
\State Compute the dominant $N$ eigenvalues of $\tilde{\mathbb{K}}$ and stack the corresponding eigenvectors column-by-column into $Z\in\mathbb{C}^{M'\times N}$. 
\State Apply a QR decomposition to orthogonalise $Z$ to $Q=\begin{bmatrix}Q_1 & \cdots& Q_{N} \end{bmatrix}\in\mathbb{C}^{M'\times N}$.
\State Apply \cref{alg:res_EDMD,alg:mod_EDMD} with $\{\hat{\pmb{x}}^{(m)},\hat{\pmb{y}}^{(m)}\}_{m=1}^{M''}$ and the dictionary $\{\psi_j\}_{j=1}^{N}$, where
$$
\psi_j(\pmb{x})=\begin{bmatrix}
\mathcal{S}(\pmb{x},\tilde{\pmb{x}}^{(1)})&\mathcal{S}(\pmb{x},\tilde{\pmb{x}}^{(2)})&\cdots & \mathcal{S}(\pmb{x},\tilde{\pmb{x}}^{(M')})
\end{bmatrix}(U_{M'}\Sigma_{M'}^\dagger)Q_j, \qquad 1\leq j\leq N.
$$
\end{algorithmic} \textbf{Output:} Spectral properties of Koopman operator according to \cref{alg:res_EDMD,alg:mod_EDMD}.
\caption{: \textbf{ResDMD with kEDMD selected observables.}}\label{alg:kern_algs}\linespread{1}\selectfont{}
\end{algorithm}

In the case of DMD, the two stage process is summarised in \cref{alg:DMD2}. Often a suitable choice of $N$ can be obtained by studying the decay of the singular values of the data matrix.

In the case of kEDMD, we follow~\citet{colbrook2021rigorous} and the two stage process is summarised in \cref{alg:kern_algs}. The choice of kernel $\mathcal{S}$ determines the dictionary and the best choice depends on the application. In the following experiments, we use the Laplacian kernel $\mathcal{S}(\pmb{x},\pmb{y})=\exp\left(-\gamma{\|\pmb{x}-\pmb{y}\|}\right)$, where $\gamma$ is the reciprocal of the average $\ell^2$-norm of the snapshot data after it is shifted to have mean zero.

We can now apply the theory of~\cref{sec:RES_spec_comp1} in the limit $M''\rightarrow\infty$. It is well-known that the eigenvalues computed by DMD and kEDMD may suffer from spectral pollution. However, and crucially in our setting, we do not directly use these methods to compute spectral properties of $\mathcal{K}$. Instead, we are only using them to select a reasonable dictionary of size $N$, after which our rigorous ResDMD algorithms can be used. Moreover, we use $\{\hat{\pmb{x}}^{(m)},\hat{\pmb{y}}^{(m)}\}_{m=1}^{M''}$ to check the quality of the constructed dictionary. By studying the residuals and using the error control in ResDMD, we can tell a-posteriori whether the dictionary is satisfactory and whether $N$ is sufficiently large.

Finally, it is worth pointing out that the above choices of dictionaries are certainly not the only choices. ResDMD can be applied to any suitable choice. For example, one could use other data-driven methods such as diffusion kernels~\citep{giannakis2018koopman} or trained neural networks~\citep{li2017extended,murata2020nonlinear}.

\section{Spectral measures of measure-preserving systems}
\label{sec:spectral_measures_new_and_old}

Many physical systems of interest described by \eqref{eq:DynamicalSystem} are measure-preserving (preserve volume). Examples include Hamiltonian flows~\citep{arnold1989mathematical}, geodesic flows on Riemannian manifolds~\citep[Chapter 5]{dubrovin2012modern}, Bernoulli schemes in probability theory~\citep{shields1973theory}, and ergodic systems~\citep{walters2000introduction}. The Koopman operator $\mathcal{K}$ associated with a measure-preserving dynamical system  is an isometry, i.e., $\|\mathcal{K}g\|=\|g\|$ for all observables $g\in \mathcal{D}(\mathcal{K})=L^2(\Omega,\omega)$. In this case, spectral measures provide a way of including continuous spectra in the Koopman mode decomposition. This is beneficial in the case of turbulent flows where a-priori knowledge of the spectra (i.e., does it contain a continuous part) may be unknown. The methods described in this section allow us to compute continuous spectra. 

For completeness, we provide a mathematical description of spectral measures in \cref{sec:unitarySpectralMeasure}. The reader should think of these measures as supplying a diagonalisation of $\mathcal{K}$, and hence of the dynamical system. This is made precise in \cref{sec:spec_meas_KMD_rel} through the Koopman mode decomposition. Another interpretation, discussed in \cref{sec:spec_meas_auto_rel}, is that the Fourier coefficients of spectral measures are exactly the forward-time dynamical autocorrelations. In \cref{sec:computing_spec_measures}, we discuss one of the methods of computing spectral measures from~\citet{colbrook2021rigorous}, which makes use of autocorrelations. Finally, we make a connection between spectral measures and power spectra in \cref{new_ps_connection}, which relates directly to experimentally measuring turbulent flows.

\subsection{Spectral measures and Koopman mode decompositions}
\label{sec:spec_meas_KMD_rel}

Given an observable $g\in L^2(\Omega,\omega)$ of interest, the spectral measure of $\mathcal{K}$ with respect to $g$ is a measure $\nu_g$ defined on the periodic interval $[-\pi,\pi]_{\mathrm{per}}$. If $g$ is normalised so that $\|g\|=1$, then $\nu_g$ is a probability measure, otherwise $\nu_g$ is a positive measure of total mass $\|g\|^2$. Lebesgue's decomposition~\citep{stein2009real} allows us to split $\nu_g$ into discrete and continuous parts:
\begin{equation}\label{eqn:spec_meas}
d\nu_g(y)= \underbrace{\sum_{\lambda=\exp(i\theta)\in\mathrm{\sigma}_{{\rm p}}(\mathcal{K})}\langle\mathcal{P}_\lambda g,g\rangle\,\delta({y-\theta})dy}_{\text{discrete part}}+\underbrace{\rho_g(y)\,dy +d\nu_g^{(\mathrm{sc})}(y)}_{\text{continuous part}}.
\end{equation}
Throughout this paper, we use the term ``discrete spectra'' to mean the eigenvalues of $\mathcal{K}$, also known as the point spectrum. This also includes eigenvalues embedded in the continuous spectrum, in contrast to the usual definition of discrete spectrum. The discrete (or atomic) part of $\nu_g$ is a sum of Dirac delta distributions, supported on $\mathrm{\sigma}_{{\rm p}}(\mathcal{K})$, the set of eigenvalues of $\mathcal{K}$. The coefficient of each $\delta$ in the sum is $\langle\mathcal{P}_\lambda g,g\rangle=\|\mathcal{P}_\lambda g\|^2$, where $\mathcal{P}_\lambda$ is the orthogonal spectral projection associated with the eigenvalue $\lambda$. The continuous part of $\nu_g$ consists of a part that is absolutely continuous with density $\rho_g\in L^1([-\pi,\pi]_{\rm per})$, and a singular continuous component $\smash{\nu_g^{(\mathrm{sc})}}$.

The decomposition in~\eqref{eqn:spec_meas} provides important information on the evolution of dynamical systems. For example, suppose that there is no singular continuous spectrum, then any $g\in L^2(\Omega,\omega)$ can be written as
\begin{equation}
\label{koopman_mode_decomp_re0}
g=\sum_{\lambda\in\mathrm{\sigma}_{{\rm p}}(\mathcal{K})}c_\lambda\varphi_\lambda+\int_{[-\pi,\pi]_{\mathrm{per}}}\phi_{\theta,g}\, d\theta,
\end{equation}
where the $\varphi_\lambda$ are the eigenfunctions of $\mathcal{K}$, $c_\lambda$ are expansion coefficients and $\rho_g(\theta)=\langle \phi_{\theta,g},g\rangle$. One should think of $\phi_{\theta,g}$ as a ``continuously parametrised'' collection of eigenfunctions. Using \eqref{koopman_mode_decomp_re0}, one obtains the Koopman mode decomposition~\citep{mezic2005spectral}
\begin{equation}
\label{koopman_mode_decomp_re}
g(\pmb{x}_n)=[\mathcal{K}^ng](\pmb{x}_0)=\sum_{\lambda\in\mathrm{\sigma}_{{\rm p}}(\mathcal{K})}c_{\lambda}\lambda^n\varphi_\lambda(\pmb{x}_0)+\int_{[-\pi,\pi]_{\mathrm{per}}}e^{in\theta}\phi_{\theta,g}(\pmb{x}_0)\, d\theta.
\end{equation}
For characterisations of the dynamical system in terms of these decompositions, see, for example,~\citet{halmos2017lectures,mezic2013analysis,zaslavsky2002chaos}. Generally speaking, the eigenvalues correspond to isolated frequencies of oscillation present in the fluid flow and the growth rates of stable and unstable modes, whilst the continuous spectrum corresponds to chaotic motion. Computing the measures $\nu_g$ provides us with a diagonalisation of the non-linear dynamical system in \eqref{eq:DynamicalSystem}.

\subsection{Spectral measures and autocorrelations}
\label{sec:spec_meas_auto_rel}

The Fourier coefficients of $\nu_g$ are given by
\begin{equation}
\widehat{\nu_g}(n):=\frac{1}{2\pi}\int_{[-\pi,\pi]_{\mathrm{per}}}e^{-in\theta}\,d\nu_g(\theta),\quad n\in\mathbb{Z}.
\label{eq:autoCorrelations} 
\end{equation}
These Fourier coefficients can be expressed in terms of correlations $\langle \mathcal{K}^n g,g\rangle$ and $\langle g,\mathcal{K}^n g\rangle$~\citep{colbrook2021rigorous}. That is, for $g\in L^2(\Omega,\omega)$,
\begin{equation}
\label{F_to_prod_final}
\widehat{\nu_g}(n)=\frac{1}{2\pi}\langle \mathcal{K}^{-n}g,g\rangle,\quad n<0, \qquad \widehat{\nu_g}(n)=\frac{1}{2\pi}\langle g,\mathcal{K}^ng\rangle,\quad n\geq 0.
\end{equation}
From~\eqref{F_to_prod_final}, we see that $\widehat{\nu_{g}}(-n)=\overline{\widehat{\nu_{g}}(n)}$ for $n\in\mathbb{Z}$. In particular, $\nu_g$ is completely determined by the forward-time dynamical \textit{autocorrelations} $\langle  g,\mathcal{K}^ng\rangle$ with $n\geq 0$. Equivalently, the spectral measure of $\mathcal{K}$ with respect to $g\in L^2(\Omega,\omega)$ is a signature for the forward-time dynamics of~\eqref{eq:DynamicalSystem}. This connection allows us to interpret spectral measures as an infinite-dimensional version of power spectra in \cref{new_ps_connection}.

\subsection{Computing spectral measures}
\label{sec:computing_spec_measures}

To approximate spectral measures, we make use of the relation \eqref{F_to_prod_final} between the Fourier coefficients of $\nu_g$ and the autocorrelations $\langle g,\mathcal{K}^ng\rangle$. There are typically three ways to compute the autocorrelations, corresponding to the three scenarios discussed in \cref{sec:three_quad_conv_scen}:

\vspace{2mm}

\begin{enumerate}
\item \textbf{Random sampling:} The autocorrelations can be approximated as
$$
\langle g,\mathcal{K}^ng\rangle\approx\frac{1}{M}\sum_{m=1}^{M}g(\pmb{x}^{(m)})\overline{[\mathcal{K}^ng](\pmb{x}^{(m)})}.
$$
\item \textbf{High-order quadrature:} The autocorrelations can be approximated as 
$$
\langle g,\mathcal{K}^ng\rangle = \int_{\Omega}g(\pmb{x})\overline{[\mathcal{K}^ng](\pmb{x})}\,d\omega(\pmb{x}) \approx\sum_{m=1}^{M}w_mg(\pmb{x}^{(m)})\overline{[\mathcal{K}^ng](\pmb{x}^{(m)})}.
$$
\item \textbf{Ergodic sampling:} The autocorrelations can be approximated as
\begin{equation}
\label{autocorel_arg}
\langle g,\mathcal{K}^ng\rangle\approx \frac{1}{M-n}\!\!\!\! \sum_{m=0}^{M-n-1}\!\!\!\! g(\pmb{x}_{m})\overline{[\mathcal{K}^ng](\pmb{x}_{m})}= \frac{1}{M-n}\!\!\!\! \sum_{m=0}^{M-n-1}\!\!\!\! g(\pmb{x}_{m})\overline{g(\pmb{x}_{m+n})}, 
\end{equation}
\end{enumerate}

\vspace{2mm}

The first two methods require multiple snapshots of the form $\{x_0^{(m)},\ldots, x_{n}^{(m)}\}$, with $[\mathcal{K}^ng](\pmb{x}_0^{(m)})=g(x_{n}^{(m)})$. Ergodic sampling only requires a single trajectory, and the ergodic averages in \eqref{autocorel_arg} can be re-written in terms of discrete (non-periodic) convolutions. Thus, all of the averages are simultaneously and rapidly computed using the FFT.

We suppose now that one has already computed the autocorrelations $\langle g,\mathcal{K}^ng\rangle$ for $0\leq n\leq N_{\mathrm{ac}}$. In practice, given a fixed data set of $M$ snapshots, we choose a suitable $N_{\mathrm{ac}}$ by checking for convergence of the autocorrlations by comparing with smaller values of $M$. We would like to recover an approximation of $\nu_g$ from the computed autocorrelations. Because of~\eqref{eq:autoCorrelations}, the task is similar to Fourier recovery~\citep{gottlieb1997gibbs,adcock2012stable}. Following~\citet{colbrook2021rigorous}, we define an approximation to $\nu_g$ as 
\begin{equation}
\label{fejer_sum}
\nu_{g,N_{\mathrm{ac}}}(\theta)=\sum_{n=-N_{\mathrm{ac}}}^{N_{\mathrm{ac}}} \varphi\!\left(\frac{n}{N_{\mathrm{ac}}}\right)\widehat{\nu_g}(n)e^{in\theta}. 
\end{equation}
The function $\varphi : [-1,1]\rightarrow \mathbb{R}$ is often called a filter function~\citep{tadmor2007gibbs,hesthaven2017numerical} and $\varphi(x)$ is close to $1$ when $x$ is close to $0$, but tapers to $0$ near $x=\pm 1$. \cref{alg:spec_meas_poly} summarises the approach and the choice of $\varphi$ effects the convergence. For $m\in\mathbb{N}$, suppose that $\varphi$ is an even continuous function that is compactly supported on $[-1,1]$ such that
\vspace{1mm}
\begin{itemize}
	\item[(a)] $\varphi \in \mathcal{C}^{m-1}([-1,1])$,
	\item[(b)] $\varphi(0)=1$ and $\varphi^{(n)}(0)=0$ for any integer $1\leq n\leq m-1$,
	\item[(c)] $\varphi^{(n)}(1)=0$ for any integer $0\leq n\leq m-1$,
	\item[(d)] $\varphi|_{[0,1]}\in \mathcal{C}^{m+1}([0,1])$.
\end{itemize}\vspace{1mm}
Then, we have the following forms of convergence:
\vspace{1mm}
\begin{itemize}
	\item under suitable smoothness assumptions, $\nu_{g,N_{\mathrm{ac}}}(\theta)$ approximates the spectral density $\rho_g(\theta)$ in \eqref{eqn:spec_meas} to order $\mathcal{O}(N_{\mathrm{ac}}^{-m}\log(N_{\mathrm{ac}}))$,
	\item a suitably rescaling of $\nu_{g,N_{\mathrm{ac}}}(\{\theta_0\})$ approximates the point spectrum at $\theta_0$,
	\item for any $\phi\in\mathcal{C}^{n,\alpha}([-\pi,\pi]_{\mathrm{per}})$,
\begin{equation}\label{weak_bd_m_ord}
	\left|\int_{[-\pi,\pi]_{\mathrm{per}}}\!\!\!\!\!\! \phi(\theta)\nu_{g,N_{\mathrm{ac}}}(\theta)\,d\theta - \int_{[-\pi,\pi]_{\mathrm{per}}}\!\!\!\!\!\!\phi(\theta)\,d\nu_g(\theta)\right|\!\lesssim\! \|\phi\|_{\mathcal{C}^{n,\alpha}}\!\!\left(N_{\mathrm{ac}}^{-(n+\alpha)}\!+\! N_{\mathrm{ac}}^{-m}\log(N_{\mathrm{ac}})\right).
	\end{equation}
\end{itemize}\vspace{1mm}
This last property is known as weak convergence. Explicit statements of these results and proofs can be found in~\citet[Section 3]{colbrook2021rigorous}. In general, one should think of $\nu_{g,N_{\mathrm{ac}}}$ as a smooth function that approximates the spectral measure $\nu_g$ to order $\mathcal{O}(N_{\mathrm{ac}}^{-m}\log(N_{\mathrm{ac}}))$, with a frequency smoothing scale of $\mathcal{O}(N_{\mathrm{ac}}^{-m})$.

\begin{algorithm}[t]\linespread{1.1}\selectfont{}
\textbf{Input:} Trajectory data, a filter $\varphi$, and an observable $g\in L^2(\Omega,\omega)$. \\
\vspace{-4mm}
\begin{algorithmic}[1]
\State Approximate the autocorrelations $\widehat{\nu_g}(n) = \frac{1}{2\pi}\langle g,\mathcal{K}^ng\rangle$ for $0\leq n\leq N_{\mathrm{ac}}$. (The precise value of $N_{\mathrm{ac}}$ and the approach depends on the trajectory data.)
\State Set $\widehat{\nu_g}(-n) = \overline{\widehat{\nu_g}(n)}$ for $1\leq n\leq N_{\mathrm{ac}}$.
\end{algorithmic} \textbf{Output:} The function $\nu_{g,N_{\mathrm{ac}}}(\theta) = \sum_{n=-N_{\mathrm{ac}}}^{N_{\mathrm{ac}}} \varphi\left(\frac{n}{N_{\mathrm{ac}}}\right)\widehat{\nu_g}(n) e^{in\theta}$.
\caption{\textbf{: Approximating spectral measures from autocorrelations.}}\linespread{1}\selectfont{}
\label{alg:spec_meas_poly}
\end{algorithm}

One can also compute spectral measures without the autocorrelations. \citet{colbrook2021rigorous} place \cref{alg:spec_meas_poly} in the wider framework of convolution kernels. One can develop high-order methods based on rational kernels that approximate spectral measures using the ResDMD matrices. Computing suitable residuals allows an adaptive and rigorous selection of the smoothing parameter used in the convolution. In particular, this setup allows us to deal with general snapshot data $\{\pmb{x}^{(m)},\pmb{y}^{(m)}=F(\pmb{x}^{(m)})\}_{m=1}^M$ without the need for long trajectories. For brevity, we omit the details.

\subsection{Interpretation of Koopman spectral measures as power spectra}
\label{new_ps_connection}

We can interpret \cref{alg:spec_meas_poly} as an approximation of the power spectrum of the signal $g(\pmb{x}(t))$, given by \cite[Chapter 8]{glegg2017aeroacoustics}
\begin{equation}
\label{power_def}
S_{gg}(f)=\int_{-T}^TR_{gg}(t)e^{2\pi i ft}\, dt,
\end{equation}
over a time window $[-T,T]$ and for frequency $f$ (measured in Hertz). Here, $R_{gg}(t)$ is the delay autocorrelation function, defined for $t\geq 0$ as
$$
R_{gg}(t)=\langle g,g\circ F_t\rangle,
$$
where $F_t$ is the forward time propagator for a timestep $t$. In particular, we have
$$
R_{gg}(n\Delta t)=\begin{dcases}\langle g,\mathcal{K}^ng\rangle, \quad&\text{if }n\geq 0\\
\overline{\langle g,\mathcal{K}^{-n}g\rangle}=\langle\mathcal{K}^{-n} g,g\rangle,\quad &\text{otherwise.}
\end{dcases}
$$
We apply windowing and multiply the integrand in \eqref{power_def} by the filter function $\varphi(t/T)$. We then discretise the resulting integral using the trapezoidal rule (noting that the endpoint contributions vanish) with step size $\Delta t=T/N_{\mathrm{ac}}$ to obtain the approximation
\begin{equation}
\label{power_def2}
\frac{{S}_{gg}(f)}{2\pi\Delta t}\approx\sum_{n=-N_{\mathrm{ac}}}^{N_{\mathrm{ac}}} \varphi\left(\frac{n}{N_{\mathrm{ac}}}\right)\frac{R_{gg}(n\Delta t)}{2\pi}e^{in(2\pi f\Delta t)}=\sum_{n=-N_{\mathrm{ac}}}^{N_{\mathrm{ac}}} \varphi\left(\frac{n}{N_{\mathrm{ac}}}\right)\widehat{\nu_g}(n)e^{in(2\pi f\Delta t)}.
\end{equation}
It follows that $\nu_{g,N_{\mathrm{ac}}}(2\pi f\Delta t)$ can be understood as a discretised version of ${S}_{gg}(f)/(2\pi\Delta t)$ over the time window $[-T,T]$. Taking the limit $N_{\mathrm{ac}}\rightarrow\infty$ with $(2\pi \Delta t) f=\theta$, we see that $\nu_g$ is an appropriate limit of the power spectrum with time resolution $\Delta t$. There are two key benefits of using $\nu_g$. First, we do not explicitly periodically extend the signal to compute autocorrelations, and thus avoid the problem of broadening. Instead we window in the frequency domain. Second, we have rigorous convergence theory as $N_{\mathrm{ac}}\rightarrow\infty$. We compare spectral measures to power spectra in \cref{sec:verif_method_example_2}.

\section{Example I: Flow past a cylinder wake}
\label{sec:verif_method_example_1}

We first verify our method by considering the classic example of low Reynolds number flow past a circular cylinder. Due to its simplicity and its relevance in engineering, this is one of the most studied examples in modal-analysis techniques~\citep[Table 3]{rowley2017model}, \citep{taira2020modal,chen2012variants}. We consider the post-transient regime with $\Rey=100$, corresponding to periodic oscillations on a stable limit cycle. The Koopman operator of the flow has pure point spectrum with a lattice structure \citep{bagheri2013koopman}.

\subsection{Computational setup}

The flow around a circular cylinder of diameter $D$ is obtained using an incompressible, two-dimensional lattice-Boltzmann computational fluid mechanics (CFD) flow solver. The solver uses the D2Q9 lattice model along with the BGKW collision model to calculate the velocity field. The temporal resolution (timestep) of the flow simulations is such that the Courant--Friedrichs--Lewy  number is unity on the uniform numerical grid. However, storing each timestep results in a very finely resolved flow field and a large volume of data. An initial down-sampling study revealed that storing the simulation data such that keeping 12 snapshots of flow field data within a period of vortex shedding still enables us to use our analysis tools without affecting the DMD results. The size of the computational domain is $18D$ in length and $5D$ in height. The cylinder is positioned $2D$ downstream of the inlet at the mid-height of the domain. The cylinder walls and the side-walls are defined as bounce-back no-slip walls, and a parabolic velocity inlet profile is defined at the inlet of the domain such that $\Rey=100$. The outlet is defined as non-reflecting outflow. For a detailed description of the solver, we refer the reader to~\cite{jozsa2016validation} and~\cite{szHoke2017performance}.

\subsection{Results}

We collect snapshot data of the velocity field along a single trajectory in the post-transient regime of the flow and split the data into two data sets according to \cref{sec:prob_sol_2_np_data}. The first set $\{\tilde{\pmb{x}}^{(m)},\tilde{\pmb{y}}^{(m)}\}_{m=1}^{M'}$ corresponds to $M'=500$ snapshots. We then collect $M''=1000$ further snapshots $\{\hat{\pmb{x}}^{(m)},\hat{\pmb{y}}^{(m)}\}_{m=1}^{M''}$, where $\hat{\pmb{x}}^{(1)}$ corresponds to approximately 40 time periods of vortex shedding later than $\tilde{\pmb{x}}^{(M')}$. We use \cref{alg:DMD2,alg:kern_algs}, which we refer to as a linear dictionary (obtained using DMD) and a non-linear dictionary (obtained using kEDMD), respectively. For the linear dictionary we use $N=200$ functions. For the non-linear dictionary, we use $N=400$ functions. We use a larger number of functions for the non-linear dictionary since we found this choice of dictionary to be better conditioned than the linear dictionary. Similar results and the same conclusions as ours are obtained using different choices of $N,M'$ and $M''$.

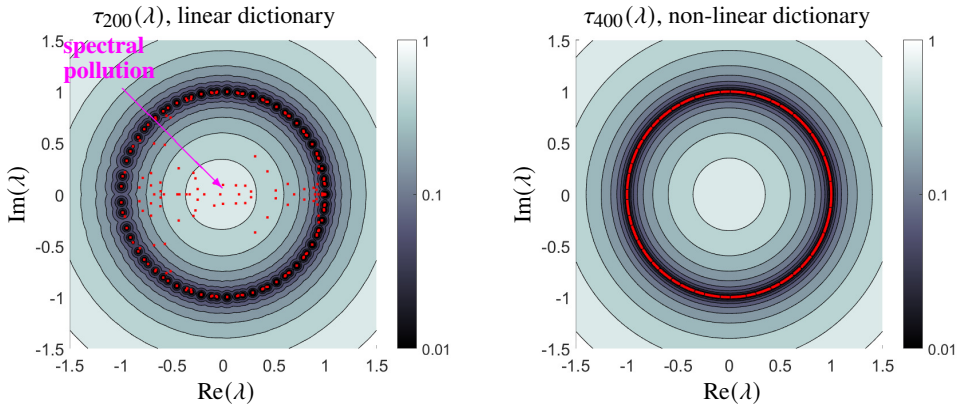
\begin{figure}
 \centering
 \begin{minipage}[b]{0.49\textwidth}
  \begin{overpic}[width=\textwidth,trim={0mm 0mm 0mm 0mm},clip]{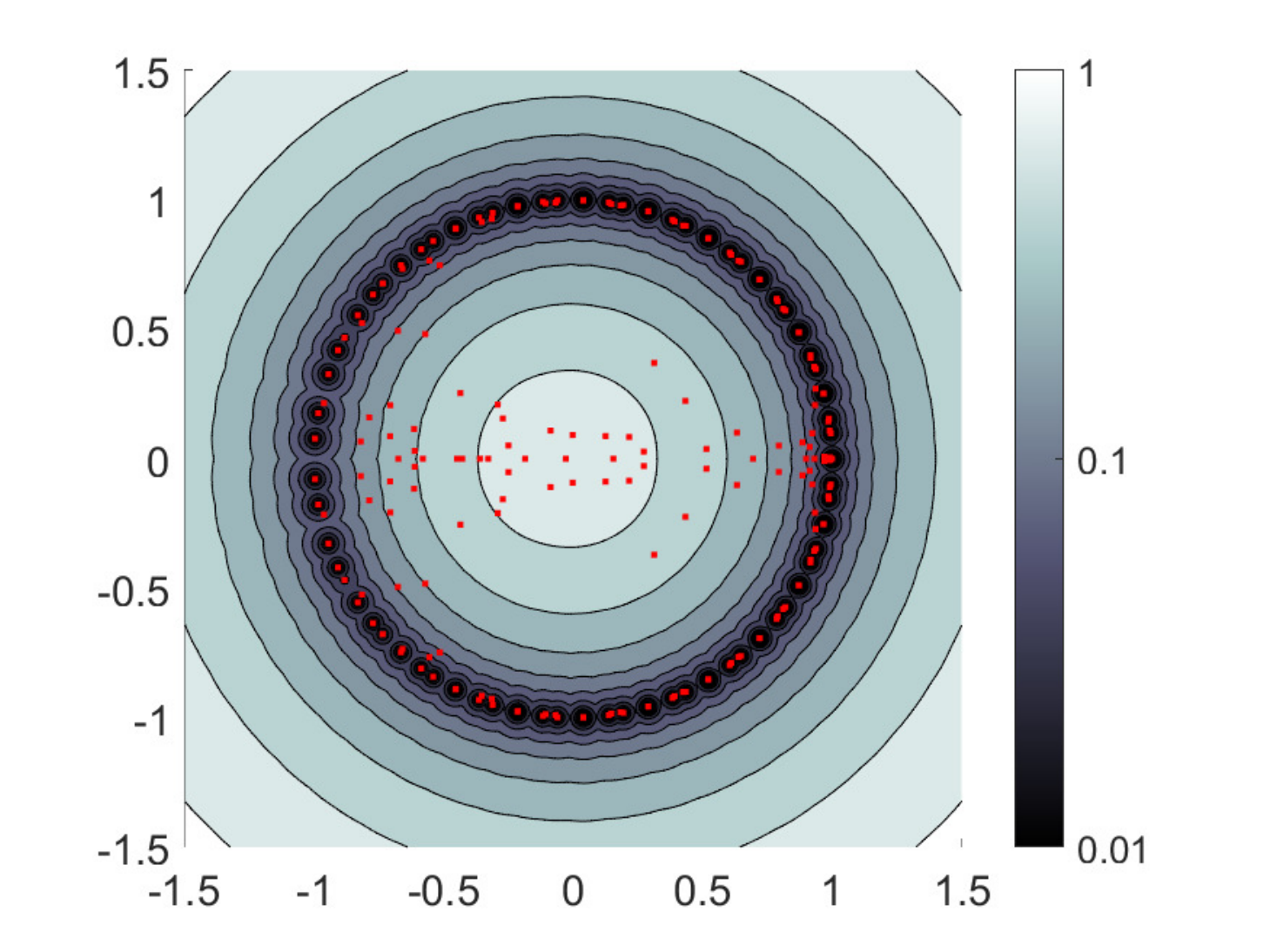}
		\put (19,73) {$\tau_{200}(\lambda)$, linear dictionary}
   \put (40,-2) {$\mathrm{Re}(\lambda)$}
		\put (2,33) {\rotatebox{90}{$\mathrm{Im}(\lambda)$}}
		\put (12,67) { {\textcolor[rgb]{1,0,1}{\textbf{spectral}}}}
		\put (12,62) { {\textcolor[rgb]{1,0,1}{\textbf{pollution}}}}
	\put(25,60)  {\color[rgb]{1,0,1}\vector(1,-1){20}}
   \end{overpic}
 \end{minipage}
\begin{minipage}[b]{0.49\textwidth}
  \begin{overpic}[width=\textwidth,trim={0mm 0mm 0mm 0mm},clip]{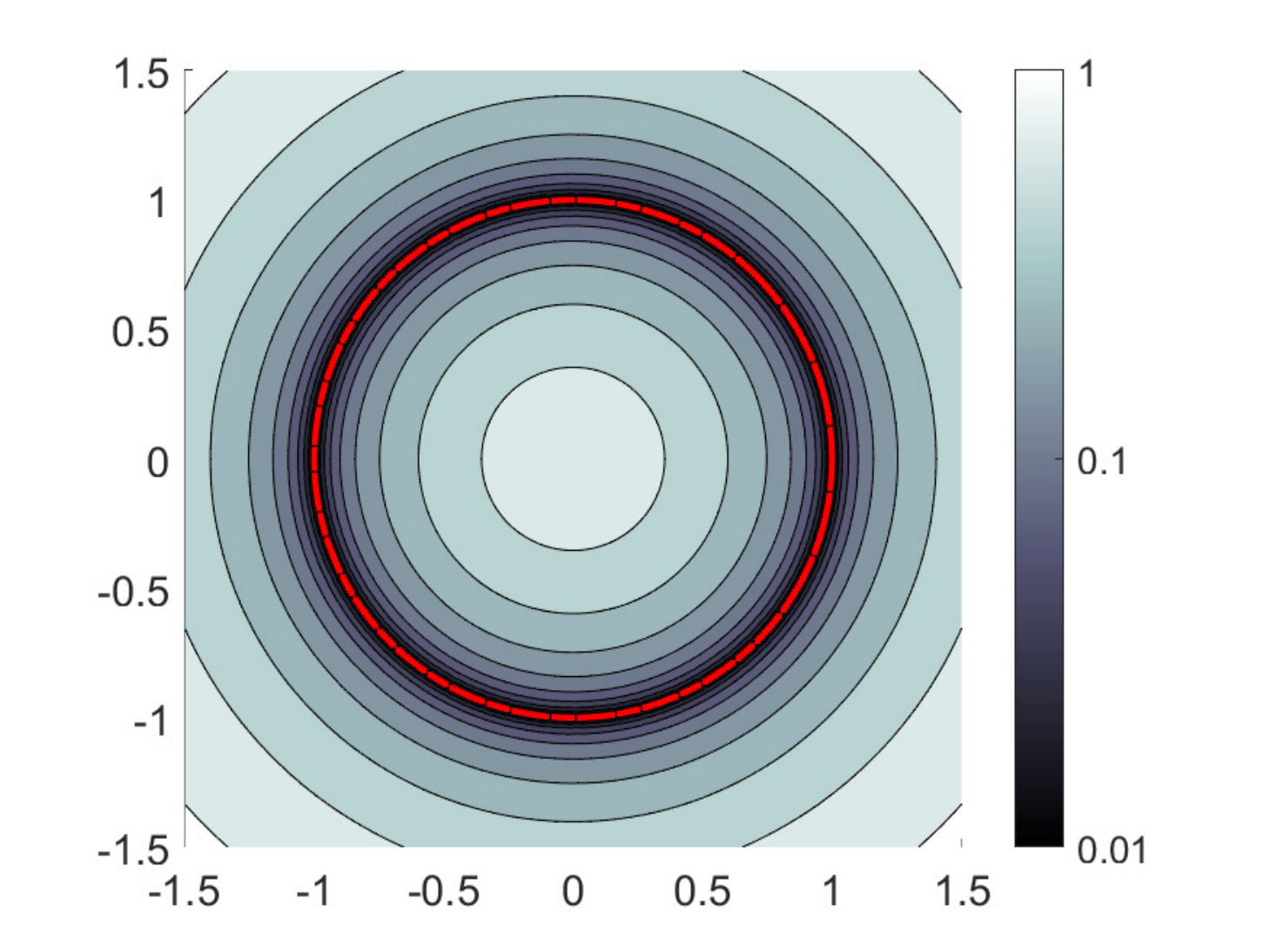}
		\put (16,73) {$\tau_{400}(\lambda)$, non-linear dictionary}
   \put (40,-2) {$\mathrm{Re}(\lambda)$}
		\put (2,33) {\rotatebox{90}{$\mathrm{Im}(\lambda)$}}
   \end{overpic}
 \end{minipage}
\vspace{1mm}
\caption{Pseudospectral contours computed using \cref{alg:res_EDMD} for the cylinder wake example, using a linear dictionary (left) and a non-linear dictionary (right). The eigenvalues of the finite Galerkin matrix $\mathbb{K}$ are shown as red dots. The computed residuals allow ResDMD to detect spectral pollution (spurious modes).}
\label{fig:cylinder1}
\end{figure}

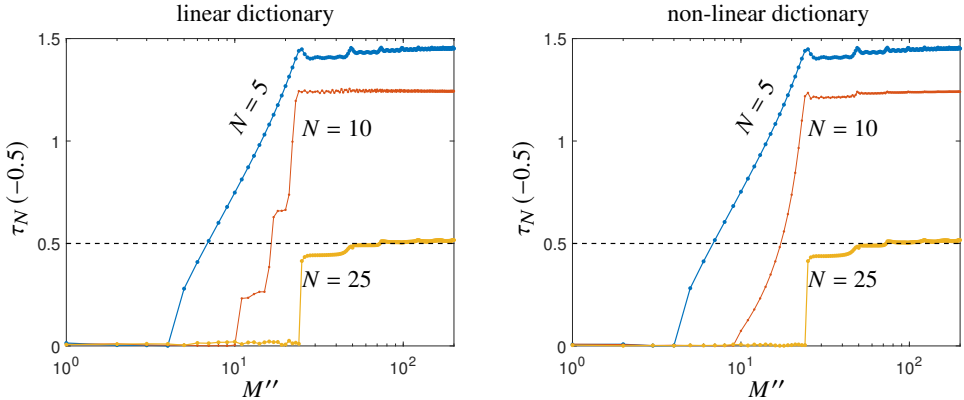
\begin{figure}
 \centering
 \begin{minipage}[b]{0.49\textwidth}
  \begin{overpic}[width=\textwidth,trim={0mm 0mm 0mm 0mm},clip]{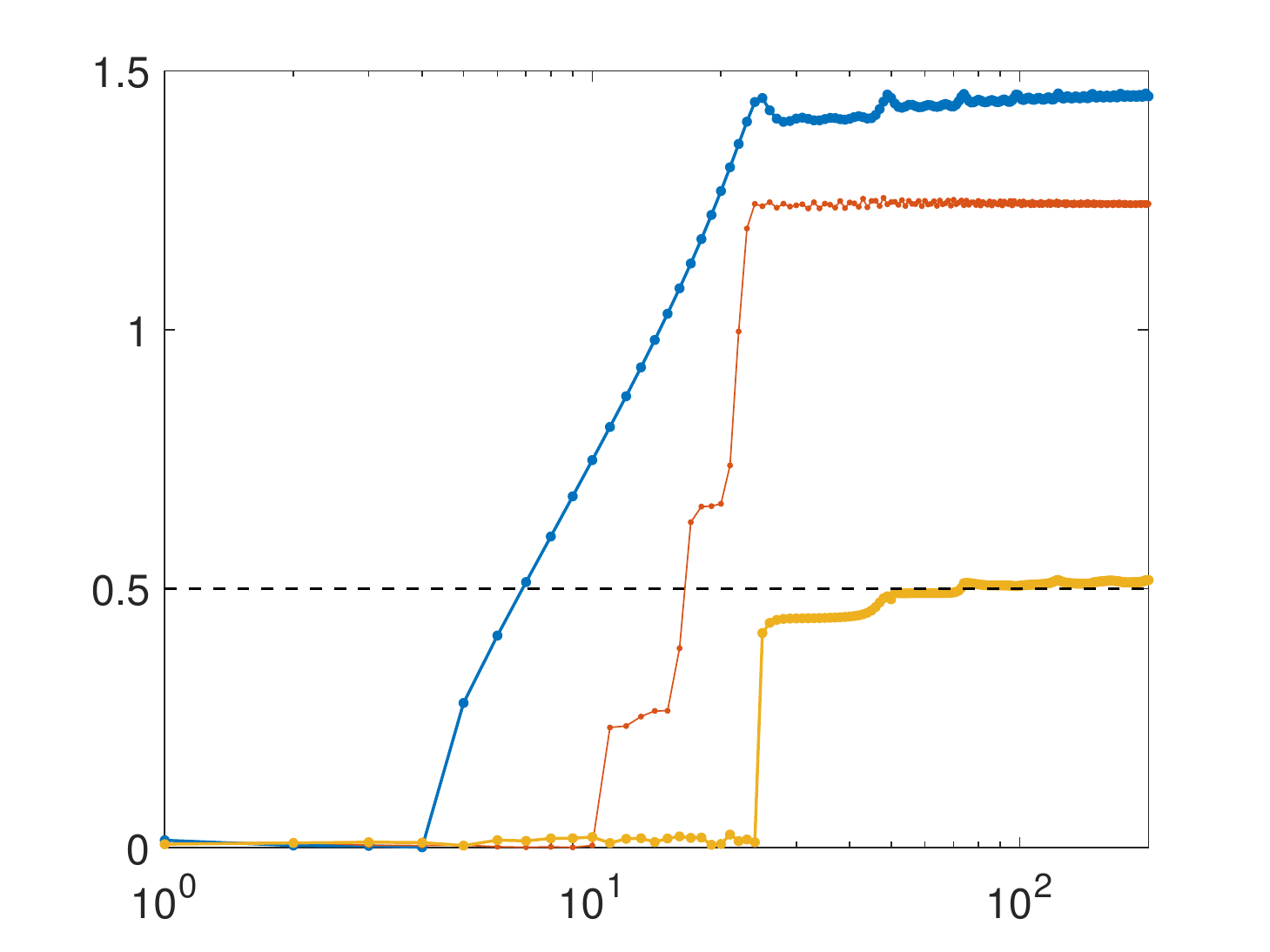}
		\put (35,73) {linear dictionary}
   \put (48,-2) {$M''$}
	\put (45,50) {\rotatebox{60}{$N=5$}}
	\put (1,30) {\rotatebox{90}{$\tau_N(-0.5)$}}
	\put (60,50) {$N=10$}
	\put (60,20) {$N=25$}
   \end{overpic}
 \end{minipage}
\begin{minipage}[b]{0.49\textwidth}
  \begin{overpic}[width=\textwidth,trim={0mm 0mm 0mm 0mm},clip]{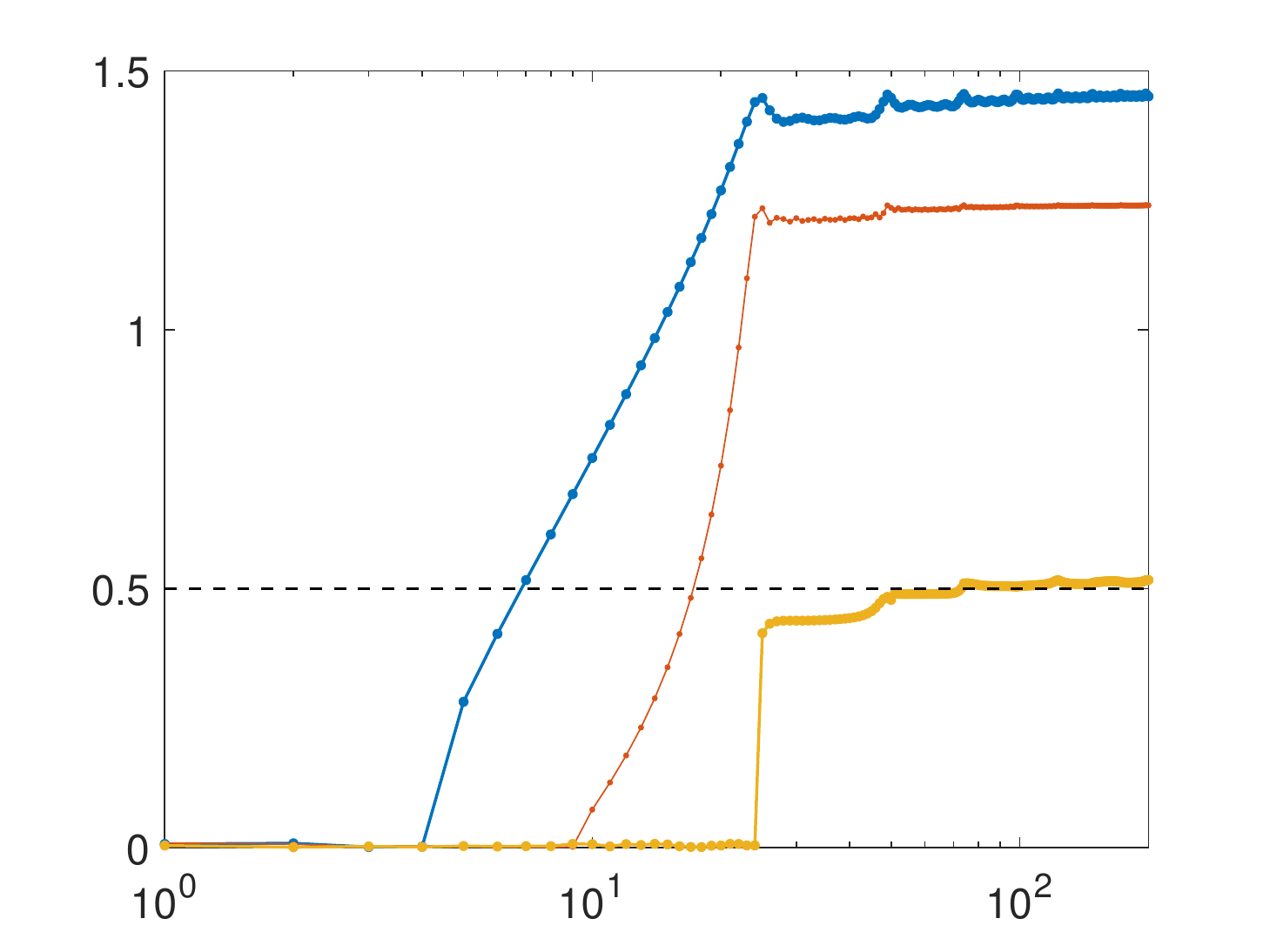}
		\put (32,73) {non-linear dictionary}
   \put (48,-2) {$M''$}
	\put (45,50) {\rotatebox{60}{$N=5$}}
	\put (1,30) {\rotatebox{90}{$\tau_N(-0.5)$}}
	\put (60,50) {$N=10$}
	\put (60,20) {$N=25$}
   \end{overpic}
 \end{minipage}
\vspace{1mm}
\caption{Convergence of $\tau_N(-0.5)$ as $M''$ (the number of snapshots used to build the ResDMD matrices) increases. The plots show the clear monotonicity of $\tau_N(-0.5)$, which decreases to $\|(\mathcal{K}+0.5)^{-1}\|^{-1}=0.5$ as $N$ increases.}
\label{fig:cylinder2}
\end{figure}

\cref{fig:cylinder1} shows values of $\tau_N$ (pseudospectral contours) computed using \cref{alg:res_EDMD}, where $\tau_N$ is the minimal residual in \eqref{tau_factor}. The circular contours show excellent agreement with the distance to the spectrum, which is the unit circle in this example. The spectrum corresponds to the closure of the set of eigenvalues which wrap around the circle. The eigenvalues of the finite $N\times N$ Galerkin matrix $\mathbb{K}$ in each case are shown as red dots. The linear dictionary demonstrates spectral pollution, i.e., ``spurious modes'', which are easily detected by ResDMD (e.g., \cref{alg:mod_EDMD}). \cref{fig:cylinder2} shows the convergence of $\tau_N(-0.5)$ (see \eqref{tau_factor}) as we vary $N$ and $M''$. As expected, we see convergence as $M''\rightarrow\infty$. Moreover, $\tau_N(-0.5)$ decreases and converges monotonically down to $\|(\mathcal{K}+0.5)^{-1}\|^{-1}=0.5$ as $N$ increases.

\begin{figure}
 \centering
\begin{minipage}[b]{0.49\textwidth}
  \begin{overpic}[width=\textwidth,trim={0mm 0mm 0mm 0mm},clip]{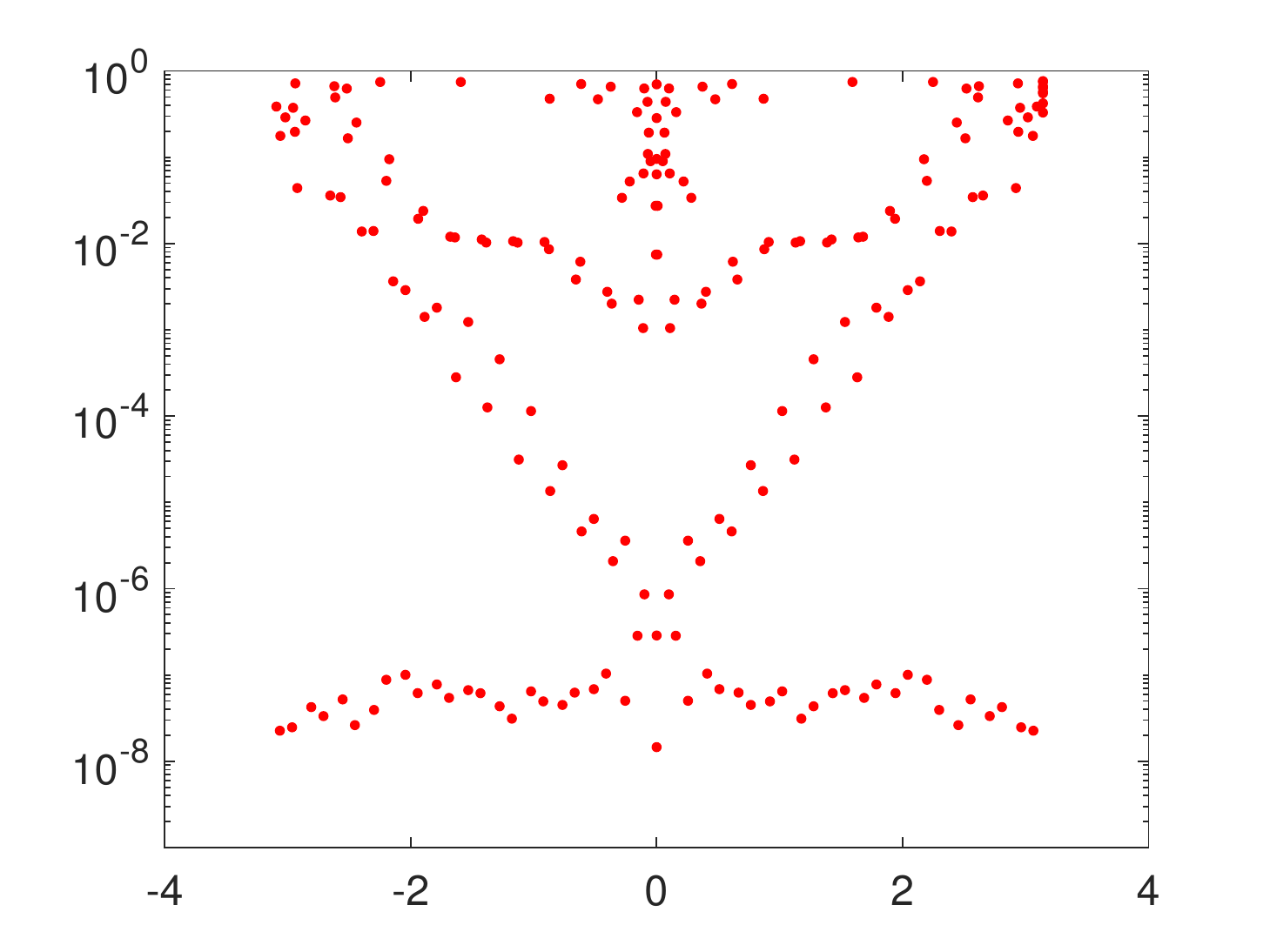}
		\put (24,73) {$\tau_{200}(\lambda_j)$, linear dictionary}
   \put (42,-2) {$\mathrm{phase}(\lambda_j)$}
	\put (67,37) {{\textcolor[rgb]{1,0,1}{\textbf{spectral}}}}
		\put (67,32) {{\textcolor[rgb]{1,0,1}{\textbf{pollution}}}}
	\put(66,39)  {\color[rgb]{1,0,1}\vector(-1,1){26}}
	\put(68,40)  {\color[rgb]{1,0,1}\vector(-1,3){8}}
   \end{overpic}
 \end{minipage}
\begin{minipage}[b]{0.49\textwidth}
  \begin{overpic}[width=\textwidth,trim={0mm 0mm 0mm 0mm},clip]{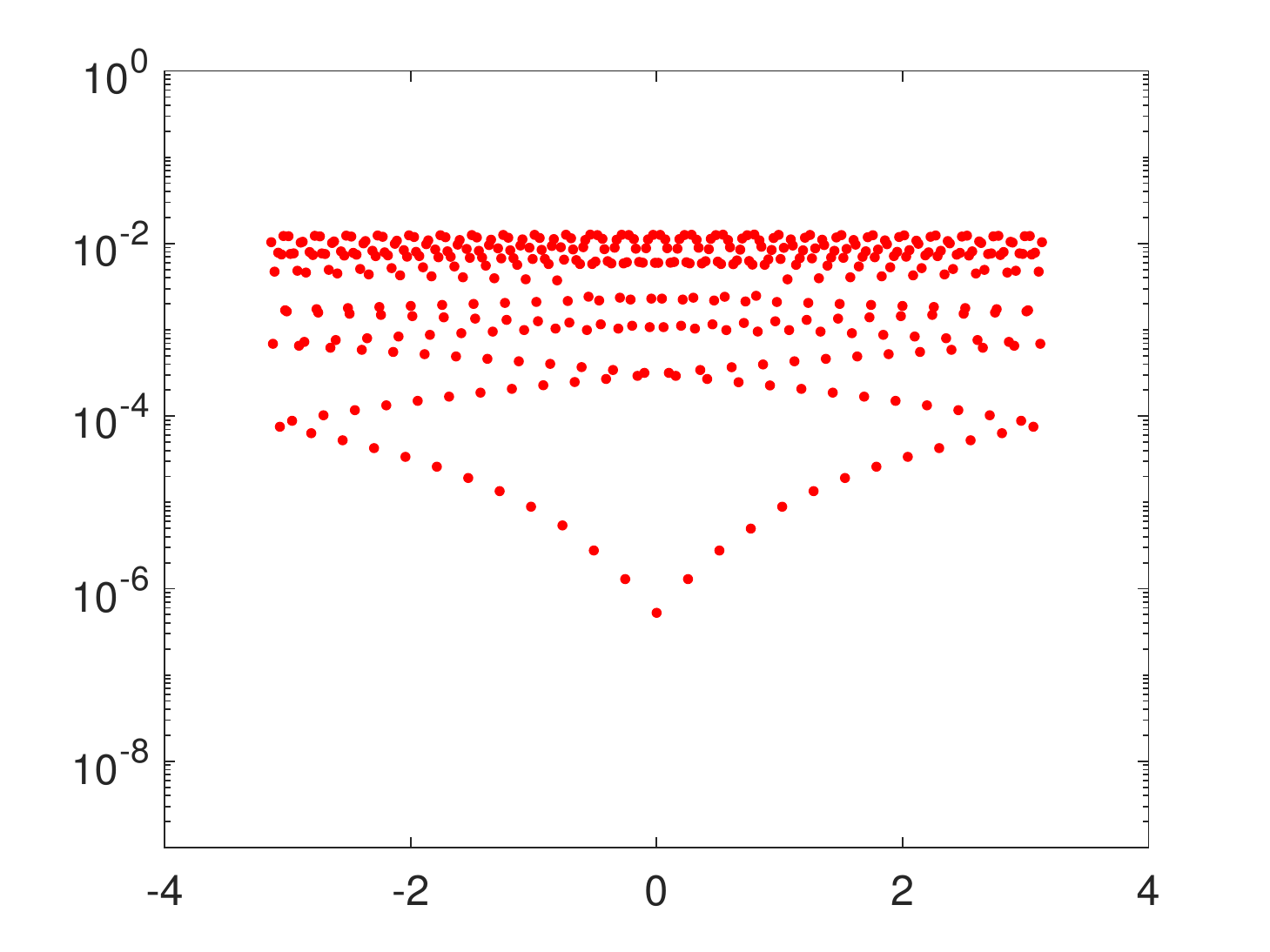}
		\put (20,73) {$\tau_{400}(\lambda_j)$, non-linear dictionary}
   \put (42,-2) {$\mathrm{phase}(\lambda_j)$}
	\put (40,15) {{\textcolor[rgb]{0.75,0.5,0}{\textbf{branch $0$}}}}
	\put(45,19)  {\color[rgb]{0.75,0.5,0}\vector(0,2){14}}
	\put(59,19)  {\color[rgb]{0.75,0.5,0}\vector(0,2){14}}

	\put (68,28) {{\textcolor[rgb]{0,0,1}{\textbf{branch $1$}}}}
	\put(76,32)  {\color[rgb]{0,0,1}\vector(-3,1){33}}
	\put(80,32)  {\color[rgb]{0,0,1}\vector(-1,1){14}}
	
	\put (15,28) {{\textcolor[rgb]{0,0.58,0}{\textbf{branch $-1$}}}}
	\put(30,32)  {\color[rgb]{0,0.58,0}\vector(3,1){33}}
	\put(24,32)  {\color[rgb]{0,0.58,0}\vector(1,1){14}}
   \end{overpic}
 \end{minipage}
\vspace{2mm}
	  \caption{Phase-residual plot of the eigenvalues of $\mathbb{K}$. The linear dictionary has very small residuals for the lower modes, yet also suffers from severe spectral pollution. The non-linear dictionary demonstrates the lattice structure of the Koopman eigenvalues. Branches of the phase as the eigenvalues wrap around the unit circle are labelled.}
\label{fig:cylinder3}
\end{figure}

\cref{fig:cylinder3} shows the eigenvalues of the finite $N\times N$ Galerkin $\mathbb{K}$ as a phase-residual plot. Some of the eigenvalues computed using the linear dictionary have very small relative residuals of approximately $10^{-8}$. Due to the squaring involved in computing the relative residual in \eqref{eq:abs_res}, these small residuals are effectively the level of machine precision. The linear dictionary also has severe spectral pollution. In contrast, the non-linear dictionary captures the lattice structure of the eigenvalues much better. We have labelled the different branches of the phase as the eigenvalues (powers of the fundamental eigenvalues) wrap around the unit circle.

\begin{figure}
 \centering
 \begin{minipage}[b]{0.49\textwidth}
  \begin{overpic}[width=\textwidth,trim={0mm 0mm 0mm 0mm},clip]{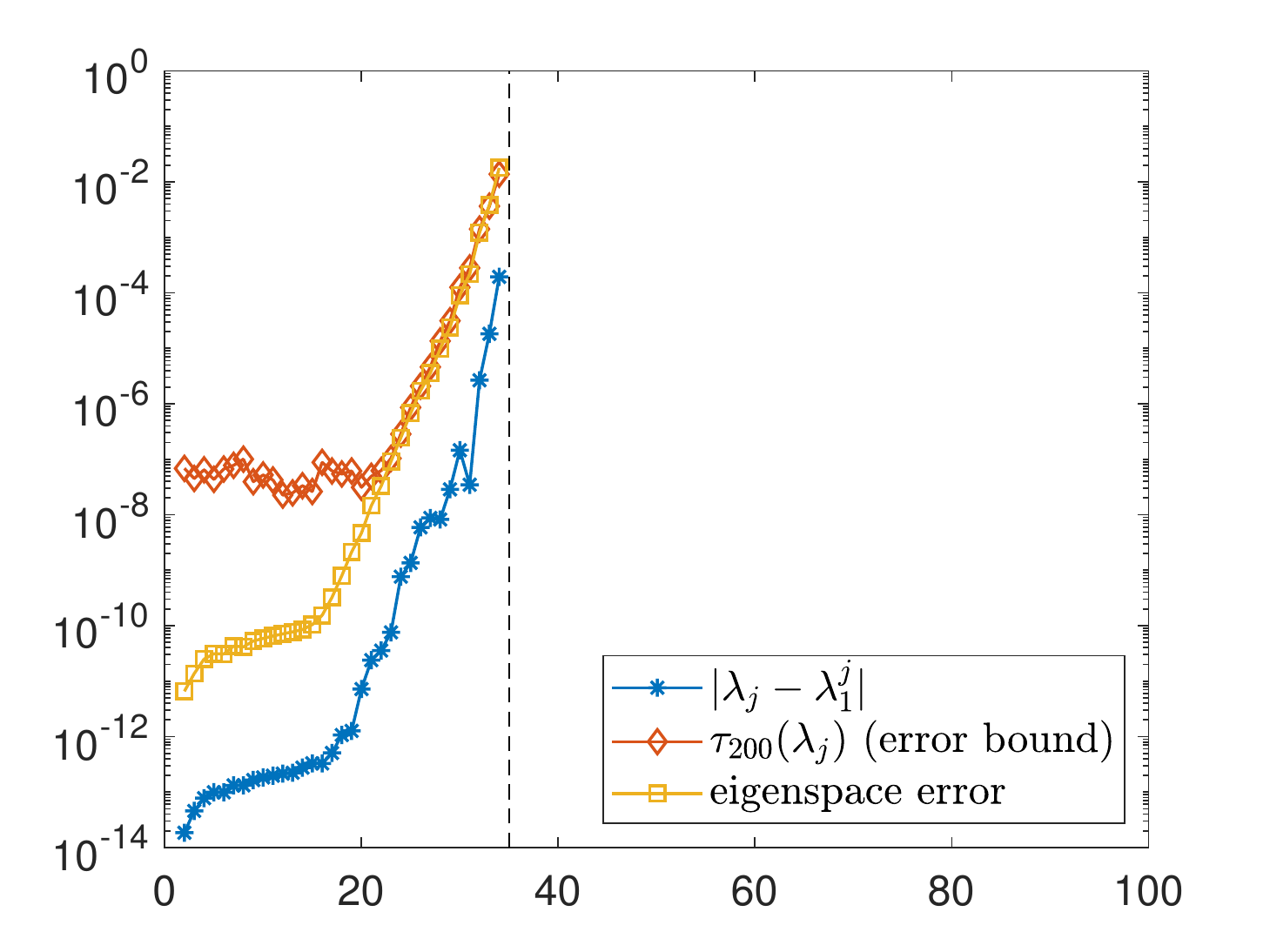}
		\put (35,73) {linear dictionary}
   \put (33,-2) {$j$ (mode number)}
   \end{overpic}
 \end{minipage}
\begin{minipage}[b]{0.49\textwidth}
  \begin{overpic}[width=\textwidth,trim={0mm 0mm 0mm 0mm},clip]{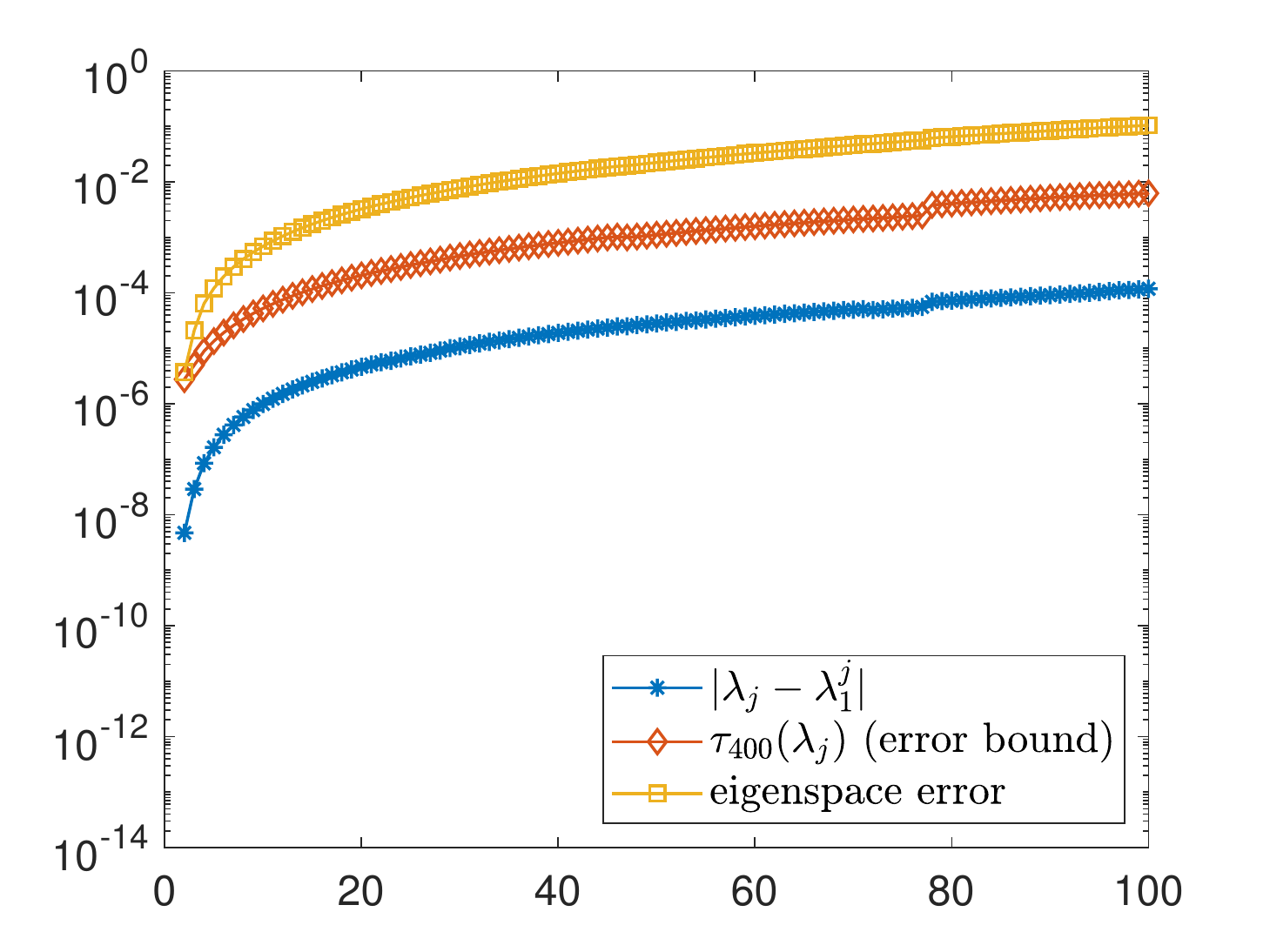}
		\put (32,73) {non-linear dictionary}
   \put (33,-2) {$j$ (mode number)}
   \end{overpic}
 \end{minipage}
\vspace{1mm}
	  \caption{Errors of the computed spectral information of each mode.}
\label{fig:cylinder4}
\end{figure}

To investigate these points further, \cref{fig:cylinder4} plots the errors of the eigenvalues, the minimal residuals, and the errors of the associated eigenspaces. For each case of dictionary, we first compute a reference eigenvalue $\lambda_1\approx0.9676 + 0.2525i$ corresponding to the first mode, and the corresponding eigenfunction $g_1$. For each $j$, we compare the computed eigenvalue $\lambda_j$ with $\lambda_1^j$. The computed residual satisfies $\tau_{N}(\lambda_j)\geq |\lambda_j-\lambda_1^j|$, confirming that \cref{alg:mod_EDMD} provides \textit{error bounds} on the computed spectra. We evaluate $g_1^j$ and $g_j$ at the points $\{\hat{\pmb{x}}^{(m)}\}_{m=1}^{M''}$, and the ``eigenspace error'' corresponds to the subspace angle between the linear spans of the resulting vectors~\citep[Eq. (1.4)]{colbrook2019infinite}. The linear dictionary does an excellent job at capturing the lower spectral content, up to about $j=35$, but is unable to capture the strongly non-linear eigenfunctions. In contrast, the non-linear dictionary does a much better job at capturing the higher-order spectral information. For this problem, only a few Koopman modes are needed to reconstruct the flow. However, for some problems, having non-linear functions in the dictionary is essential to capture the dynamics (e.g., see \citet{brunton2016koopman}, and our examples in \cref{sec:verif_method_example_3,sec:verif_method_example_4}). One is completely free to choose the dictionary used in ResDMD. For example, one could also use a mixture of the DMD and kEDMD computed dictionaries, or other methods entirely.

\begin{figure}
 \centering
 \begin{minipage}[b]{0.32\textwidth}
  \begin{overpic}[width=\textwidth,trim={15mm 0mm 12mm 0mm},clip]{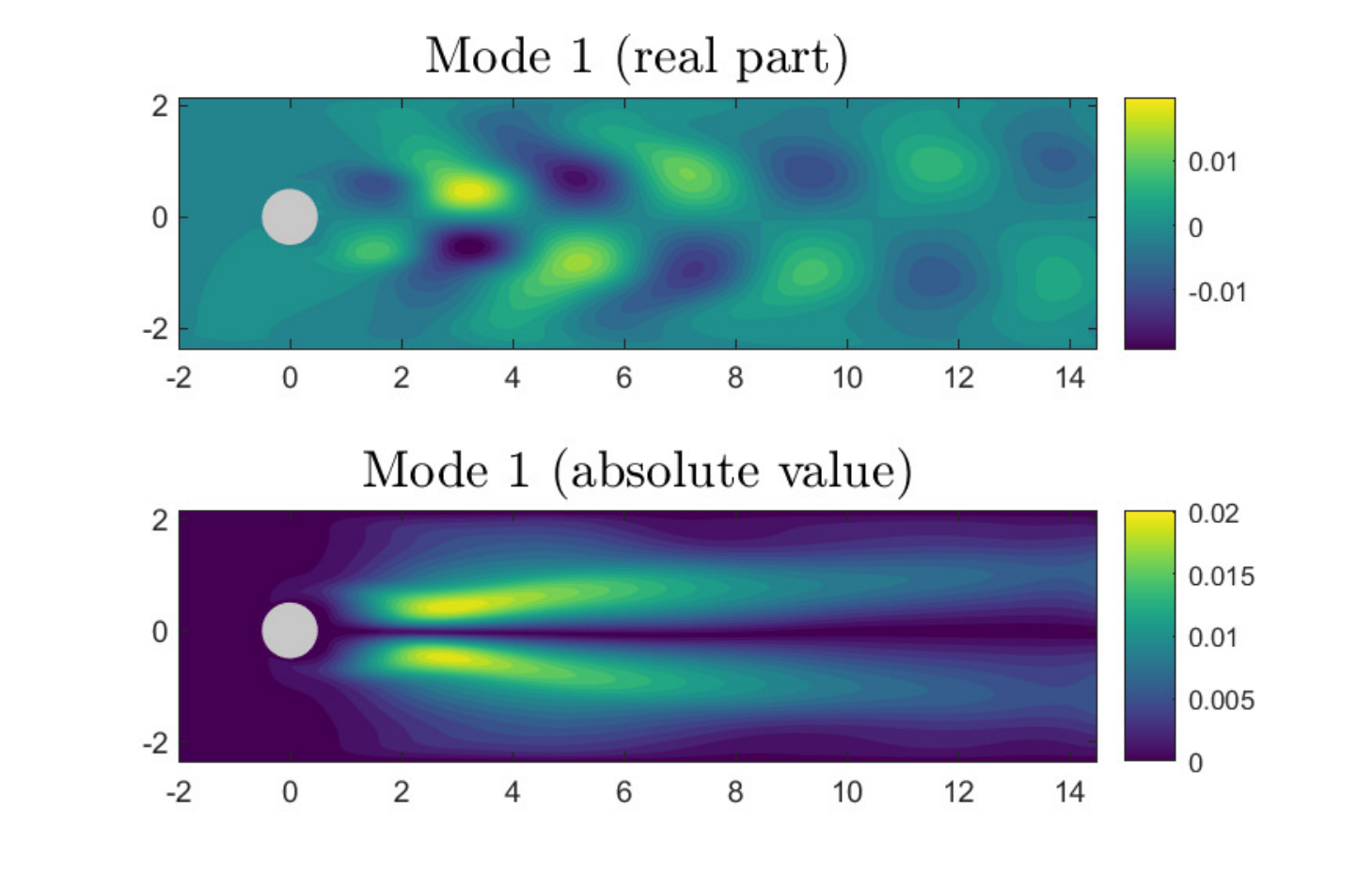}
   \end{overpic}
 \end{minipage}
\begin{minipage}[b]{0.32\textwidth}
  \begin{overpic}[width=\textwidth,trim={15mm 0mm 12mm 0mm},clip]{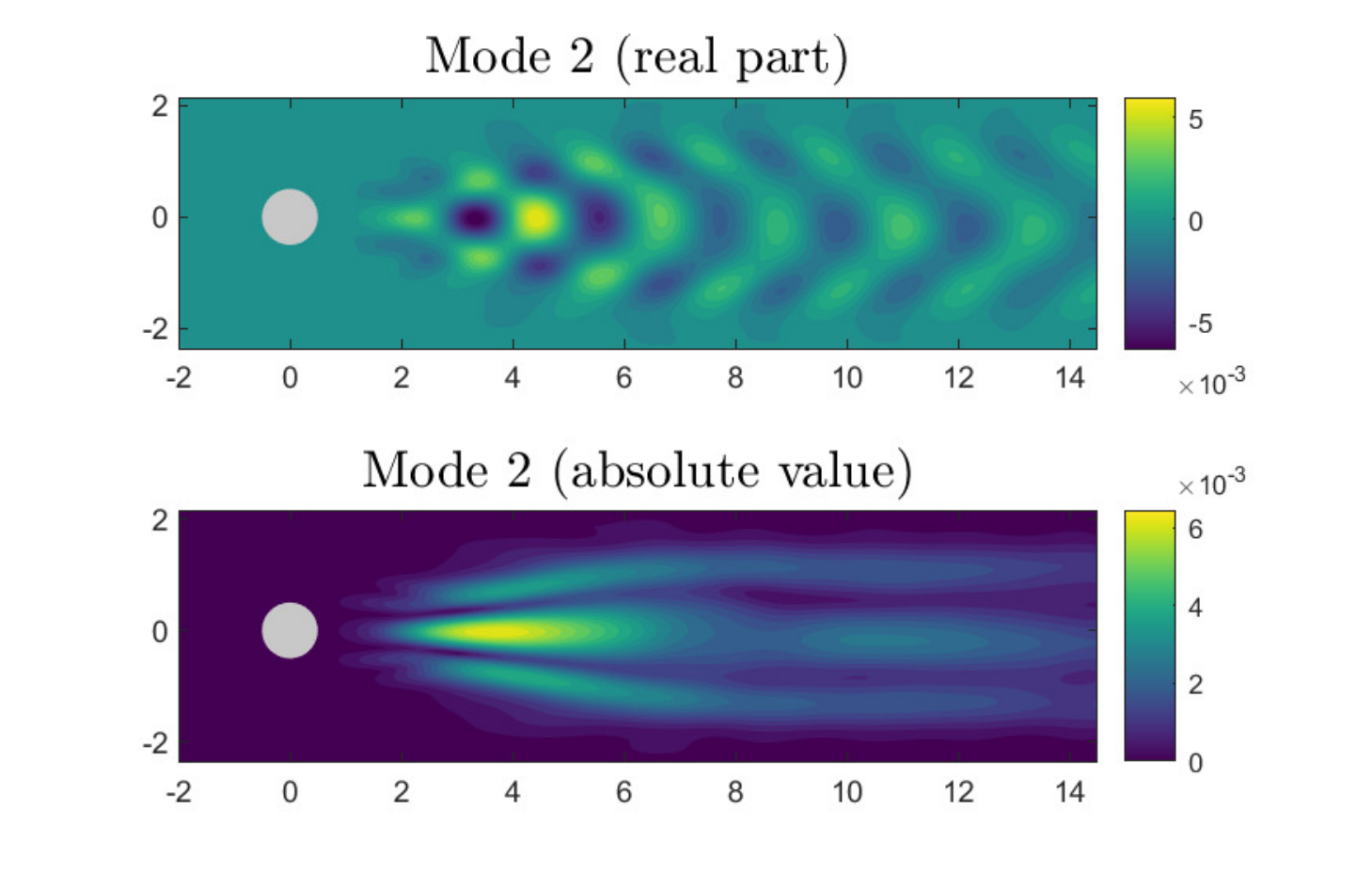}
   \end{overpic}
 \end{minipage}
\begin{minipage}[b]{0.32\textwidth}
  \begin{overpic}[width=\textwidth,trim={15mm 0mm 12mm 0mm},clip]{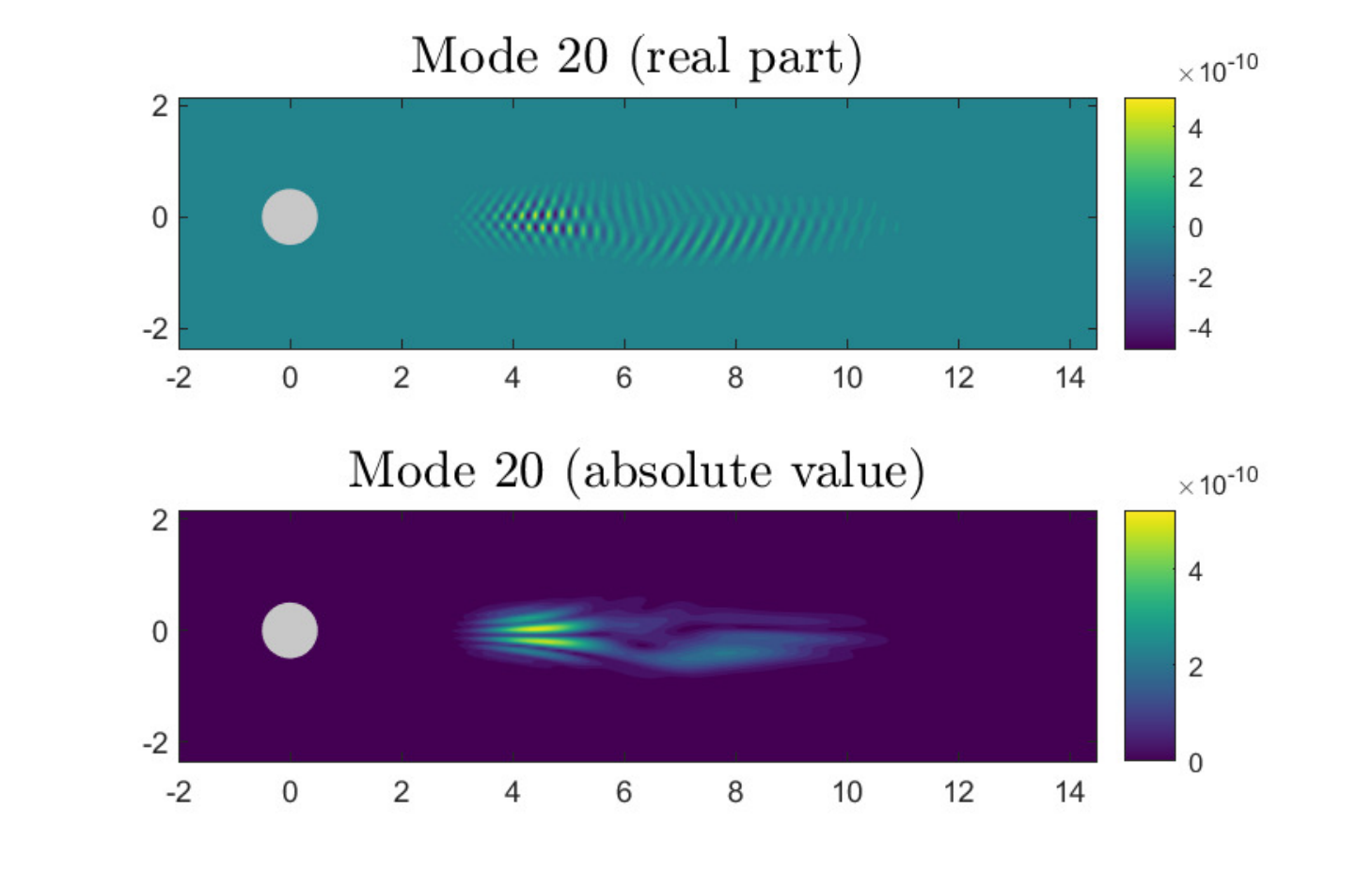}
   \end{overpic}
 \end{minipage}
	  \caption{Koopman modes for the cylinder wake, computed using the linear dictionary.}
\label{fig:cylinder5}
\end{figure}

\begin{figure}
 \centering
 \begin{minipage}[b]{0.32\textwidth}
  \begin{overpic}[width=\textwidth,trim={15mm 0mm 12mm 0mm},clip]{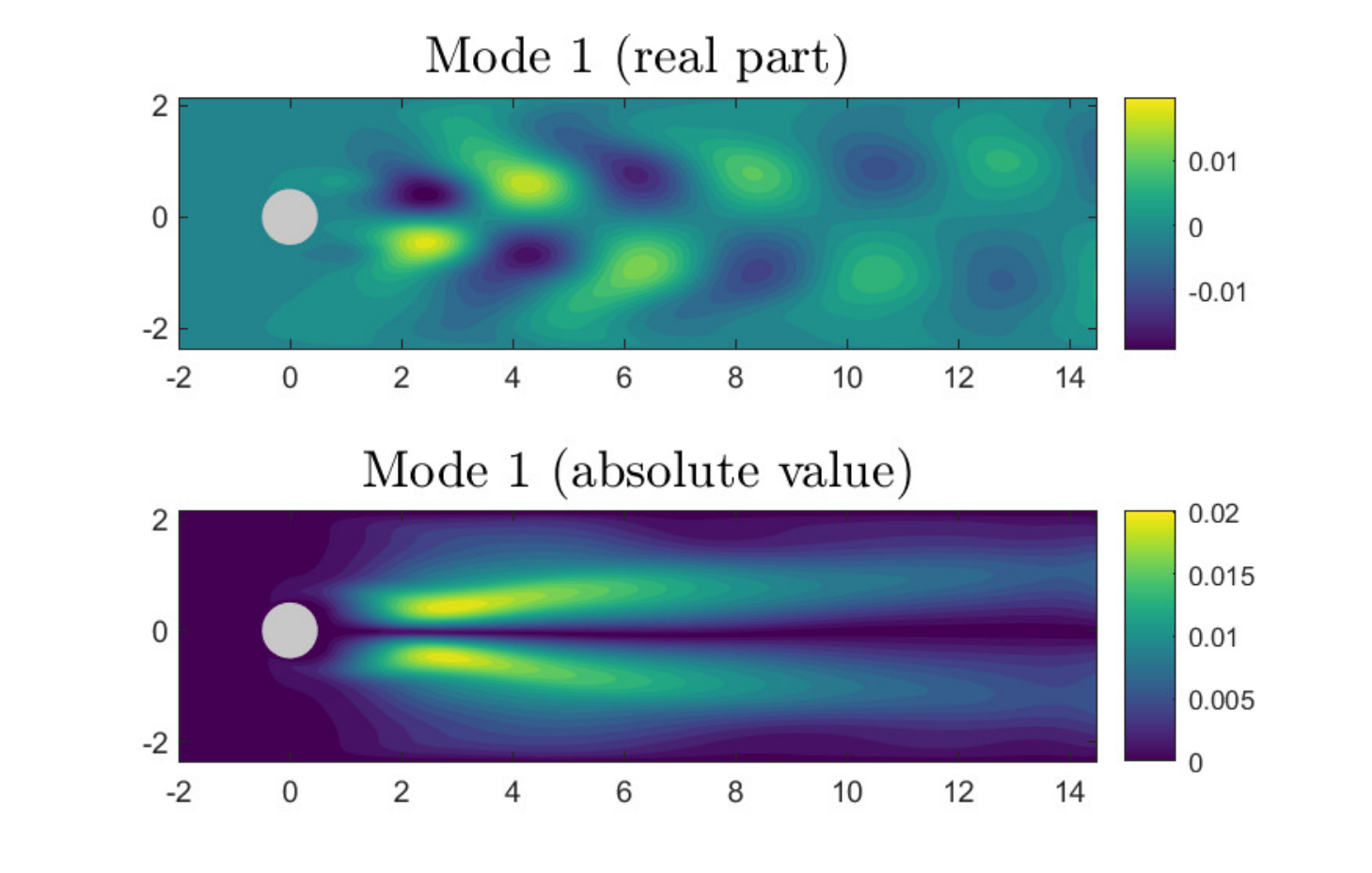}
   \end{overpic}
 \end{minipage}
\begin{minipage}[b]{0.32\textwidth}
  \begin{overpic}[width=\textwidth,trim={15mm 0mm 12mm 0mm},clip]{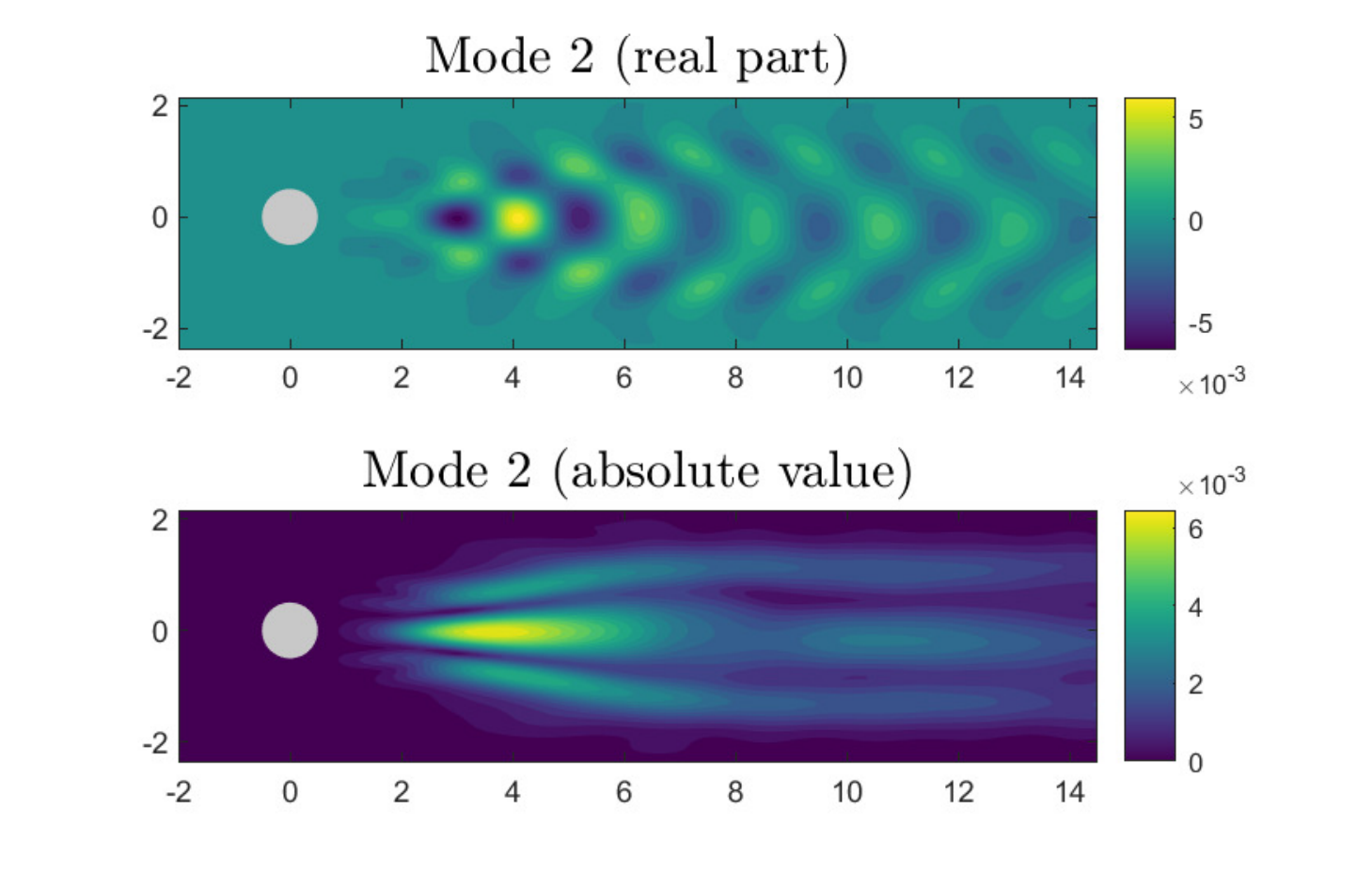}
   \end{overpic}
 \end{minipage}
\begin{minipage}[b]{0.32\textwidth}
  \begin{overpic}[width=\textwidth,trim={15mm 0mm 12mm 0mm},clip]{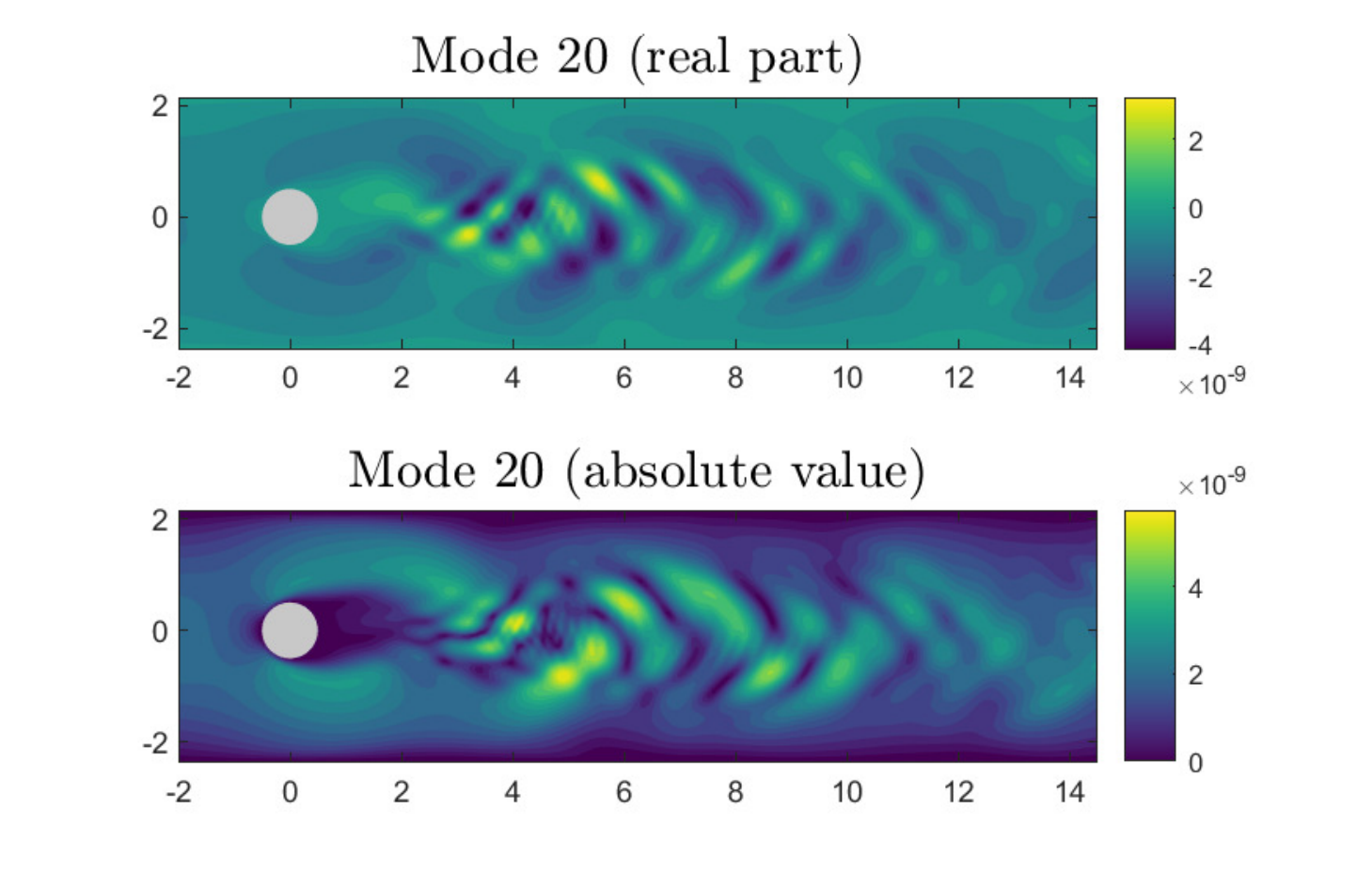}
   \end{overpic}
 \end{minipage}
	  \caption{Koopman modes for the cylinder wake, computed using the non-linear dictionary (see graphical abstract).}
\label{fig:cylinder6}
\end{figure}

\cref{fig:cylinder5} and \cref{fig:cylinder6} show examples of Koopman modes (see \eqref{koop_mode_estimate}) for the $x$ component of the velocity, computed using the linear and non-linear dictionaries, respectively. To compare the two dictionaries, each Koopman eigenfunction was normalised so that the vector $\sqrt{W}\Psi_XV(:,j)$ has norm $1$. Modes $1$ and $2$ show excellent agreement between the two choices of dictionary, and can also be compared to~\citet[Figure 3]{taira2020modal}. However for the higher mode (Mode 20), the two results bear little similarity at all. The Koopman modes correspond to the vector-valued expansion coefficients of the state in the approximate eigenfunctions as opposed to the eigenfunctions themselves. Thus, the difference indicates that for high-order modes, non-linearity becomes important in this expansion. We can be assured by the residual measure in the ResDMD algorithm that the modes arising from using the non-linear dictionary are physical features of the flow and not spurious. ResDMD therefore brings certainty to the high-order modal analysis of this classic example, which can be lacking in prior DMD or EDMD approaches.

\section{Example II: Embedded shear layer in a turbulent boundary layer flow}
\label{sec:verif_method_example_2}

We now turn to using our method to analyse a high-Reynolds number turbulent flow with a fine temporal resolution.
We consider the turbulence structure within a turbulent boundary layer over a flat plate both with and without an embedded shear profile. A shear profile is achieved by permitting a steady flow to be injected through a section of the plate. Turbulent boundary layer flow can be considered a challenging test case for DMD algorithms, particularly when assessing potential non-linear flow behaviours, and the wide range of length scales present in a high-Reynolds number flow problem. The embedded shear layer, on the other hand, is anticipated to have a set of characteristic length scales with a broadband energy distribution. 

\subsection{Experimental setup}

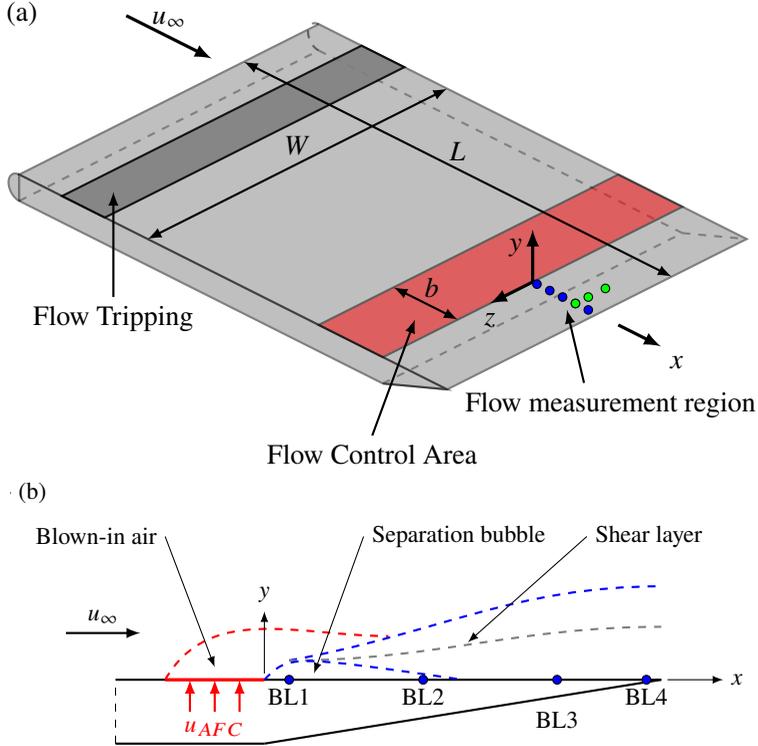
\begin{figure}
\centering
\tdplotsetmaincoords{120}{-45}
\begin{tikzpicture}[tdplot_main_coords,scale=0.8]
	\tikzstyle{every node}=[font=\normalsize]
	\def\W{7}  
	\def\L{10} 
	\def\T{0.5}  
	\def\Inst{1.5} 
	\def\csyslen{1} 
	\def\TEalpha{12} 

	\filldraw[draw=black,thick,fill=gray,opacity=0.5]%
	          (0,-\W/2,0) -- (0,\W/2,0)-- (\L,\W/2,0) -- (\L,-\W/2,0) -- cycle;
	\draw[color=gray,thick,dashed] %
	          (0,\W/2,-\T) -- (0, -\W/2,-\T)  -- (\L-\Inst, -\W/2,-\T)-- (\L-\Inst, \W/2,-\T);
	\draw[color=gray,thick] (\L-\Inst,\W/2,-\T) -- (\L, \W/2,0);
	\draw[color=gray,thick,dashed] (\L-\Inst,-\W/2,-\T) -- (\L, -\W/2,0);

	\filldraw[draw=black,thick,fill=gray,opacity=0.5,domain=-90:90] plot %
	          ({-\T/2*cos(\x)}, {\W/2}, {-\T/2+\T/2*sin(\x)});	
	\draw[draw=black,thick,opacity=0.5,domain=-90:90, dashed] plot %
	          ({-\T/2*cos(\x)},{-\W/2},{-\T/2+\T/2*sin(\x)});	
	\filldraw[draw=black,thick,fill=gray,opacity=0.5]%
	          (0,\W/2,0) -- (\L,\W/2,0) -- (\L-\Inst,\W/2,-\T) -- (0,\W/2,-\T) -- cycle;

	\filldraw[draw=black,thick,fill=red,opacity=0.5]%
	(\L-\Inst,-\W/2,0) -- (\L-\Inst,\W/2,0) -- (\L-2*\Inst,\W/2,0) -- (\L-2*\Inst,-\W/2,0) -- cycle;

	\filldraw[draw=black,thick,fill=gray,opacity=0.9](0.1*\L,-\W/2,0) -- (0.1*\L,\W/2,0) -- (0.2*\L,\W/2,0) -- (0.2*\L,-\W/2,0) -- cycle;

	\foreach \x in {8.6,8.9,...,9.9}{
		\filldraw[fill=blue](\x,0,0) circle (2pt);
		}
	\filldraw[fill=green](9.5, 0.0,0) circle (2pt);
	\filldraw[fill=green](9.5,-0.3,0) circle (2pt);
	\filldraw[fill=green](9.5,-0.7,0) circle (2pt);
		
	\draw[thick, latex-, color=black] %
	({\L-1.5*\Inst}, 2, -0.2) -- ({\L-1.5*\Inst}, 3, -2) node[below]{Flow Control Area} ;

	\draw[thick, latex-, color=black] ({0.15*\L}, 0.4*\W, -0.0) -- ({0.15*\L}, 0.4*\W, -2)node[below]{Flow Tripping};

	\draw[thick, latex-, color=black] (0.93*\L, 0.0*\W, -0.2) -- (0.9*\L, -0.1*\W, -2)node[below]{\rule{20pt}{0pt}Flow measurement region} ;

	\draw[very thick, -latex, color=black]  (0, 1, 2) -- (2, 1, 2) node[pos=0.5,above]{$u_\infty$};

	\draw[thick, latex-latex, color=black] (0, -0.25*\W, 0) -- (1.0*\L, -0.25*\W, 0) node[pos=0.5,above]{$L$};
	\draw[thick, latex-latex, color=black] (0.3*\L, -0.5*\W, 0) -- (0.3*\L, 0.5*\W, 0) node[pos=0.5,above]{$W$};
	\draw[thick, latex-latex, color=black] (\L-\Inst, 0.25*\W, 0) -- (\L-2*\Inst, 0.25*\W, 0) node[pos=0.4,above]{$b$};

	\draw[very thick, -latex, color=black] (\L-\Inst+2, 0, 0) -- (\L-\Inst+\csyslen+2, 0, 0) node[below right]{$x$};
	\draw[very thick, -latex, color=black] (\L-\Inst, 0, 0) -- (\L-\Inst, \csyslen, 0) node[below]{$z$};
	\draw[very thick, -latex, color=black] (\L-\Inst, 0, 0) -- (\L-\Inst, 0, \csyslen) node[below left]{$y$};

\draw[color=black](-4,0,0)--(-4,0,0) node[right]{(a)};
\end{tikzpicture}
\begin{tikzpicture}[scale=0.82]
  \def\Thickness{3}
  \def\TEAngle{20}
  \def\PlateLength{8}
  \draw[color=black,thick]({-\PlateLength*1.1},0)--(0,0);
  \draw[color=black,thick]({-\PlateLength*0.8},{-sin(\TEAngle)*\Thickness})--(0,0);
  \draw[color=black,thick]({-\PlateLength*0.8},{-sin(\TEAngle)*\Thickness})--(-1.1*\PlateLength,{-sin(\TEAngle)*\Thickness});
  \draw[color=black,thin,dashed](-\PlateLength*1.1,{-sin(\TEAngle)*\Thickness})--(-\PlateLength*1.1,0);
  \draw[color=red,very thick]({-1.0*\PlateLength},0)--({-0.8*\PlateLength},0);
  \draw[color=red, thick,dashed] ({-1.00*\PlateLength},   0) to [out=60, in=180] ({-0.55*\PlateLength},0.7); 
  \draw[color=blue,thick,dashed] ({-0.75*\PlateLength},0.31) to [out=10, in=180] (0,1.5); 
  \draw[color=gray,thick,dashed] ({-0.75*\PlateLength},0.31) to [out=0, in=180] (0,0.85); 
  \draw[color=blue,thick,dashed] ({-0.80*\PlateLength},0   ) .. controls({-0.75*\PlateLength},0.4) .. ({-0.40*\PlateLength}, 0); 
  %
  \node[draw,white] at (0, 2.8)   (a) {A};
  \draw[color=black,thin,latex-]({-\PlateLength*0.390},0.60)--({-\PlateLength*0.15},2.3) node[right]{Shear layer};
  \draw[color=black,thin,latex-]({-\PlateLength*0.7},0.1)--({-\PlateLength*0.6},2.3) node[right, align=left]{Separation bubble};
  \draw[color=black,thin,latex-]({-\PlateLength*0.9},0.20)--({-\PlateLength*1.0},2.3) node[left]{Blown-in air};
  %
  \filldraw[fill=blue,draw=black]({-\PlateLength*0.75},0) circle (2pt) node[below]{BL1};
  \filldraw[fill=blue,draw=black]({-\PlateLength*0.48},0) circle (2pt) node[below]{BL2};
  \filldraw[fill=blue,draw=black]({-\PlateLength*0.21},0) circle (2pt) node[below=0.3cm]{BL3};
  \filldraw[fill=blue,draw=black]({-\PlateLength*0.03},0) circle (2pt) node[below]{BL4};

  %
  %
  \draw[color=black,thick,-latex]({-\PlateLength*1.2},0.25*\Thickness)--({-\PlateLength*1.05},0.25*\Thickness) node[pos=0.5,above]{$u_\infty$};
  \draw[color=red,thick,latex-]({-\PlateLength*0.85},0)--({-\PlateLength*0.85},{-0.9*sin(35)});
  \draw[color=red,thick,latex-]({-\PlateLength*0.95},0)--({-\PlateLength*0.95},{-0.9*sin(35)});
  \draw[color=red,thick,latex-]({-\PlateLength*0.90},0)--({-\PlateLength*0.90},{-0.9*sin(35)}) node[below]{$u_{AFC}$};
  \draw({-\PlateLength*1.0},0);
  \draw[color=black,thin,-latex](0.1,0)--(1,0) node[right]{$x$};
  \draw[color=black,thin,-latex](-\PlateLength*0.8,0.1)--(-\PlateLength*0.8,1.1) node[above]{$y$};

\draw[color=black](-10.5,3.0)--(-10.5,3.0) node[right]{(b)};

\end{tikzpicture}
\caption{Schematic of the experimental setup for a boundary layer with embedded shear profile, and side view of the boundary layer near the flow control region.}
\label{fig:BLschematic}
\end{figure}

We consider two cases of turbulent boundary layer flow, a canonical (baseline) and one with an embedded shear layer. Both cases are generated experimentally and detailed in \cite{szHoke2020uniform}. For clarity, we briefly recall the key features and illustrate the experimental setup in \cref{fig:BLschematic}. The baseline case consists of a canonical, zero pressure gradient turbulent boundary layer (TBL) passing over a flat plate (no flow control turned on). The friction Reynolds number is $\Rey_\tau = 1400$. In the experiments, the development of a shear layer is triggered using perpendicular flow injection (active flow control) through a finely perforated sheet, denoted by $u_{AFC}$. The velocity field is sensed at several positions downstream of the flow control area by traversing a hot-wire probe across the entire boundary layer. The data gathered using the hot-wire sensor provides a fine temporal description of the flow. For a more detailed description of the experimental setup and the flow field, see~\citet{szHoke2020uniform}. A wide range of downstream positions and flow control severity values were considered in the original study, with the purpose of assessing the effects of flow control on trailing edge noise. Here, the same data is used, but we focus our attention to one streamwise position (labeled as BL3 in \cref{fig:BLschematic}) and consider a flow injection case where a shear layer develops as a result of perpendicular flow injection.

\subsection{Results}

The fine temporal resolution of the flow field enables us to assess the spectral properties of the flow. We use \cref{alg:spec_meas_poly} to calculate the spectral measures of the flow and compare our results to the power spectral density obtained using Welch's power spectral density estimate as described in~\cite{szHoke2020uniform}. Note that for this example, DMD-type methods cannot robustly approximate the Koopman operator of the underlying dynamical system since data is only collected along a single line. However, \cref{alg:spec_meas_poly} can still be used to compute the spectral measures of the Koopman operator.

\begin{figure}
 \centering
 \begin{minipage}[b]{0.49\textwidth}
  \begin{overpic}[width=\textwidth,trim={0mm 0mm 0mm 0mm},clip]{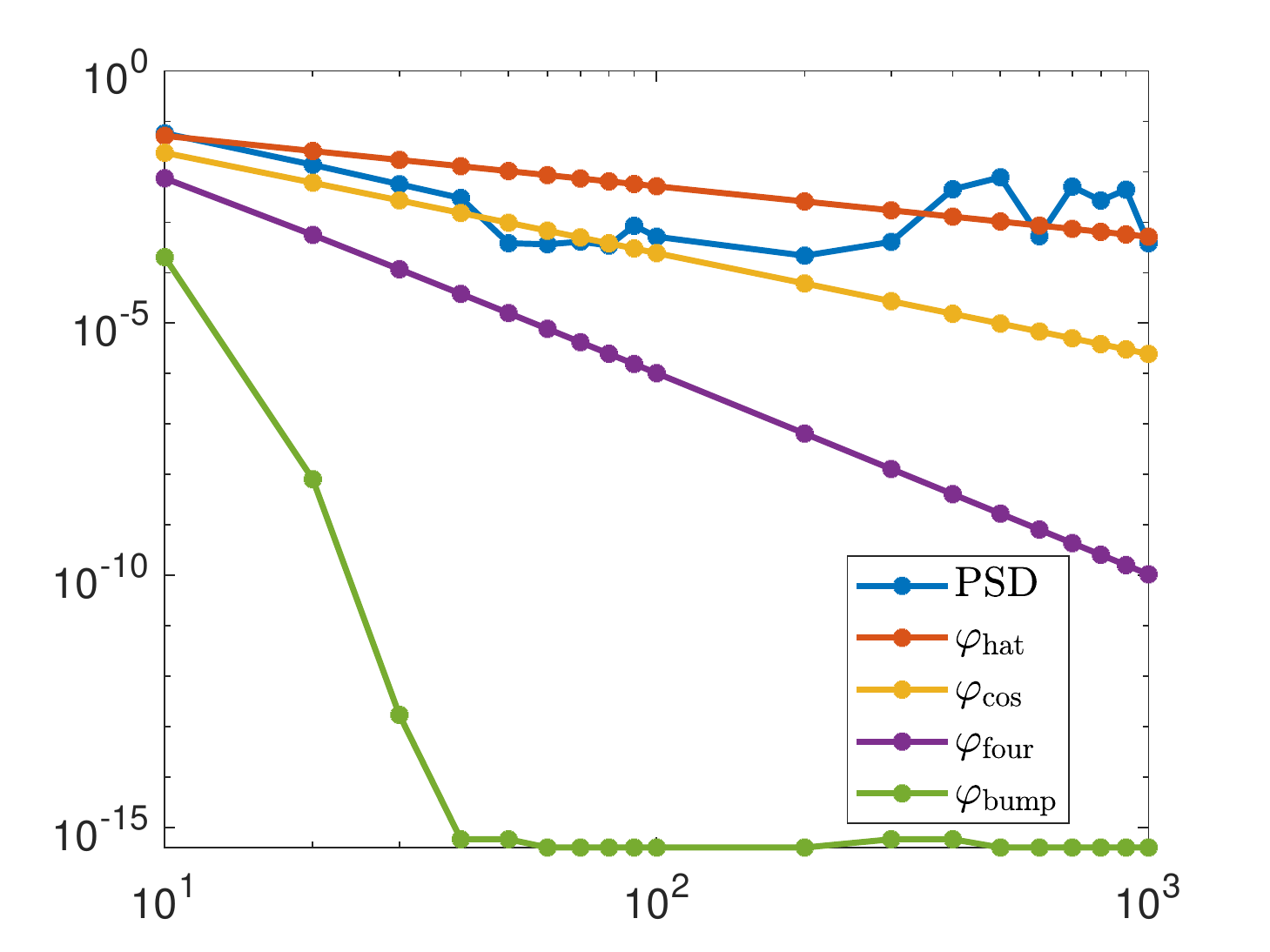}
	\put (27,73) {relative error (injection)}
   \put (47,-2) {$N_{\mathrm{ac}}$}
   \end{overpic}
 \end{minipage}
\begin{minipage}[b]{0.49\textwidth}
  \begin{overpic}[width=\textwidth,trim={0mm 0mm 0mm 0mm},clip]{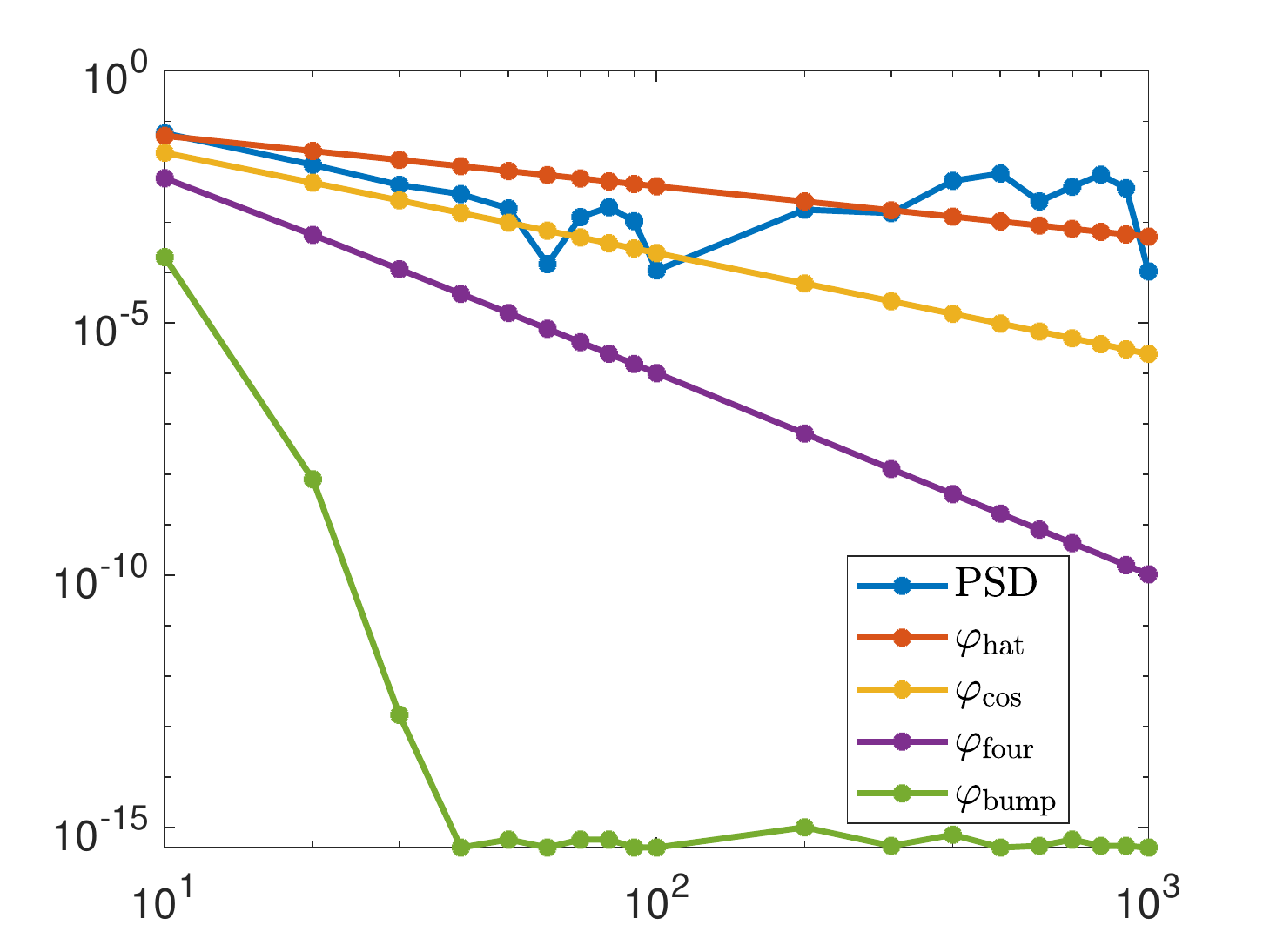}
		\put (27,73) {relative error (baseline)}
   \put (47,-2) {$N_{\mathrm{ac}}$}
   \end{overpic}
 \end{minipage}
\vspace{1mm}
\caption{Relative error in integration against the test function $\phi(\theta)=\exp(\sin(\theta))$ for data collected at $y/\delta_0=0.1000$. \cref{alg:spec_meas_poly} converges at the expected rates for various filter functions. In contrast, the PSD approximation appears to initially converge at a second order rate, but then stagnates.}
\label{fig:hotwire_res0}
\end{figure}

As a first numerical test, we compute spectral measures of the data collected at $y/\delta_0=0.1000$. We then integrate this spectral measure against the test function $\phi(\theta)=\exp(\sin(\theta))$. Rather than being a physical example, this is chosen just to demonstrate the convergence of our method. We consider the integral computed using the power spectral density \eqref{power_def} with a window size of $N_{\mathrm{ac}}$ for direct comparison, and \cref{alg:spec_meas_poly} for choices of filter $\varphi_{\rm hat}(x) = 1-|x|$, $\varphi_{{\rm cos}}(x) =\frac{1}{2}(1-\cos(\pi x))$, $\varphi_{{\rm four}}(x) = 1- x^4(-20|x|^3 + 70x^2 - 84|x| + 35)$ and $\varphi_{{\rm bump}}(x)=\exp\left(-\frac{2}{1-|x|}\exp\left(-\frac{c}{x^4}\right)\right)$ ($c\approx 0.109550455106347$). These filters are first, second, fourth and infinite order, respectively. \cref{fig:hotwire_res0} shows the results, where we see the expected rates of convergence of \cref{alg:spec_meas_poly}. In contrast, the PSD approximation initially appears to converge at a second order rate, and then stagnates.

\begin{figure}
 \centering
 \begin{minipage}[b]{0.49\textwidth}
  \begin{overpic}[width=\textwidth,trim={0mm 0mm 0mm 0mm},clip]{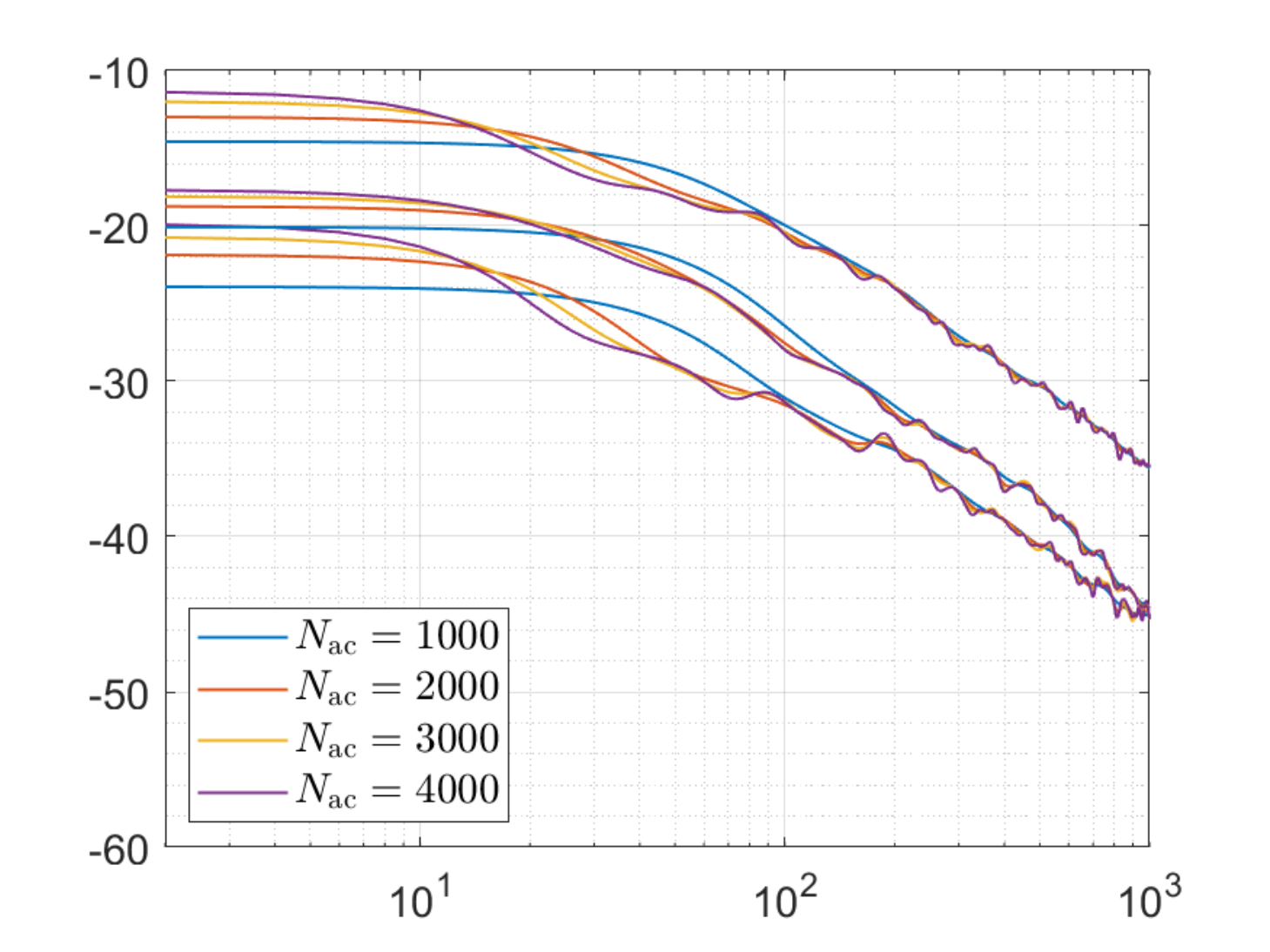}
		\put (38,73) {PSD (injection)}
   \put (37,-2) {frequency (Hz)}
	\put (55,63) {\rotatebox{-30}{$y/\delta_0=0.1000$}}
	\put (50,44) {\rotatebox{-30}{$y/\delta_0=0.0088$}}
	\put (65,48) {\rotatebox{-33}{$y/\delta_0=1.3529$}}
   \end{overpic}
 \end{minipage}
\begin{minipage}[b]{0.49\textwidth}
  \begin{overpic}[width=\textwidth,trim={0mm 0mm 0mm 0mm},clip]{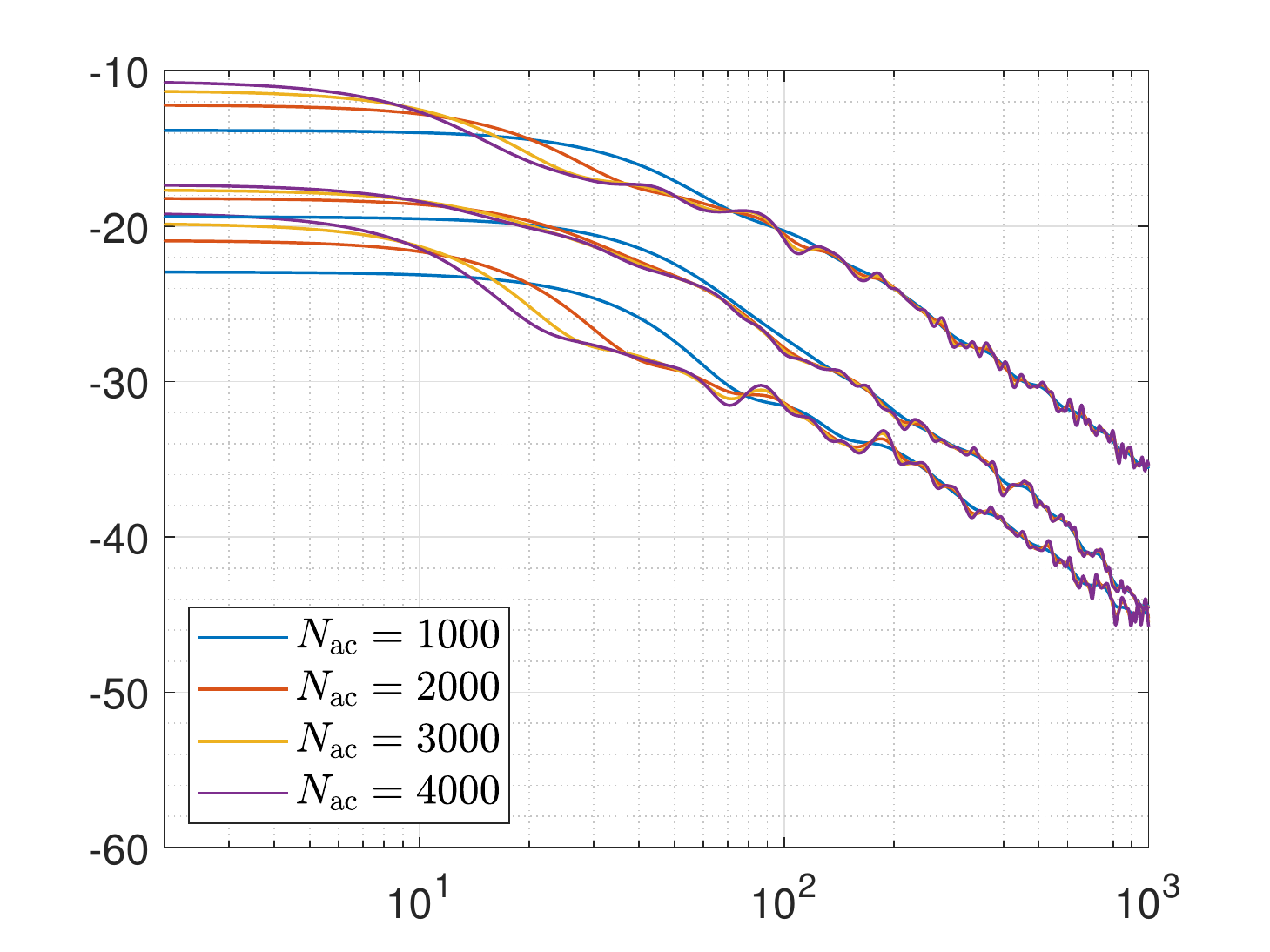}
		\put (25,73) {spectral measure (injection)}
   \put (37,-2) {frequency (Hz)}
	\put (55,63) {\rotatebox{-30}{$y/\delta_0=0.1000$}}
	\put (50,44) {\rotatebox{-30}{$y/\delta_0=0.0088$}}
	\put (65,48) {\rotatebox{-33}{$y/\delta_0=1.3529$}}
   \end{overpic}
 \end{minipage}\\
\vspace{5mm}
\begin{minipage}[b]{0.49\textwidth}
  \begin{overpic}[width=\textwidth,trim={0mm 0mm 0mm 0mm},clip]{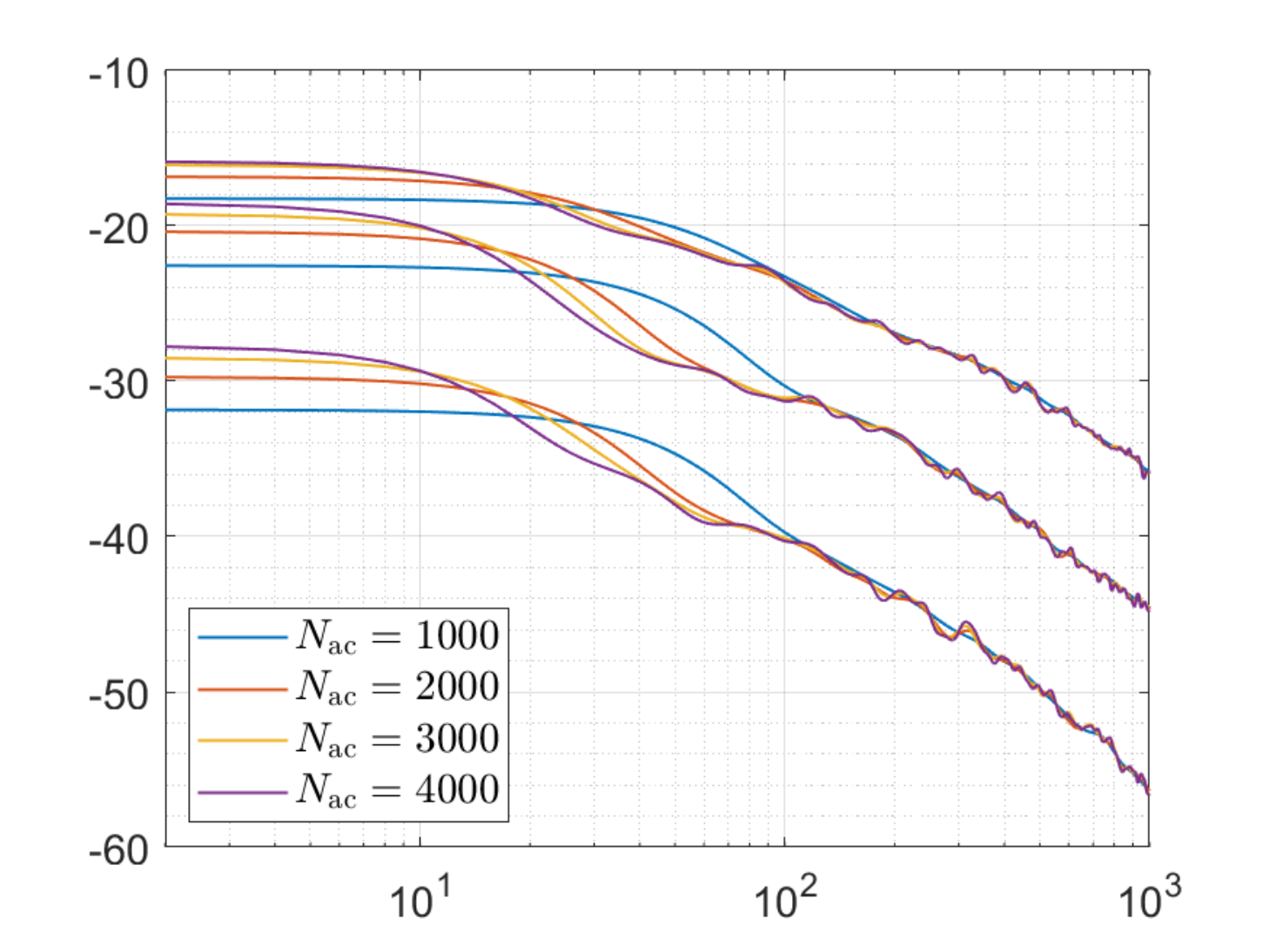}
		\put (38,73) {PSD (baseline)}
   \put (37,-2) {frequency (Hz)}
	\put (62,56) {\rotatebox{-30}{$y/\delta_0=1.3529$}}
	\put (62,46) {\rotatebox{-30}{$y/\delta_0=0.1000$}}
	\put (62,34) {\rotatebox{-30}{$y/\delta_0=0.0088$}}
   \end{overpic}
 \end{minipage}
\begin{minipage}[b]{0.49\textwidth}
  \begin{overpic}[width=\textwidth,trim={0mm 0mm 0mm 0mm},clip]{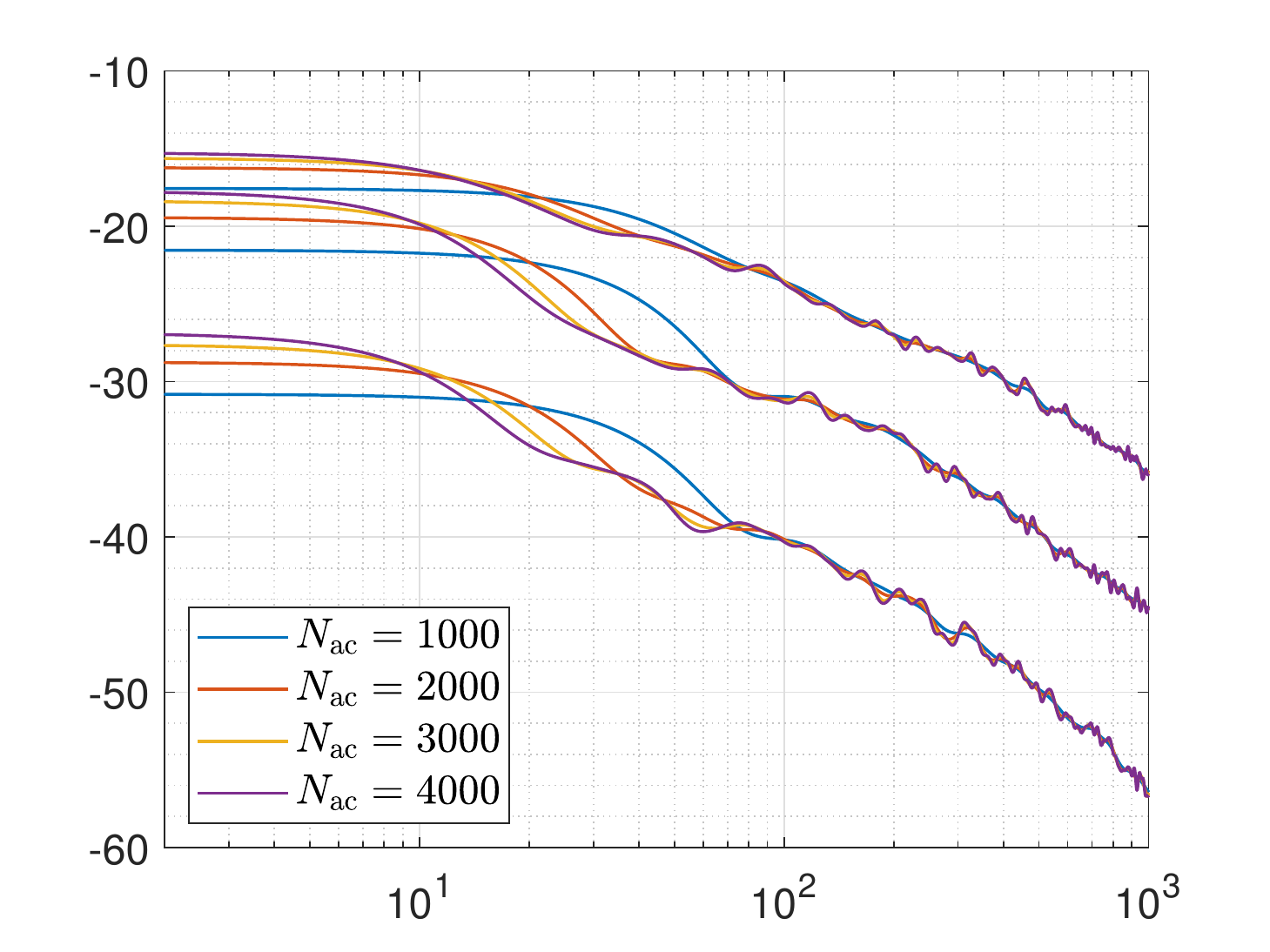}
		\put (25,73) {spectral measure (baseline)}
   \put (37,-2) {frequency (Hz)}
	\put (62,56) {\rotatebox{-30}{$y/\delta_0=1.3529$}}
	\put (62,46) {\rotatebox{-30}{$y/\delta_0=0.1000$}}
	\put (62,34) {\rotatebox{-30}{$y/\delta_0=0.0088$}}
   \end{overpic}
 \end{minipage}
\vspace{1mm}
\caption{Comparison of the traditional PSD definition \eqref{power_def} with spectral measure \eqref{power_def2}, for baseline and injection flows, at a range of vertical heights within the boundary layer.}
\label{fig:hotwire_res}
\end{figure}

\begin{figure}
 \centering
 \begin{minipage}[b]{0.49\textwidth}
  \begin{overpic}[width=\textwidth,trim={0mm 0mm 0mm 0mm},clip]{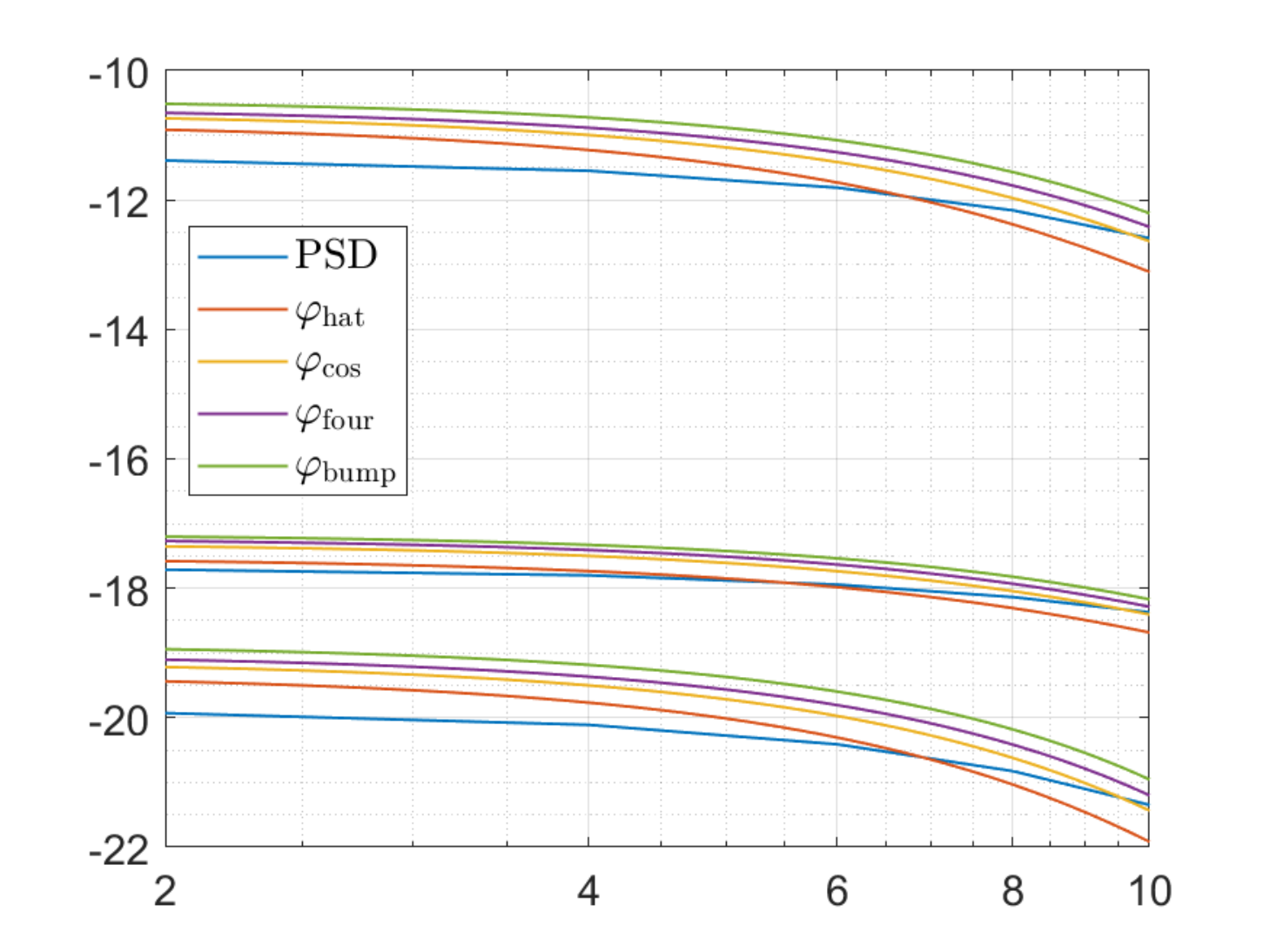}
	\put (42,73) {injection}
	\put (37,-2) {frequency (Hz)}
   \end{overpic}
 \end{minipage}
\begin{minipage}[b]{0.49\textwidth}
  \begin{overpic}[width=\textwidth,trim={0mm 0mm 0mm 0mm},clip]{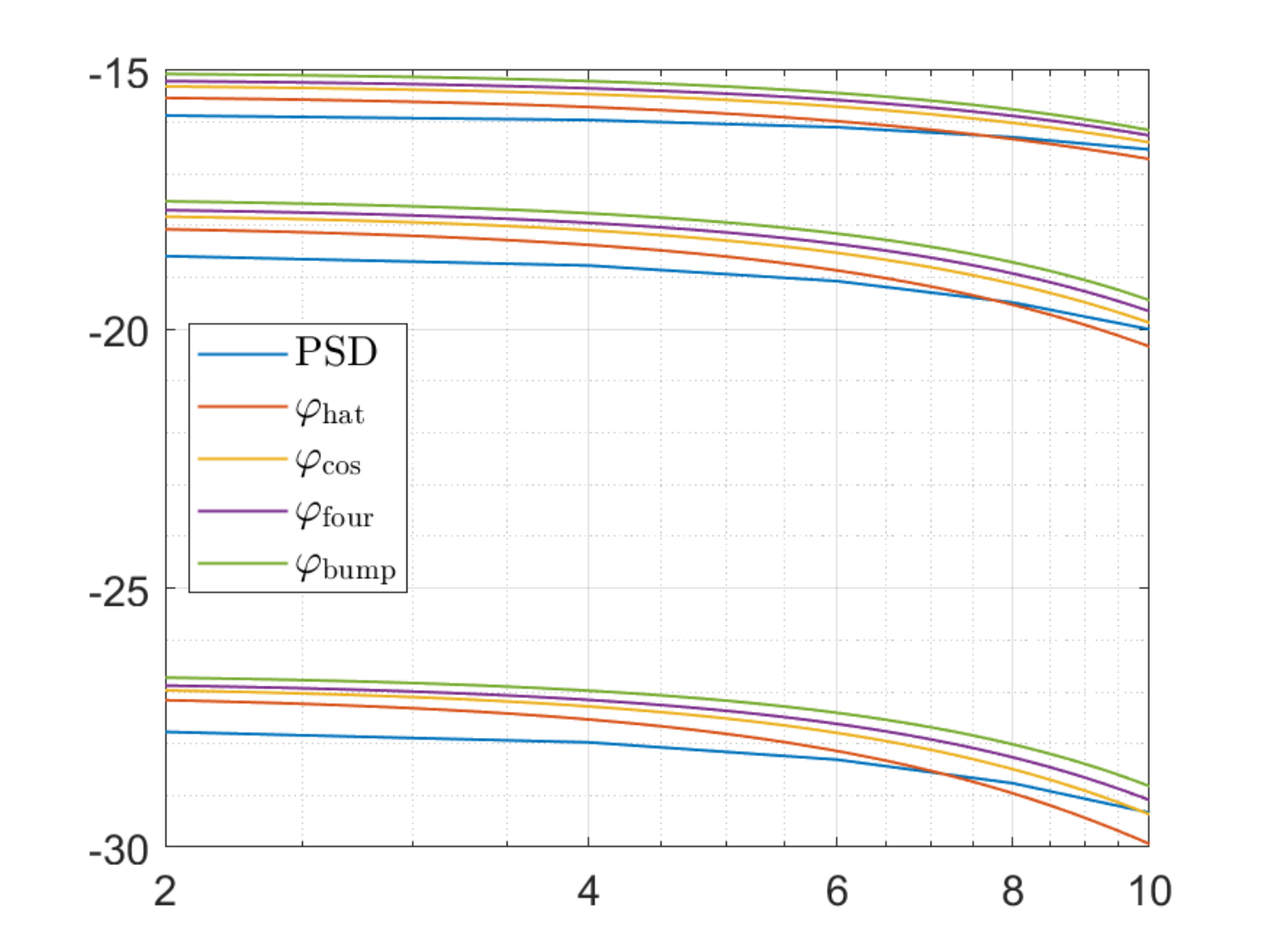}
	\put (42,73) {baseline}
		\put (37,-2) {frequency (Hz)}
   \end{overpic}
 \end{minipage}
\vspace{1mm}
\caption{Zoomed in values for low frequencies computed using $N_{\mathrm{ac}}=4000$ and various choices of filter function. In general, higher order filters lead to a sharper peak at low frequencies.}
\label{fig:hotwire_res10}
\end{figure}

\cref{fig:hotwire_res} compares the spectral measure, as found using \cref{alg:spec_meas_poly}, to the power spectral density \eqref{power_def} obtained from direct processing of the experimental data, where we have used a window size of $N_{\mathrm{ac}}$ for direct comparison. To directly compare to PSD, we use \cref{alg:spec_meas_poly} with the second order filter $\varphi_{{\rm cos}}$. A range of different vertical locations $y/\delta_0$ are considered, where $\delta_0$ is the boundary layer thickness of the baseline case (i.e., $u_{AFC}=0$). Whilst the high-frequency behaviour is almost identical between the two methods, at low frequencies ($<10$Hz) the spectral measure returns values approximately $\sim 1$dB greater than the PSD processing because the conventional PSD calculation results observe broadening at low frequencies. This is confirmed in \cref{fig:hotwire_res10} which shows the low frequency values for $N_{\mathrm{ac}}=4000$ and various choices of filter function. In general, higher order filters lead to a sharper peak at low frequencies. As we are assured that the spectral measure rigorously converges, this new method provides the more accurate measure of the power at low frequencies as $N_\mathrm{ac}\rightarrow\infty$.

\section{Example III: Wall-jet boundary layer flow}
\label{sec:verif_method_example_3}

As a further example of turbulent boundary layer flow, we now consider a wall-jet boundary layer~\citep{Gersten2015,george2000similarity,kleinfelter2019development}. Whilst hot-wire probes enable a very fine temporal resolution of the flow field, they are usually restricted to a single point or line in space, and thus preclude the use of many DMD-type methods. On the other hand, time-resolved (TR) particle image velocimetry (PIV) offers both spatial and temporal description of the flow. For this example, we assess the performance of the ResDMD algorithm on a set of TR-PIV data. We consider the boundary layer generated by a thin jet ($h_{jet}=12.7$mm) injecting air onto a smooth flat wall. As in Example II, this case is challenging for regular DMD approaches due to multiple turbulent scales expected within the boundary layer. This section demonstrates the use of ResDMD for a high Reynolds number, turbulent, complex flow field.

\begin{figure}
 \centering
\includegraphics[width=0.85\textwidth]{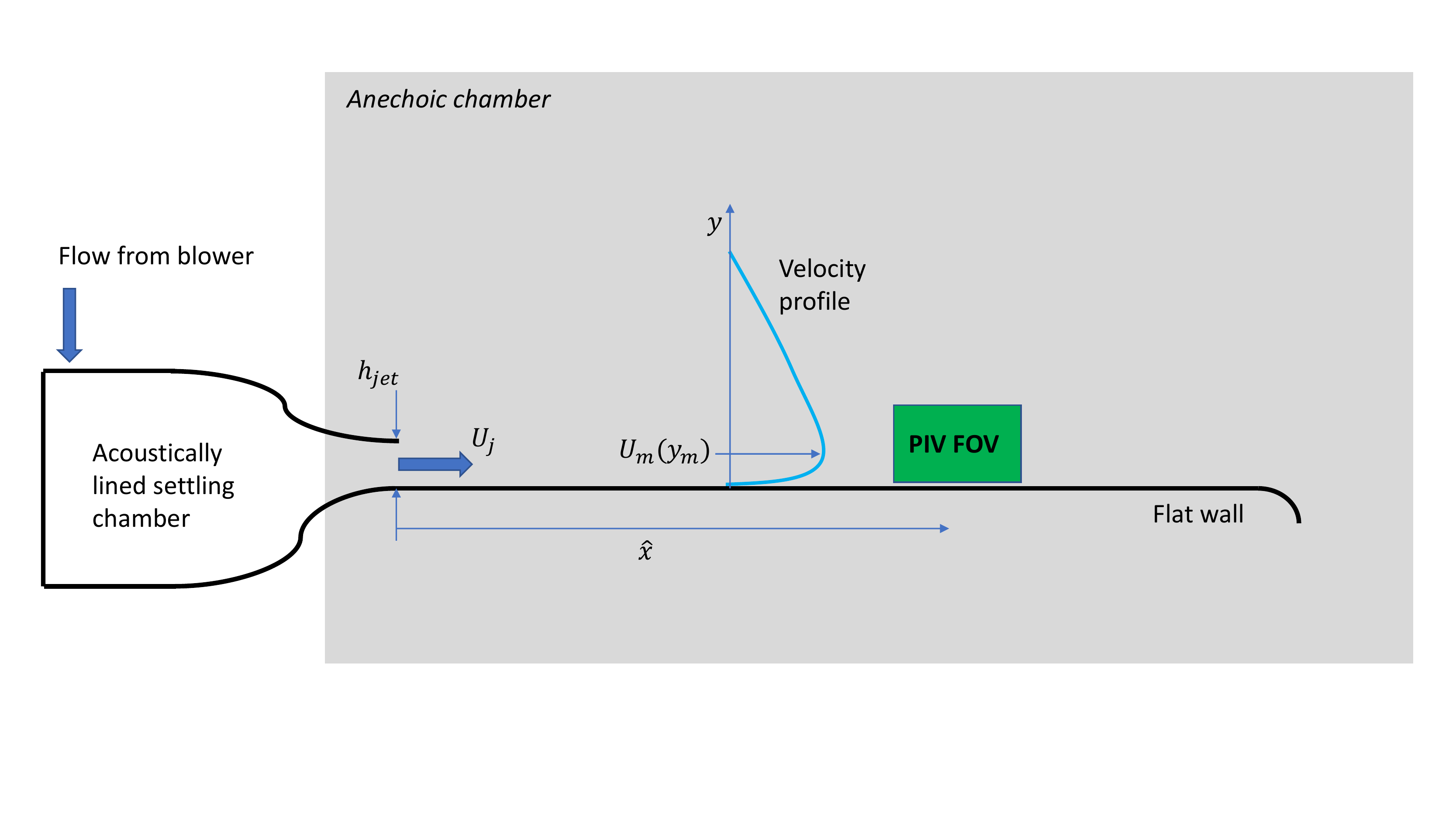}
	  \caption{The schematic of the TR-PIV experiments conducted in the Wall Jet Wind Tunnel of Virginia Tech.}
\label{fig:WJTSchematic}
\end{figure}

\subsection{Experimental setup}

Experiments using TR-PIV are performed at the Wall Jet Wind Tunnel of Virginia Tech as schematically shown in \cref{fig:WJTSchematic}. For a detailed description of the facility, we refer to \cite{kleinfelter2019development}. A two-dimensional two-component TR-PIV system is used to capture the wall-jet flow and the streamwise origin of the field-of-view (FOV) is $\hat{x} = $1282.7mm downstream of the wall-jet nozzle. We use a jet velocity of $U_j=$50m/s, corresponding to a jet Reynolds number of $\Rey_{jet} = h_{jet} U_j /\nu = 63.5 \times 10^3$. The length and height of the FOV is approximately 75mm $\times$ 45mm, and the spatial resolution of the velocity vector field is 0.25mm. The high-speed cameras are operated in a double frame mode, with a rate of 12,000 frame pairs per second, resulting in a fine temporal resolution of 0.083ms. For a more detailed description of the experimental setup and flow field, see~\cite{kleinfelter2019development,szoke2021flow}. 

The associated flow has some special properties. It is self-similar and its main characteristics (boundary layer thickness, edge velocity, skin friction coefficient, etc.) can be accurately calculated a-priori through power-law curve fits \citep{kleinfelter2019development}. The flow consists of two main regions. Within the region bounded by the wall and the peak in the velocity profile, the flow exhibits the properties of a zero pressure gradient turbulent boundary layer. Above this fluid portion, the flow is dominated by a two-dimensional shear layer consisting of rather large, energetic flow structures. While the peak in the velocity profile is $y_m \approx$ 18mm from the wall in our case, the overall thickness of the wall-jet flow is on the order of 200mm. Clearly, the PIV experiments must compromise between a good spatial resolution or capturing the entire flow field. In our case, the FOV was not tall enough to capture the entire wall-jet flow field. For this reason, the standard DMD algorithm under-predicts the energies corresponding to the shear-layer portion of the wall-jet flow as the corresponding length scales fall outside of the limits of the FOV.

\subsection{Results}

\begin{figure}
 \centering
 \begin{minipage}[b]{0.49\textwidth}
  \begin{overpic}[width=\textwidth,trim={0mm 0mm 0mm 0mm},clip]{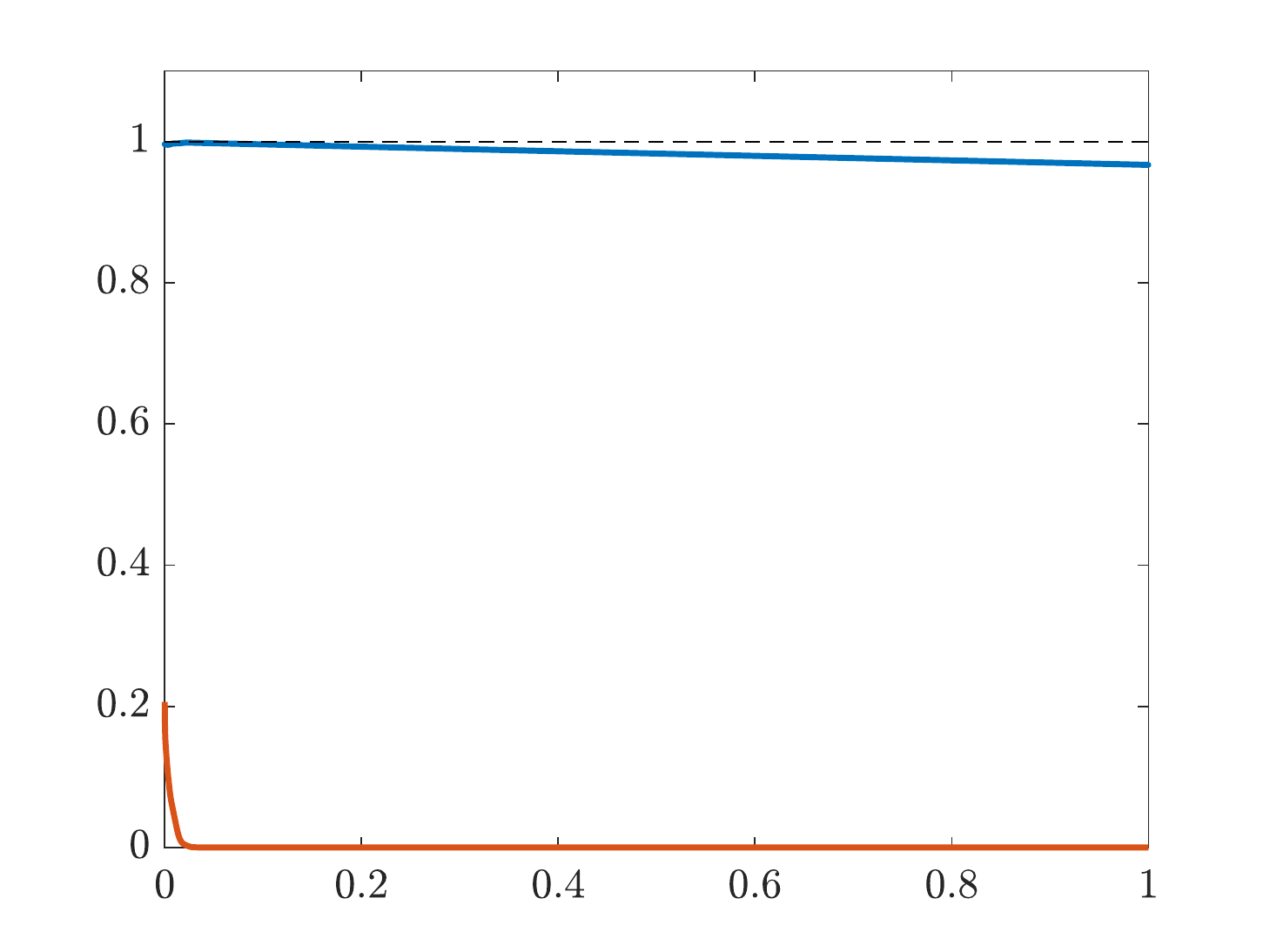}
	\put (20,73) {Forecast of total kinetic energy}
	\put (36,-2) {forecast time (s)}
	\put (45,57) {\rotatebox{-2}{non-linear dictionary}}
	\put (35,10) {\rotatebox{0}{non-linear dictionary}}
   \end{overpic}
 \end{minipage}
\begin{minipage}[b]{0.49\textwidth}
  \begin{overpic}[width=\textwidth,trim={0mm 0mm 0mm 0mm},clip]{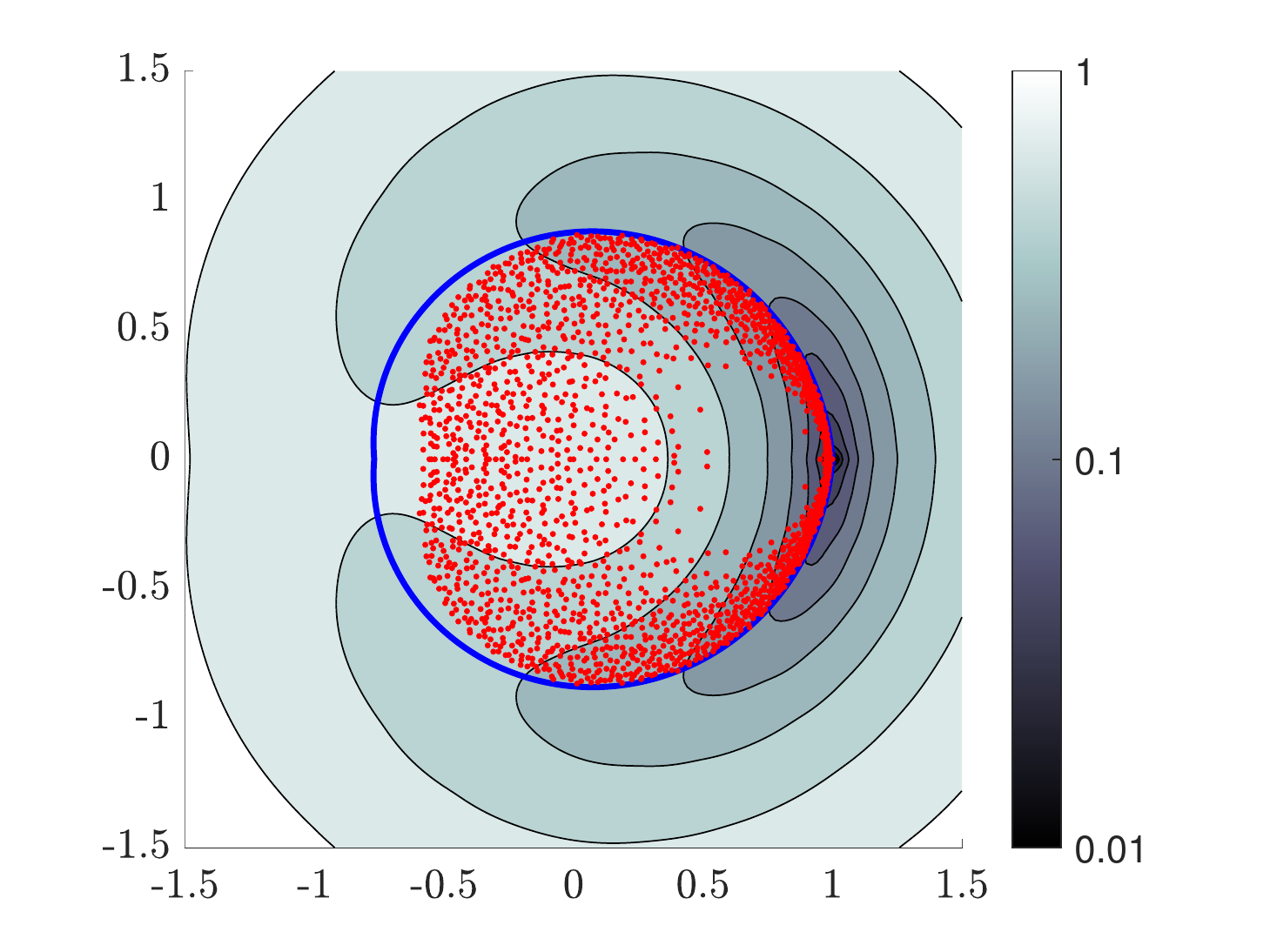}
		\put (40,73) {$\tau_{2000}(\lambda)$}
   \put (40,-2) {$\mathrm{Re}(\lambda)$}
		\put (2,33) {\rotatebox{90}{$\mathrm{Im}(\lambda)$}}
   \end{overpic}
 \end{minipage}
\vspace{1mm}
\caption{Left: Forecast of total kinetic energy (normalised by the time average of the kinetic energy), averaged over the 12000 initial conditions. Values closer to $1$ correspond to better predictions. Right: Pseudospectral contours computed using \cref{alg:res_EDMD} for the wall-jet example, using a non-linear dictionary. The eigenvalues of the finite Galerkin matrix $\mathbb{K}$ are shown as red dots. The shape of the contours reflect the transient modes. The blue curve corresponds to a fit $r=\exp(-c|\theta|)$ of these contours and the boundary of the eigenvalues, and represents successive powers of modes.}
\label{fig:canopy1}
\end{figure}
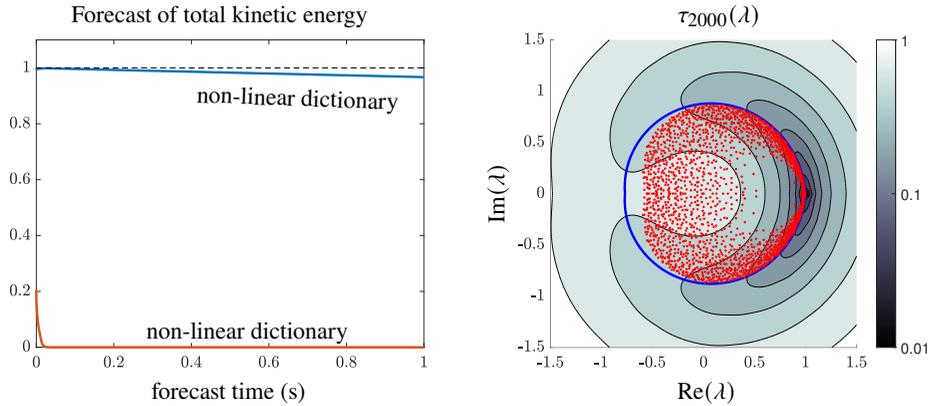

We collect snapshot data of the velocity field from \textit{two separate realisations of the experiment}. We use the first experiment to generate data $\{\tilde{\pmb{x}}^{(m)},\tilde{\pmb{y}}^{(m)}\}_{m=1}^{M'}$ with $M'=2000$, corresponding to 121 boundary layer turnover times. This data is used to select our dictionary of functions. We use the second experiment to generate data $\{\hat{\pmb{x}}^{(m)},\hat{\pmb{y}}^{(m)}\}_{m=1}^{M''}$ with $M''=12000$ (a single trajectory of one second of physical flow time and 728 boundary layer turnover times), which we use to generate the ResDMD matrices, as outlined in \cref{sec:prob_sol_2_np_data}. To demonstrate the need for non-linear functions in our dictionary, we compute the Koopman mode decomposition of the total kinetic energy of the domain using~\eqref{gen_kp_m_decm}. Using this decomposition, we compute forecasts of the total energy from a given initial condition of the system. \cref{fig:canopy1} (left) shows the results, where we average over the $12000$ initial conditions in the data set and normalise by the true time-averaged kinetic energy. We use \cref{alg:DMD2,alg:kern_algs} with $N=2000$, which we refer to as a linear dictionary and non-linear dictionary, respectively. The importance of including non-linear functions in the dictionary is clear, and corresponds to a much better approximation of $\mathcal{K}$'s spectral content near $0$. For the rest of this section, we therefore only use the non-linear dictionary. \cref{fig:canopy1} (right) shows values of $\tau_N$ (pseudospectral contours) computed using \cref{alg:res_EDMD}, where $\tau_N$ is the minimal residual in \eqref{tau_factor}. Unlike the example in \cref{sec:verif_method_example_1}, the contours are not circular. Instead, they appear to be centered around a curve of the form $r=\exp(-c|\theta|)$ (shown as blue in the plot), corresponding to successive powers of transient modes. This is reflected in the eigenvalues of the finite $N\times N$ Galerkin matrix $\mathbb{K}$, shown as red dots, some of which correspond to spectral pollution. The eigenvalues of non-normal matrices can be severely unstable to perturbations, particularly for large $N$, so we checked the computation of the eigenvalues of $\mathbb{K}$ by comparing to extended precision and predict a bound of approximately $10^{-10}$ on the error in \cref{fig:canopy1} (right).

\begin{figure}
 \centering
 \begin{minipage}[b]{0.49\textwidth}
  \begin{overpic}[width=\textwidth,trim={0mm 5mm 0mm 5mm},clip]{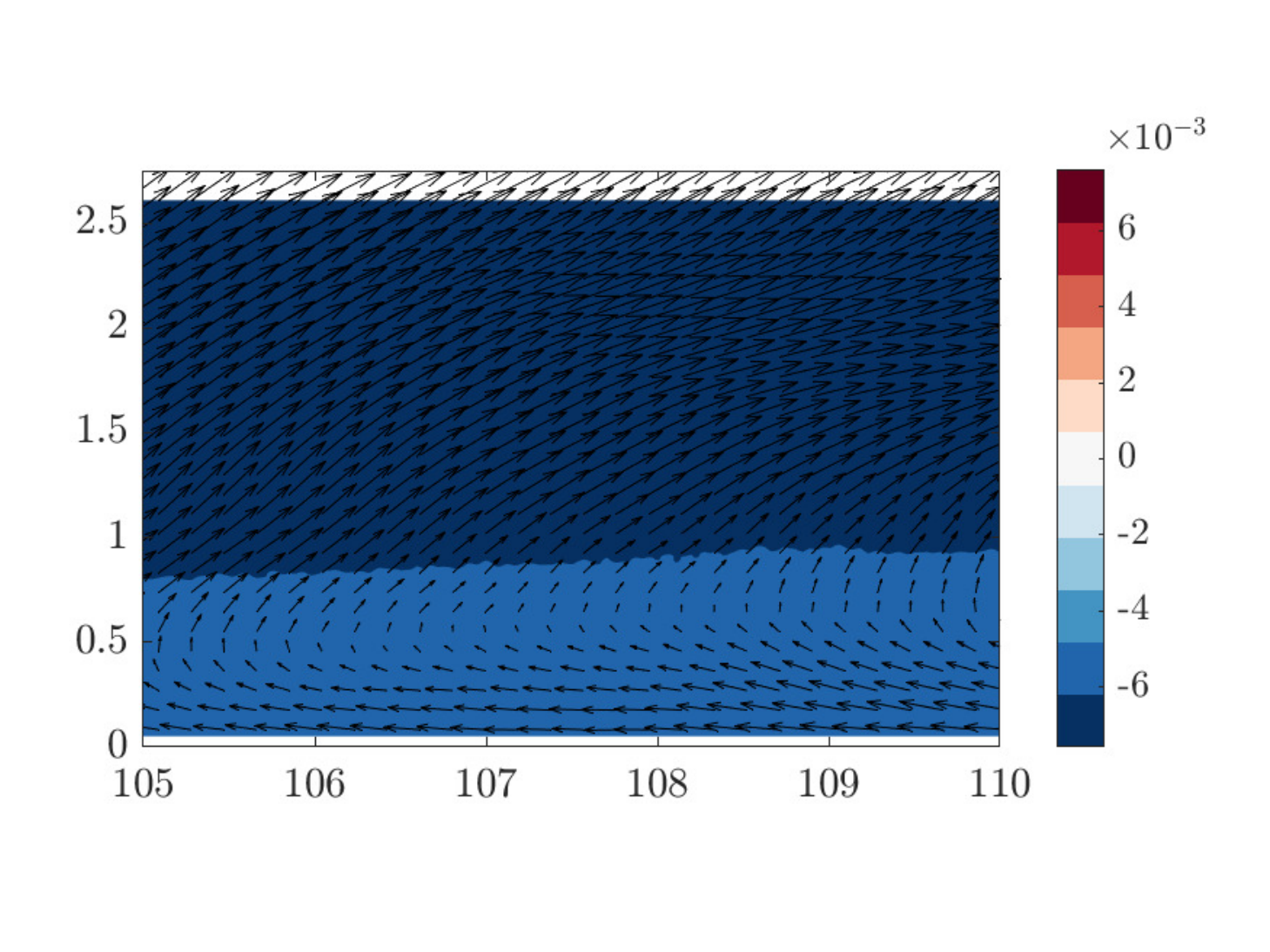}
	\put (40,2) {$\hat{x}/h_{jet}$}
	\put (-1,30) {\rotatebox{90}{$y/y_m$}}
	\put (7,62){$\lambda=1.0000 + 0.00i$, $\tau_{2000}(\lambda)=0.0024$}
   \end{overpic}
 \end{minipage}
\begin{minipage}[b]{0.49\textwidth}
  \begin{overpic}[width=\textwidth,trim={0mm 5mm 0mm 5mm},clip]{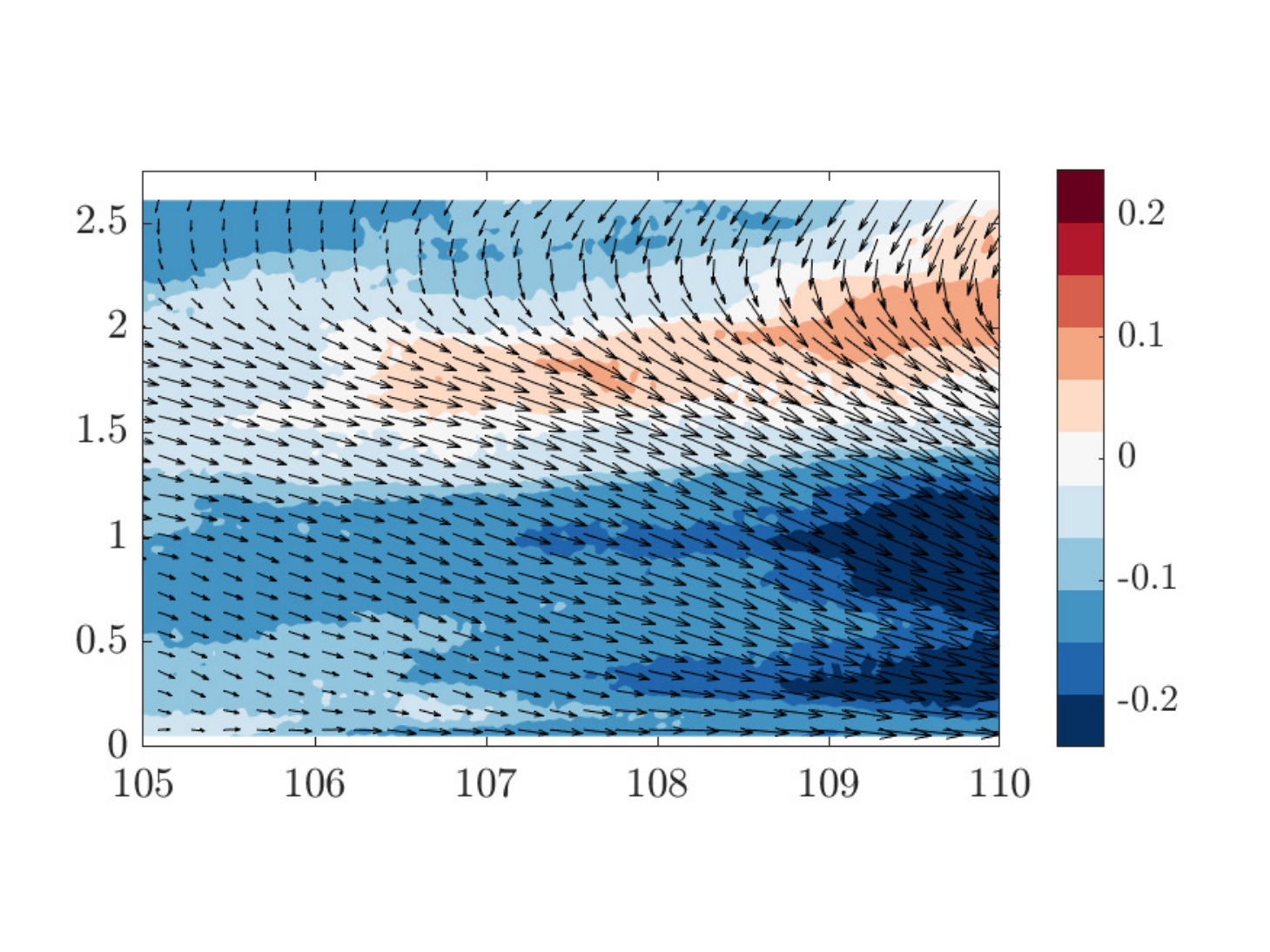}
	\put (40,2) {$\hat{x}/h_{jet}$}
	\put (-1,30) {\rotatebox{90}{$y/y_m$}}
	\put (7,62){$\lambda=0.9837 + 0.0057i$, $\tau_{2000}(\lambda)=0.0175$}
   \end{overpic}
 \end{minipage}
\vspace{2mm}
\begin{minipage}[b]{0.49\textwidth}
  \begin{overpic}[width=\textwidth,trim={0mm 5mm 0mm 5mm},clip]{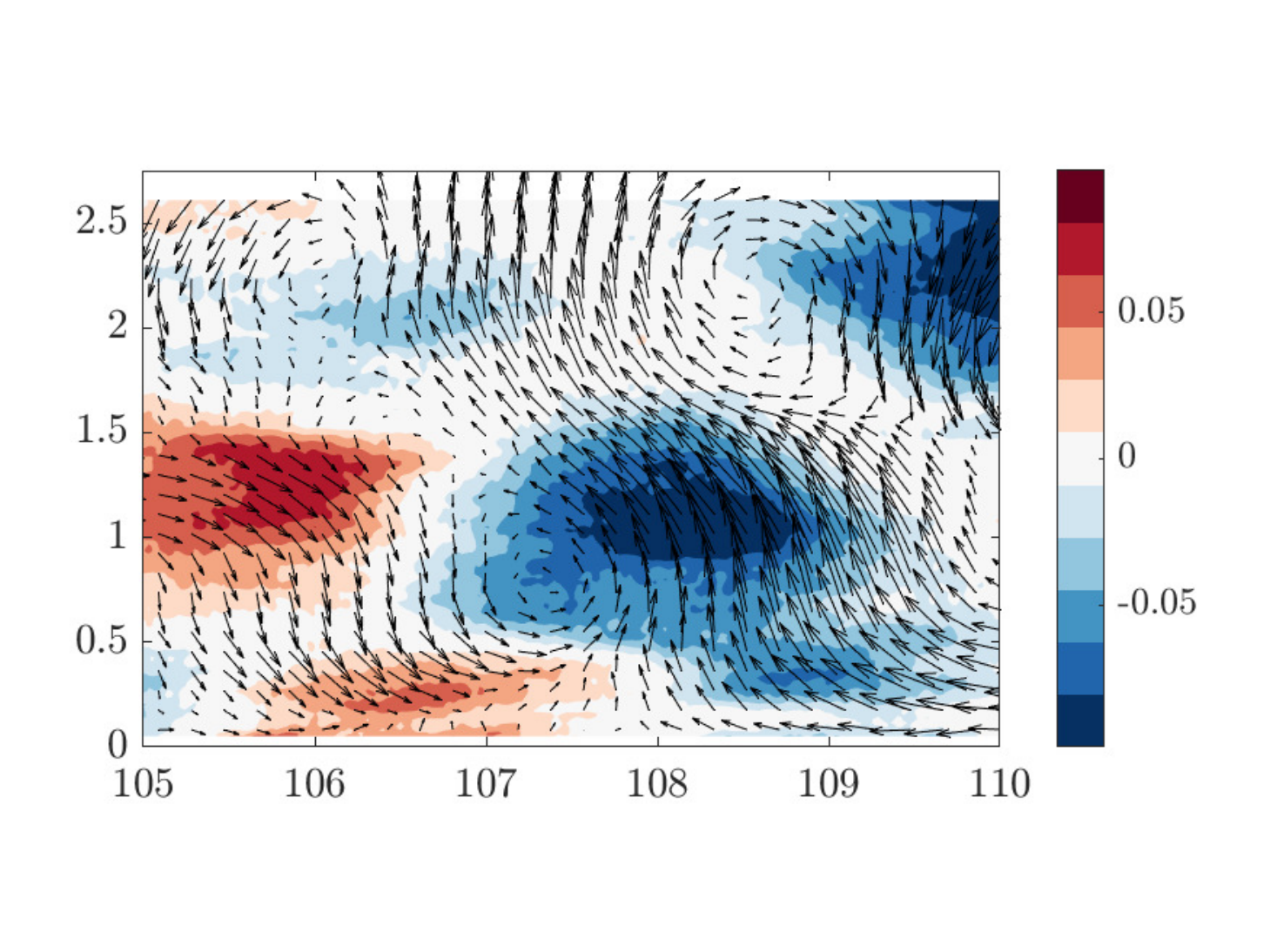}
	\put (40,2) {$\hat{x}/h_{jet}$}
	\put (-1,30) {\rotatebox{90}{$y/y_m$}}
	\put (7,62){$\lambda=0.9760 + 0.1132i$, $\tau_{2000}(\lambda)=0.0539$}
   \end{overpic}
 \end{minipage}
\begin{minipage}[b]{0.49\textwidth}
  \begin{overpic}[width=\textwidth,trim={0mm 5mm 0mm 5mm},clip]{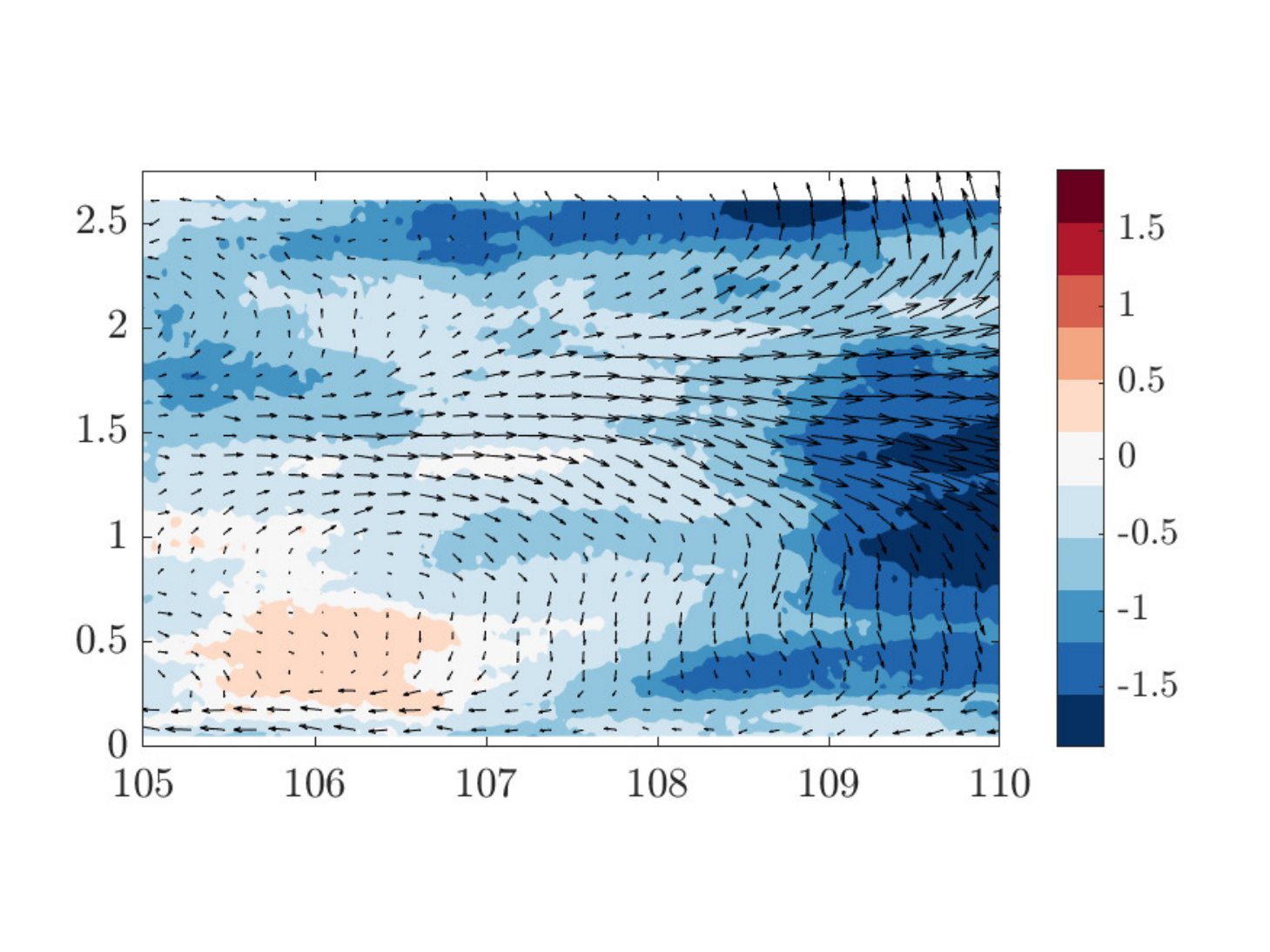}
	\put (40,2) {$\hat{x}/h_{jet}$}
	\put (-1,30) {\rotatebox{90}{$y/y_m$}}
	\put (7,62){$\lambda=0.9528+0.0000i$, $\tau_{2000}(\lambda)=0.0472$}
   \end{overpic}
 \end{minipage}
\vspace{2mm}
\begin{minipage}[b]{0.49\textwidth}
  \begin{overpic}[width=\textwidth,trim={0mm 5mm 0mm 5mm},clip]{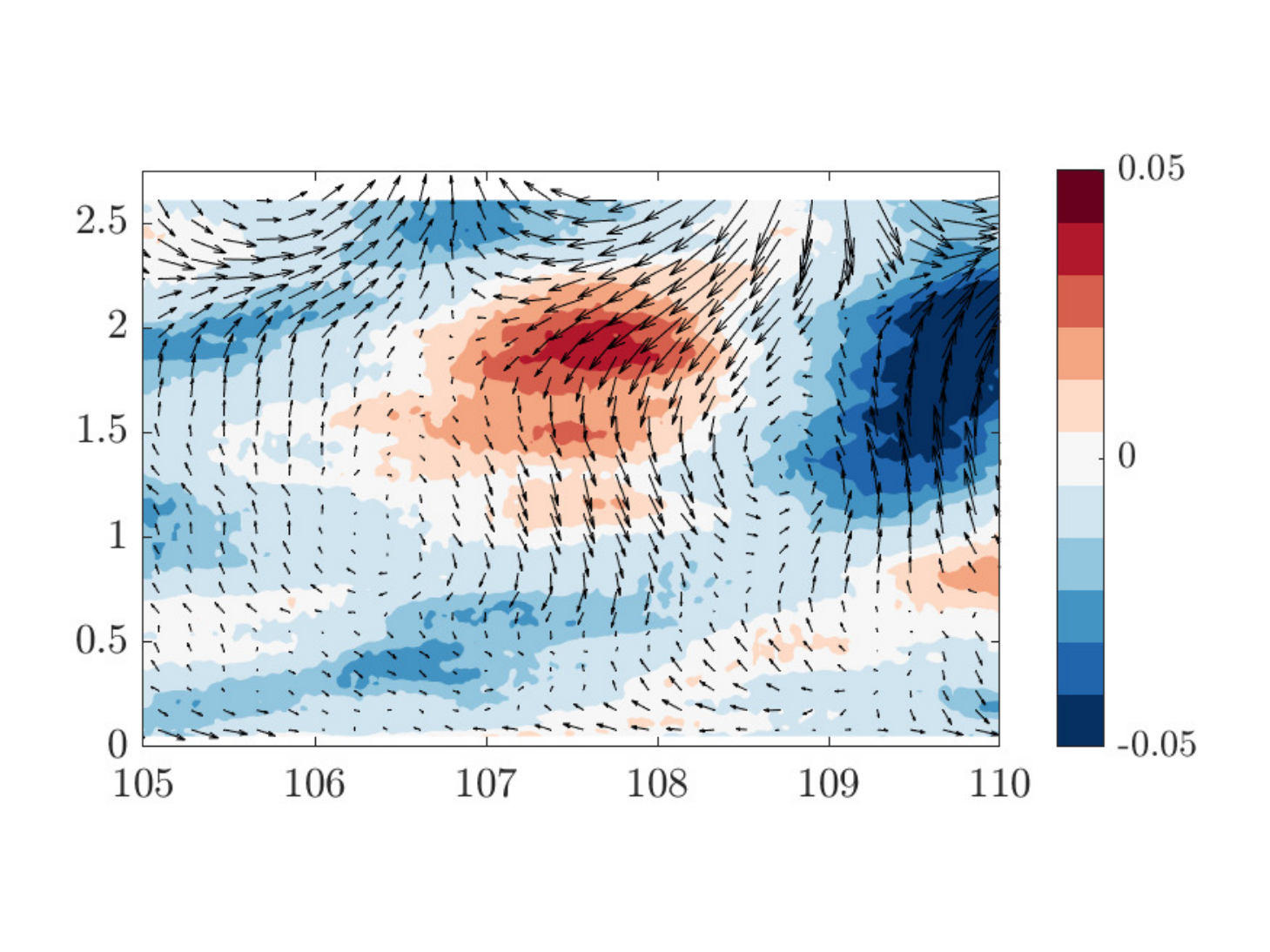}
	\put (40,2) {$\hat{x}/h_{jet}$}
	\put (-1,30) {\rotatebox{90}{$y/y_m$}}
	\put (7,62){$\lambda=0.9700 + 0.1432i$, $\tau_{2000}(\lambda)=0.0602$}
   \end{overpic}
 \end{minipage}
\begin{minipage}[b]{0.49\textwidth}
  \begin{overpic}[width=\textwidth,trim={0mm 5mm 0mm 5mm},clip]{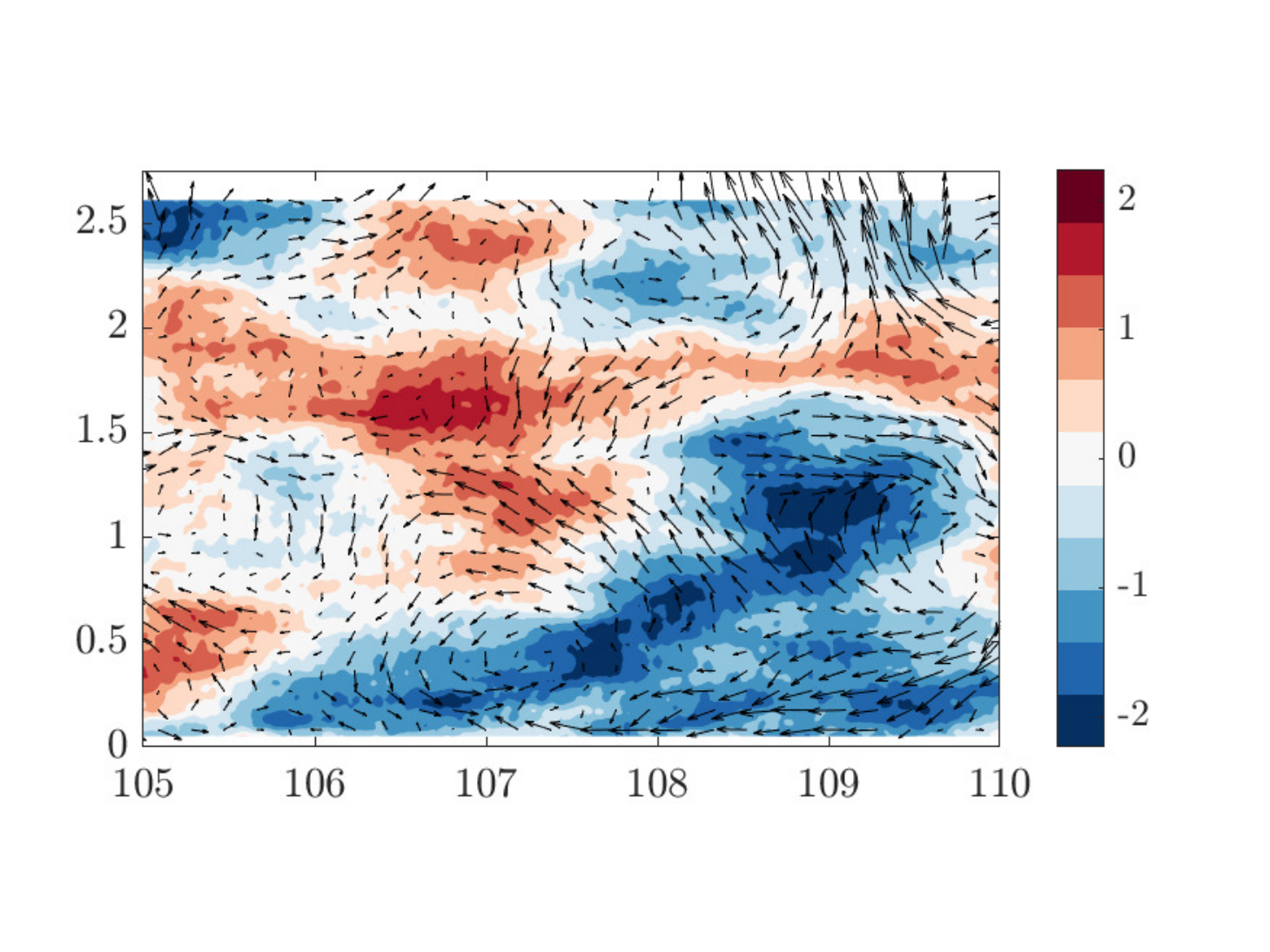}
	\put (40,2) {$\hat{x}/h_{jet}$}
	\put (-1,30) {\rotatebox{90}{$y/y_m$}}
	\put (7,62){$\lambda=0.8948 + 0.1065i$, $\tau_{2000}(\lambda)=0.1105$}
   \end{overpic}
 \end{minipage}
\vspace{2mm}
\begin{minipage}[b]{0.49\textwidth}
  \begin{overpic}[width=\textwidth,trim={0mm 5mm 0mm 5mm},clip]{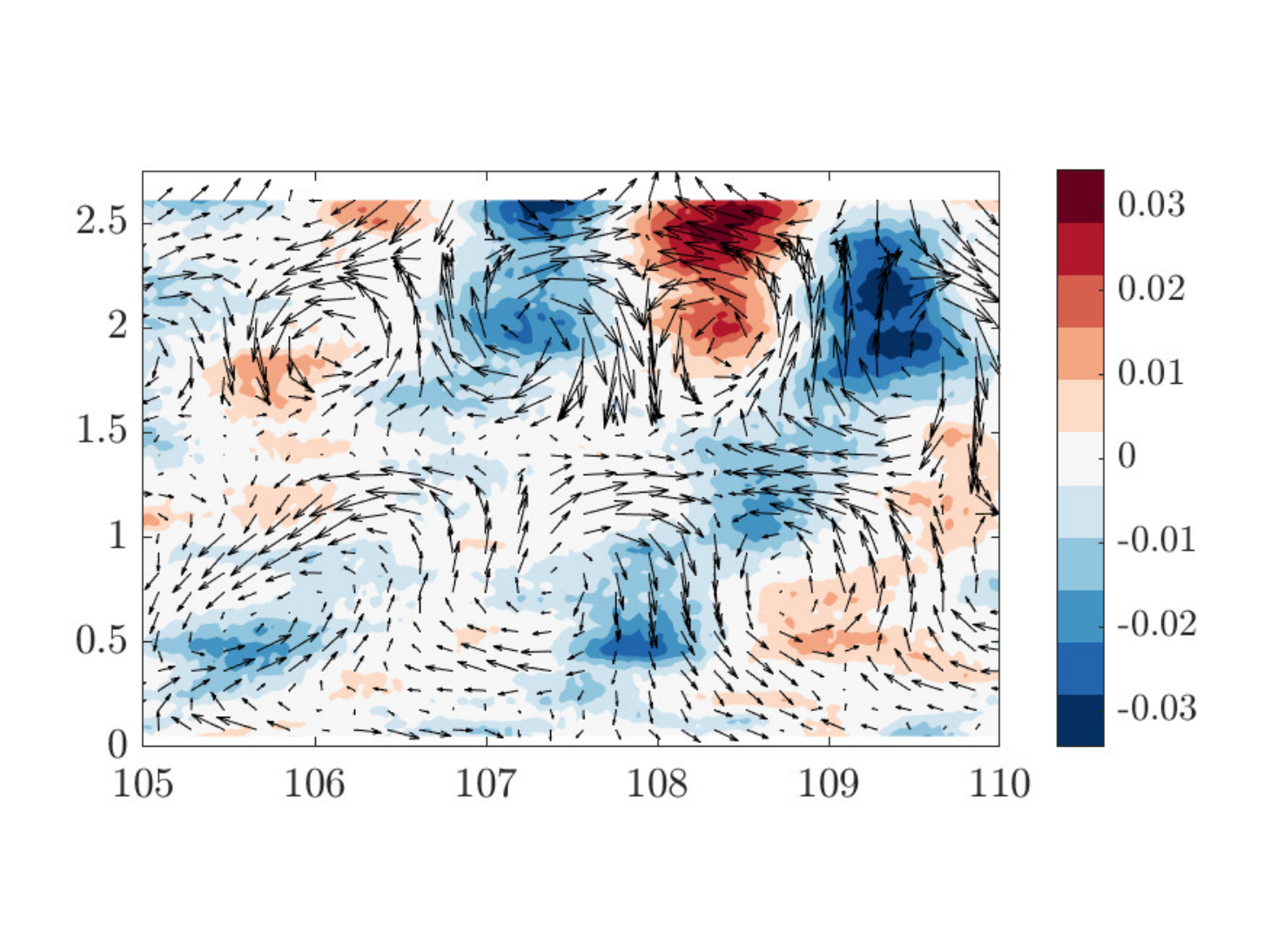}
	\put (40,2) {$\hat{x}/h_{jet}$}
	\put (-1,30) {\rotatebox{90}{$y/y_m$}}
	\put (7,62){$\lambda=0.9439 + 0.2458i$, $\tau_{2000}(\lambda)=0.0765$}
   \end{overpic}
 \end{minipage}
\begin{minipage}[b]{0.49\textwidth}
  \begin{overpic}[width=\textwidth,trim={0mm 5mm 0mm 5mm},clip]{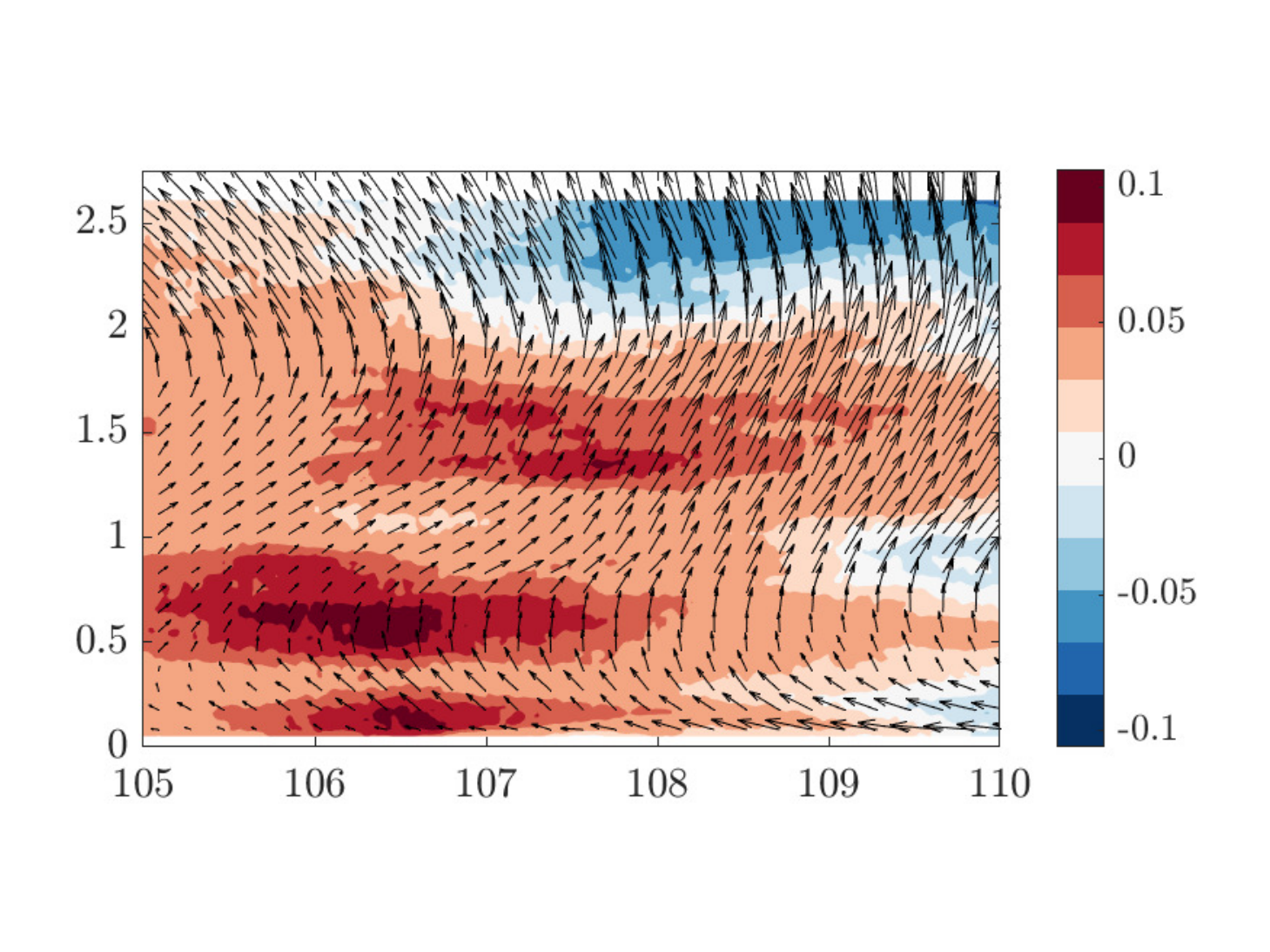}
	\put (40,2) {$\hat{x}/h_{jet}$}
	\put (-1,30) {\rotatebox{90}{$y/y_m$}}
	\put (7,62){$\lambda=0.9888 + 0.0091i$, $\tau_{2000}(\lambda)=0.0146$}
   \end{overpic}
 \end{minipage}
\vspace{1mm}
\caption{Left: A range of long-lasting modes from the ResDMD Koopman mode decomposition. Right: A range of transient modes from the ResDMD Koopman mode decomposition.The arrows dictate the unsteady fluid structure (computed from the Koopman modes of the velocity fields), with the magnitude of the arrow indicating the local flow speed, and the colourbar denotes the Koopman mode of the velocity magnitude.}
\label{fig:canopy2}
\end{figure}

To investigate the Koopman modes, we compute the ResDMD Koopman mode decomposition corresponding to~\eqref{koop_mode_estimate2} with the error tolerance $\epsilon=0.5$ to get rid of the most severe spectral pollution. The total number of modes used is 656. \cref{fig:canopy2} illustrates a range of Koopman modes which are long-lasting (left-hand column) and transient (right-hand column). Due to residual measures, we are able to accurately select physical transient modes. Within each figure, the arrows dictate the unsteady fluid structure (computed from the Koopman modes of the velocity fields), with the magnitude of the arrow indicating the local flow speed, and the colourbar denotes the Koopman mode of the velocity magnitude. The corresponding approximate eigenvalues, $\lambda$, and residual bound, $\tau_N$, are provided for each mode.

The modes in the left column of \cref{fig:canopy2} illustrate the range of rolling eddies within the boundary layer, with the smaller structures containing less energy than the largest structures. Interestingly, the third mode in the left column resembles the shape of ejection-like motions within the boundary layer flow ($y/y_m<1$) while larger-scale structures above the boundary layer ($y/y_m>1$) are also visible. This may be interpreted as a non-linear interaction in the turbulent flow field, which is efficiently captured using the ResDMD algorithm. The transient modes in the right column of \cref{fig:canopy2} show a richer structure. Based on our analysis, we interpret these modes as transient, short-lived behaviour of turbulence. The uppermost panel may be seen as the shear layer traveling over the boundary layer ($y/y_m>1$), with the following panel potentially seen as the breakdown of this transient structure into smaller structures. The third panel may be seen as an interaction between an ejection-type vortex and the shear layer, note the ejection-like shape of negative contours below $y/y_m=1.5$ with a height-invariant positive island of contour at $y/y_m\approx1.75$. Finally the bottom-most panel could be seen as a flow uplift out of the boundary layer and further turbulent streaks with counter-rotating properties.

\begin{figure}
 \centering
 \begin{minipage}[b]{0.49\textwidth}
  \begin{overpic}[width=\textwidth,trim={0mm 0mm 0mm 0mm},clip]{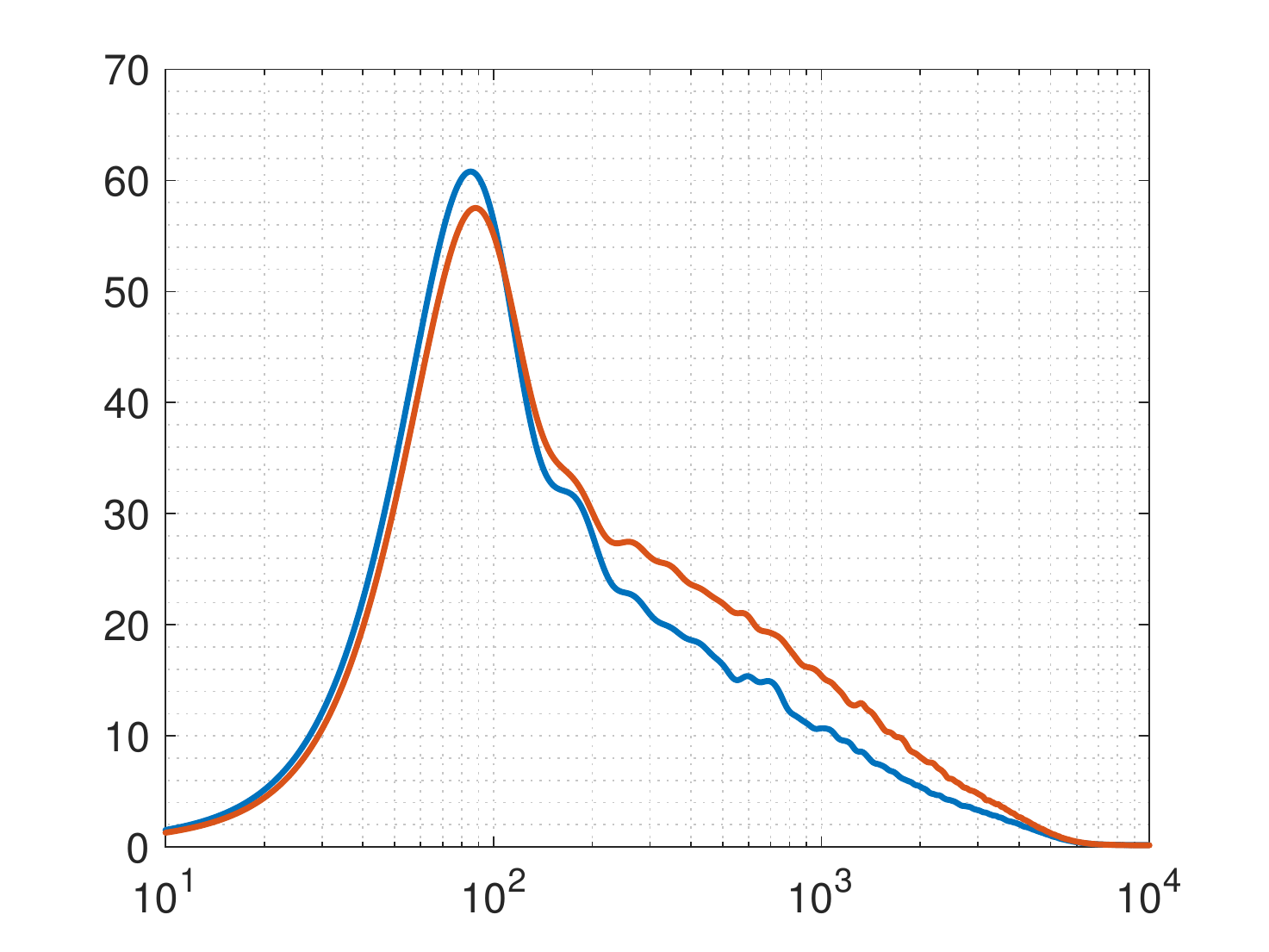}
	\put (44,-2) {$k_x$ ($\mathrm{m}^{-1}$)}
	\put (40,27) {\rotatebox{-32}{$y/y_m=0.99$}}
	\put (55,30) {\rotatebox{-32}{$y/y_m=0.13$}}
	\put (42,73) {{$k_x|\phi_{uu}|$}}
   \end{overpic}
 \end{minipage}
\begin{minipage}[b]{0.49\textwidth}
  \begin{overpic}[width=\textwidth,trim={0mm 0mm 0mm 0mm},clip]{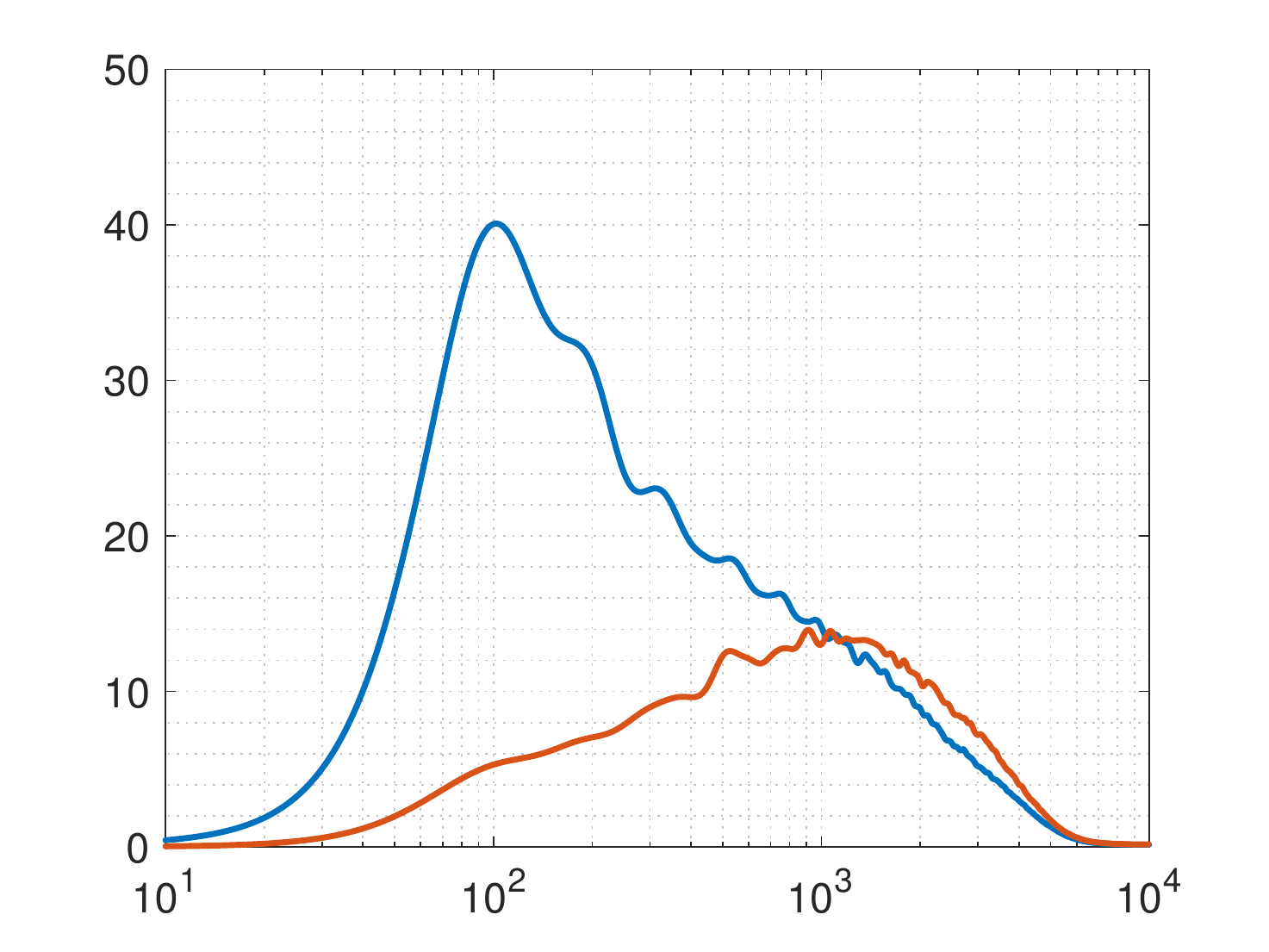}
	\put (44,-2) {$k_x$ ($\mathrm{m}^{-1}$)}
	\put (30,15) {\rotatebox{15}{$y/y_m=0.13$}}
	\put (47,45) {\rotatebox{-46}{$y/y_m=0.99$}}
	\put (42,73) {{$k_x|\phi_{vv}|$}}
   \end{overpic}
 \end{minipage}
 \begin{minipage}[b]{0.49\textwidth}
\vspace{5mm}
  \begin{overpic}[width=\textwidth,trim={0mm 0mm 0mm 0mm},clip]{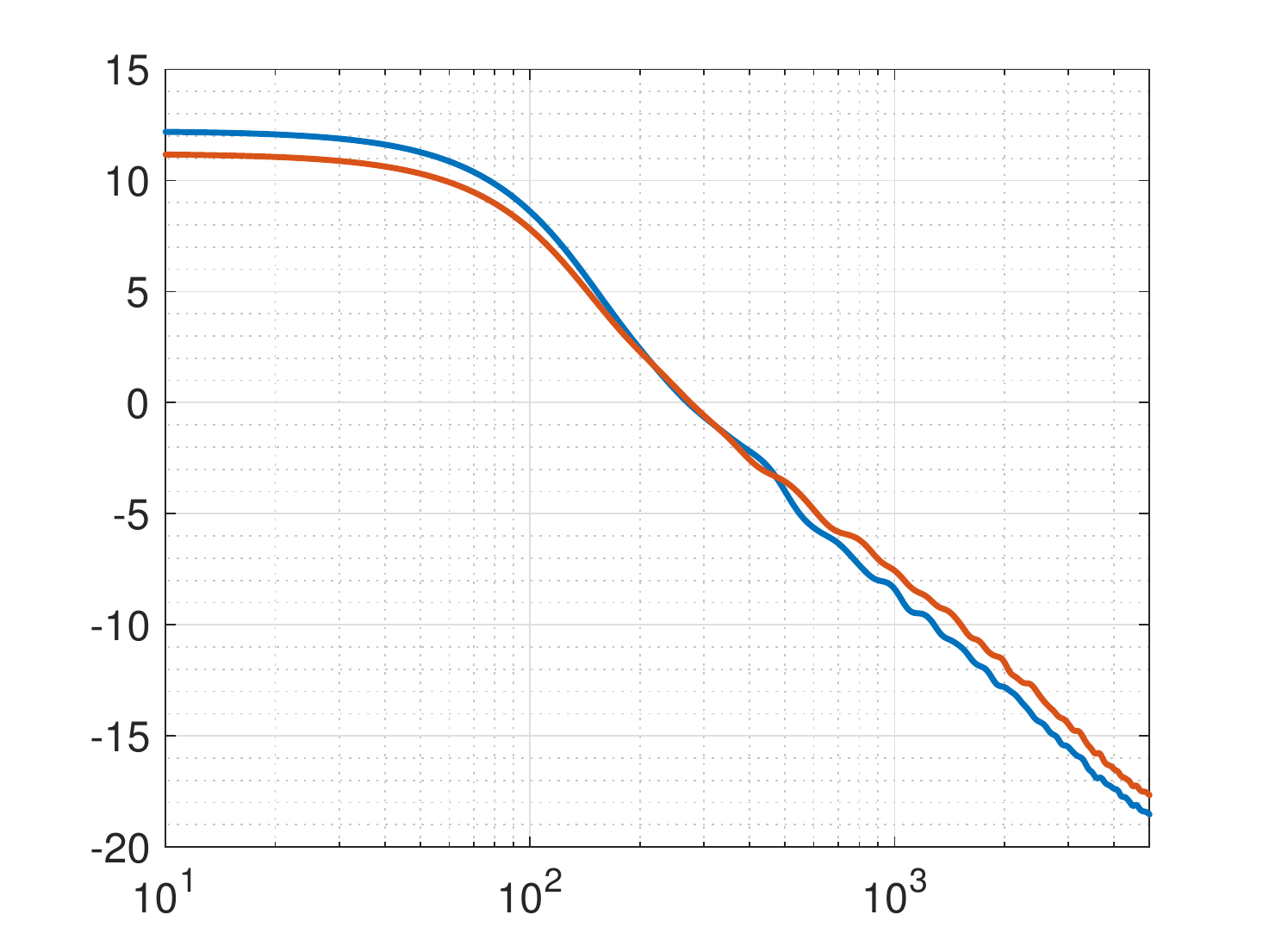}
	\put (25,73) {spectral measure ($u$ field)}
   \put (37,-2) {frequency (Hz)}
	\put (15,65) {{$y/y_m=0.99$}}
	\put (15,55) {{$y/y_m=0.13$}}
   \end{overpic}
 \end{minipage}
\begin{minipage}[b]{0.49\textwidth}
\vspace{5mm}
  \begin{overpic}[width=\textwidth,trim={0mm 0mm 0mm 0mm},clip]{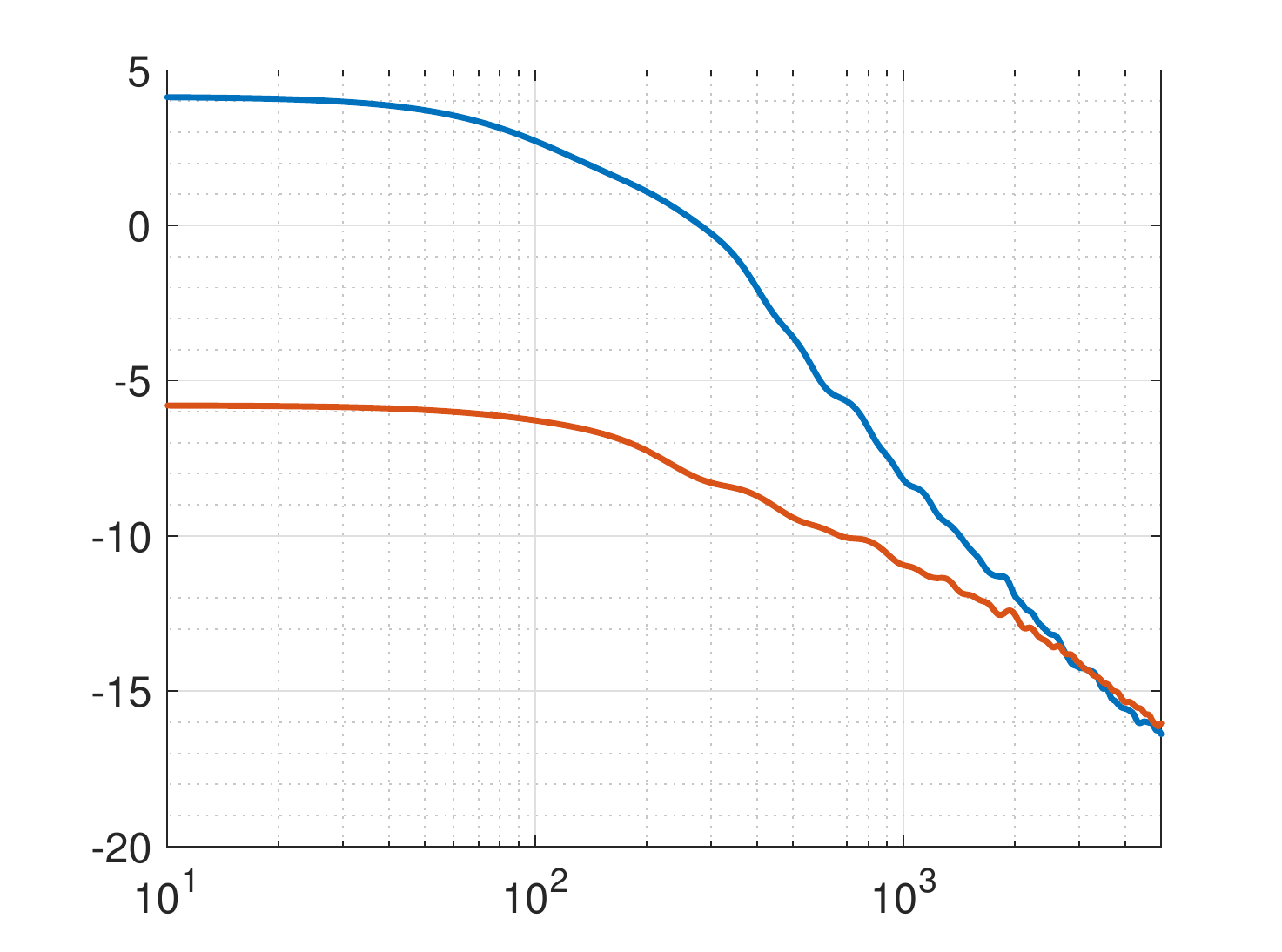}
	\put (25,73) {spectral measure ($v$ field)}
   \put (37,-2) {frequency (Hz)}
	\put (15,60) {{$y/y_m=0.99$}}
	\put (15,35) {{$y/y_m=0.13$}}
   \end{overpic}
 \end{minipage}
\vspace{1mm}
\caption{Pre-multiplied streamwise wavenumber spectra (top row) and spectral measures (bottom row) for streamwise ($u$) and wall-normal ($v$) velocity components.}
\label{fig:canopy3}
\end{figure}

Using the autocorrelation functions of both streamwise ($u$) and wall normal ($v$) velocity components and \cref{alg:spec_meas_poly}, we obtain the streamwise wavenumber spectra of the velocity fluctuations, shown in the top row of \cref{fig:canopy3}. The streamwise averaged spectral measure of the ergodic component of the flow for both streamwise and vertical unsteady velocities are shown in the bottom row of \cref{fig:canopy3}. When observing either the wavenumber spectra or the spectral measures of the streamwise velocity component, we see consistent levels of spectra across the different boundary layer heights, which is due to the dominance of the strong shear layer in the wall jet flow. For the vertical velocity spectra, as one might expect, more energy is observed for both wavenumber spectra and spectral measures with increasing $y/y_m$.

\section{Example IV: Acoustic signature of laser-induced plasma}
\label{sec:verif_method_example_4}

So far, our examples of ResDMD have focused on fluid flows. However, the capability of ResDMD capturing non-linear mechanics can be applied more broadly. Moreover, the computation of residuals allows an efficient compression of the Koopman mode decomposition through \eqref{koop_mode_estimate2}. As our final example, we demonstrate the use of ResDMD on an acoustic example where the sound source of interest exhibits highly non-linear properties. 

\subsection{Experimental setup}

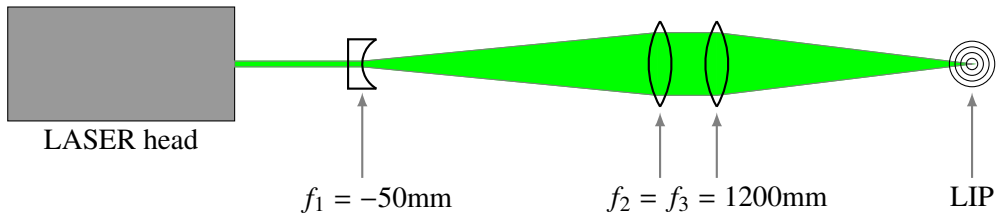
\begin{figure}
\begin{tikzpicture}[scale=0.75]
	\tikzstyle{every node}=[font=\large]

	\filldraw[fill=black!40!white, draw=black] (0,0) rectangle (4,2);

	\filldraw[fill=green, draw=green] (4,0.95) rectangle (7,1.05);
	\filldraw[fill=green, draw=green] (6+2*0.1339,0.95) -- (6+2*0.1339,1.05) -- (11.45,1.55) -- (11.45, 0.45) -- cycle;

	\filldraw[fill=green, draw=green] (11.45,0.45) rectangle (12.5,1.55);

	\filldraw[fill=green, draw=green] (12.5,1.55) -- (17, 1) -- (12.5, 0.45) -- cycle;
	
	\draw[black, thin, gray] (4,1.05) --(6.25,1.05) -- (11.45,1.55)-- (12.5,1.55) -- (17, 1) -- (12.5,0.45) -- (11.45, 0.45) -- (6.25, 0.95) -- (4, 0.95) -- cycle;

    \draw[black, thick] (6.5, 1+0.866/2) arc (120:240:0.5);
    \draw[black, thick] (6.5, 1-0.866/2) -- (6, 1-0.866/2) -- (6, 1+0.866/2) -- (6.5, 1+0.866/2);\\
    
	
    \draw[black, thick] (11.5, 1+0.75) arc (180-30:180+30:1.5);
    \draw[black, thick] (11.5, 1+0.75) arc (30:-30:1.5);

    \draw[black, thick] (12.5, 1+0.75) arc (180-30:180+30:1.5);
    \draw[black, thick] (12.5, 1+0.75) arc (30:-30:1.5);
	
    \draw (17, 1) circle (0.1);
    \draw (17, 1) circle (0.2);
    \draw (17, 1) circle (0.3);
    \draw (17, 1) circle (0.4);
    
	\draw[black] (2,0) node[below]{LASER head};
    \draw[black] (6.5,-1) node[below]{$f_1 = -50$mm};
    \draw[black] (12.5,-1) node[below]{$f_2=f_3 = 1200$mm};
    \draw[black] (17,-1) node[below]{LIP};
    
    \draw[gray, thick,latex-] (6.25, 0.5) -- (6.25,-1);
    \draw[gray, thick,latex-] (11.5, 0.15) -- (11.5,-1);
    \draw[gray, thick,latex-] (12.5, 0.15) -- (12.5,-1);
    \draw[gray, thick,latex-] (17, 0.5) -- (17,-1);


\end{tikzpicture}
\caption{Schematic diagram of the laser beam setup to generate the laser-induced plasma.}
\label{fig:laser}
\end{figure}

We investigate a near-ideal acoustic monopole source that is generated using the laser optical setup illustrated in~\cref{fig:laser}. When a high-energy laser beam is focused into a point, the air ionizes and plasma is generated  due to the extremely high electromagnetic energy density (on the order of $10^{12}$W/m$^2$). As a result of the sudden deposit of energy, the volume of air undergoes a sudden expansion that generates a shockwave. The initial propagation characteristics can be modeled using von Neumann's point strong explosion theory~\citep{von1941point}, which was originally developed for nuclear explosion modeling. For our ResDMD analysis, we use laser-induced plasma (LIP) sound signature data measured using an 1/8inch, Bruel \& Kjaer (B\&K) type 4138 microphone operated using a B\&K Nexus module \cite{szHoke2022investigating}. The data from the microphone is acquired using an NI-6358 module at a sampling rate of $f_s=$1.25MS/s.  With this apparatus, we can resolve the high-frequency nature of the LIP up to 100kHz. For a detailed description of the laser-optical setup, experimental apparatus, and data processing procedures, see~\citet{szHoke2022investigating,szHoke2021propagation}. 

\begin{figure}
\begin{tikzpicture}[scale=1.0]
	\tikzstyle{every node}=[font=\normalsize]
	\def\w{3}
    \def\s{0.5}
    \def\centx{  0.427} 
    \def\centy{  0.212} 

\node[inner sep=0pt] (t1) at (0,0)
    {\includegraphics[width=\w cm]{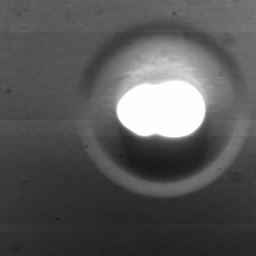}};		
	\draw[black](-0.49*\w, -0.6*\w) node[right]{a) $t = 5~\mu$s};

\node[inner sep=0pt] (t2) at (\w+\s,0)
    {\includegraphics[width=\w cm]{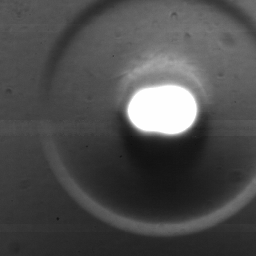}};		

	\draw[black](-0.49*\w+\w+\s, -0.6*\w) node[right]{b) $t = 10~\mu$s};

\node[inner sep=0pt] (t3) at (2*\w+2*\s,0)
    {\includegraphics[width=\w cm]{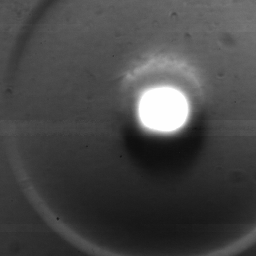}};		


	\draw[black](-0.49*\w+2*\w+2*\s, -0.6*\w) node[right]{c) $t = 15~\mu$s};


\node[inner sep=0pt] (t4) at (3*\w+3*\s,0)
    {\includegraphics[width=\w cm]{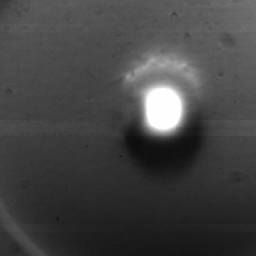}};		

        
	\draw[black](-0.49*\w+3*\s+3*\w, -0.6*\w) node[right]{d) $t = 20~\mu$s};

\end{tikzpicture}
\caption{Schlieren images of the initial laser-induced plasma illustrating the shock wave formation and propagation.}
\label{fig:Schlieren}
\end{figure}
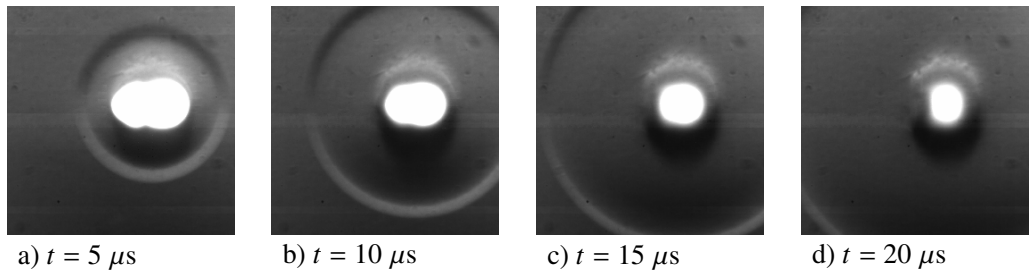

The important acoustic characteristic of the LIP is that it has a short time period of initial supersonic propagation speed, which are shown as Schlieren images taken over a $15\mu$s window in~\cref{fig:Schlieren}. When observed from the far field, this initial supersonic propagation is observed as a non-linear characteristic despite that the wavespeed is supersonic only in a short radius around the source, namely, until about 3--4mm from the optical focal point. During the experiments, 65 realisations of LIP are captured using microphones. Each realisation of LIP is then gated in time such that only the direct propagation path of the LIP remains in the signal~\citep{szHoke2021propagation}. We use this gated data for our ResDMD analysis. 

\subsection{Results}

For a positive integer $d$, we take the state at time $n$ to be
$$
\pmb{x}_n=\begin{pmatrix}p(n)&p(n-1)&\cdots &p(n-d+1)\end{pmatrix}^\top\in\mathbb{R}^d,
$$
where $p$ is acoustic pressure. This corresponds to time-delay embedding, which is a popular method for DMD-type algorithms~\citep{arbabi2017ergodic,kamb2020time,pan2020structure}. There is a further interpretation of $d$ when we make future state predictions using the Koopman mode decomposition. The value of $d$ corresponds to the initial time interval that we use to make future state prediction. This is shown as vertical dashed lines in the plots below.

We split the data into three parts. The first 10 realisations of LIP correspond to $\{\tilde{\pmb{x}}^{(m)},\tilde{\pmb{y}}^{(m)}\}_{m=1}^{M'}$ and are used to train the dictionary. The next 50 realisations correspond to $\{\hat{\pmb{x}}^{(m)},\hat{\pmb{y}}^{(m)}\}_{m=1}^{M''}$, and are used to construct the ResDMD matrices. The final 5 realisations are used to test the resulting Koopman mode decomposition. We consider two choices of dictionary. The first is a linear dictionary computed using \cref{alg:DMD2}. The second is the union of the linear dictionary and the dictionary computed using \cref{alg:kern_algs} with $N=200$. We refer to this combined dictionary as the non-linear dictionary.

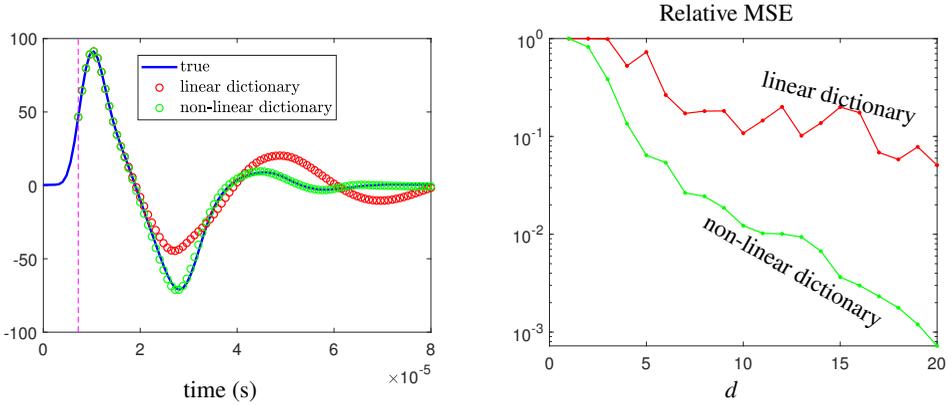
\begin{figure}
 \centering
 \begin{minipage}[b]{0.49\textwidth}
  \begin{overpic}[width=\textwidth,trim={0mm 0mm 0mm 0mm},clip]{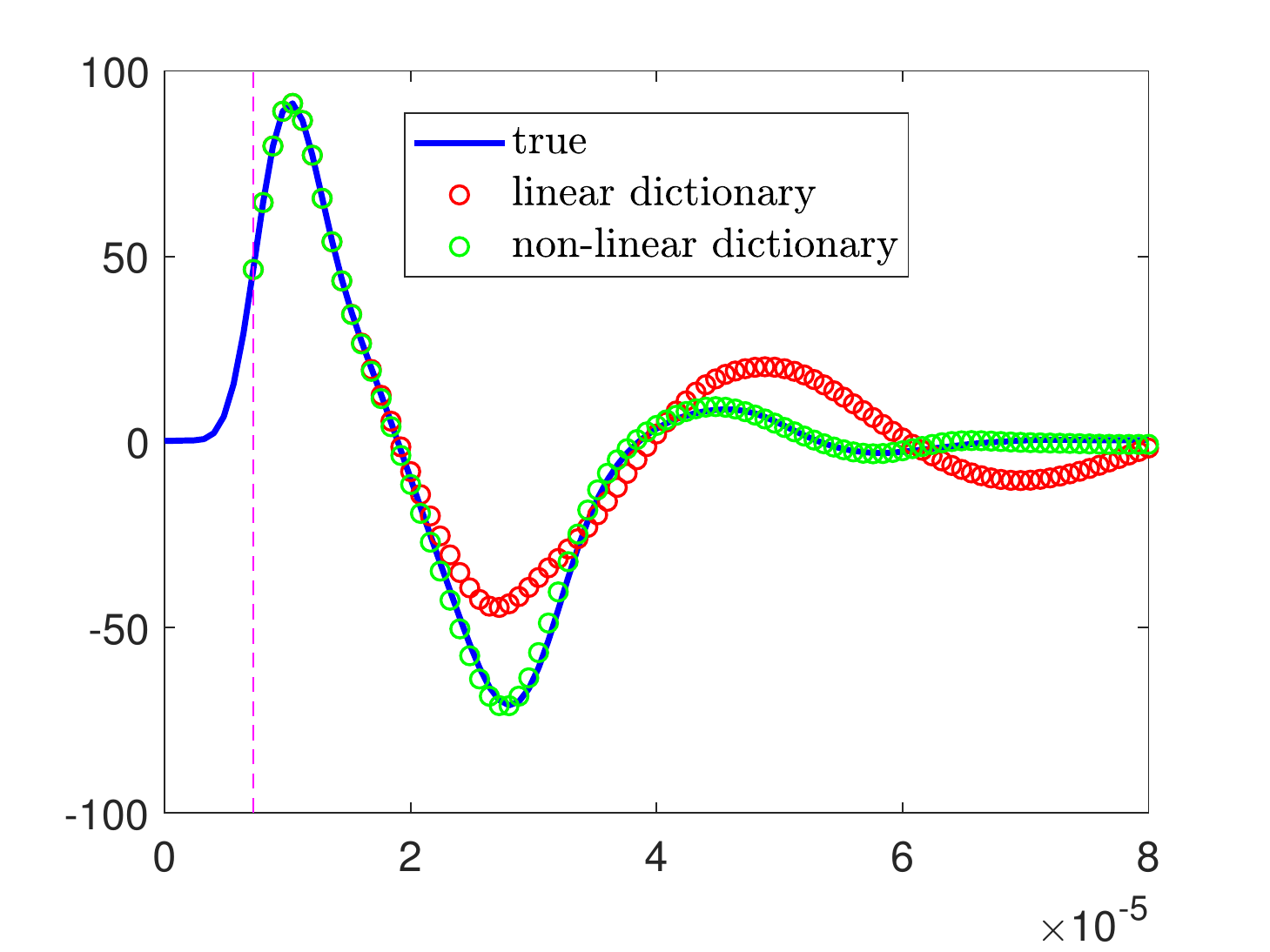}
	\put (41,-2) {time (s)}
   \end{overpic}
 \end{minipage}
\begin{minipage}[b]{0.49\textwidth}
  \begin{overpic}[width=\textwidth,trim={0mm 0mm 0mm 0mm},clip]{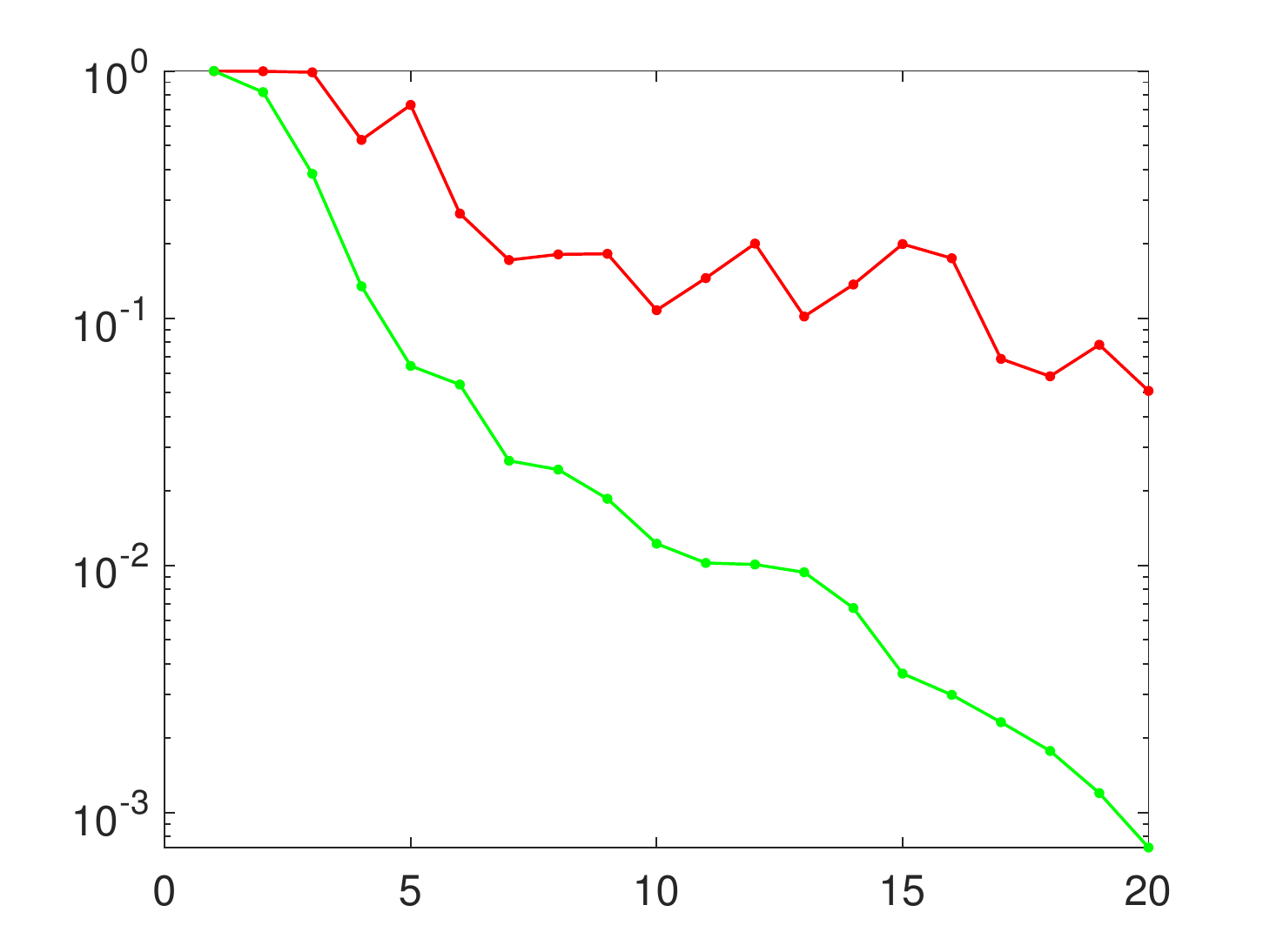}
		\put (35,73) {Relative MSE}
   \put (48,-2) {$d$}
	\put (55,60) {\rotatebox{-14}{linear dictionary}}
	\put (43,32) {\rotatebox{-30}{non-linear dictionary}}
   \end{overpic}
 \end{minipage}
\vspace{1mm}
	  \caption{Left: Prediction using \eqref{LIP_pred} on the first experiment in the test set. The values of $p$ to the left of the vertical dashed line correspond to $\pmb{x}_0$. Right: Relative mean squared error (RMSE) averaged over the test set for different values of $d$.}
\label{fig:LIP1}
\end{figure}

\cref{fig:LIP1} (left) shows the results of the Koopman mode decomposition \eqref{koop_mode_estimate}, applied to the first realisation of the experiment in the test set, with $d=10$. Namely, we approximate the state as
\begin{equation}
\begin{split}
\pmb{x}_n&\approx \mathbb{K}^n\Psi(\pmb{x}_0)V \left[V^{-1}(\sqrt{W}\Psi_X)^\dagger \sqrt{W}\begin{pmatrix}\hat{\pmb{x}}^{(1)}&\cdots&\hat{\pmb{x}}^{(M'')}\end{pmatrix}^\top\right]\\
&=\Psi(\pmb{x}_0)V\Lambda^n \left[V^{-1}(\sqrt{W}\Psi_X)^\dagger \sqrt{W}\begin{pmatrix}\hat{\pmb{x}}^{(1)}&\cdots&
\hat{\pmb{x}}^{(M'')}\end{pmatrix}^\top\right].
\end{split}
\label{LIP_pred}
\end{equation}
In particular, we test the Koopman mode decomposition on unseen data corresponding to the test set. The values of $p$ to the left of the vertical dashed line correspond to $\pmb{x}_0$. It is clear that the non-linear dictionary does a much better job of representing the non-linear behaviour of the system. While the linear dictionary can predict the positive pressure peak, it fails to predict both the magnitude and shape of the negative peak, and it also fails to capture the smaller, high-frequency oscillations following the fist two large oscillation. These discrepancies between the linear and non-linear dictionary-based results also pinpoint where non-linearity in the signal relies. In other words, the non-linear signature of the pressure wave relies in the negative portion of the wave.  \cref{fig:LIP1} (right) plots the relative mean squared error (RMSE) averaged over the test set for different values of $d$. The non-linear dictionary allows an average relative $L^2$ error of around $6\%$ for $d=15$.

\begin{figure}
 \centering
\begin{minipage}[b]{0.49\textwidth}
  \begin{overpic}[width=\textwidth,trim={0mm 0mm 0mm 0mm},clip]{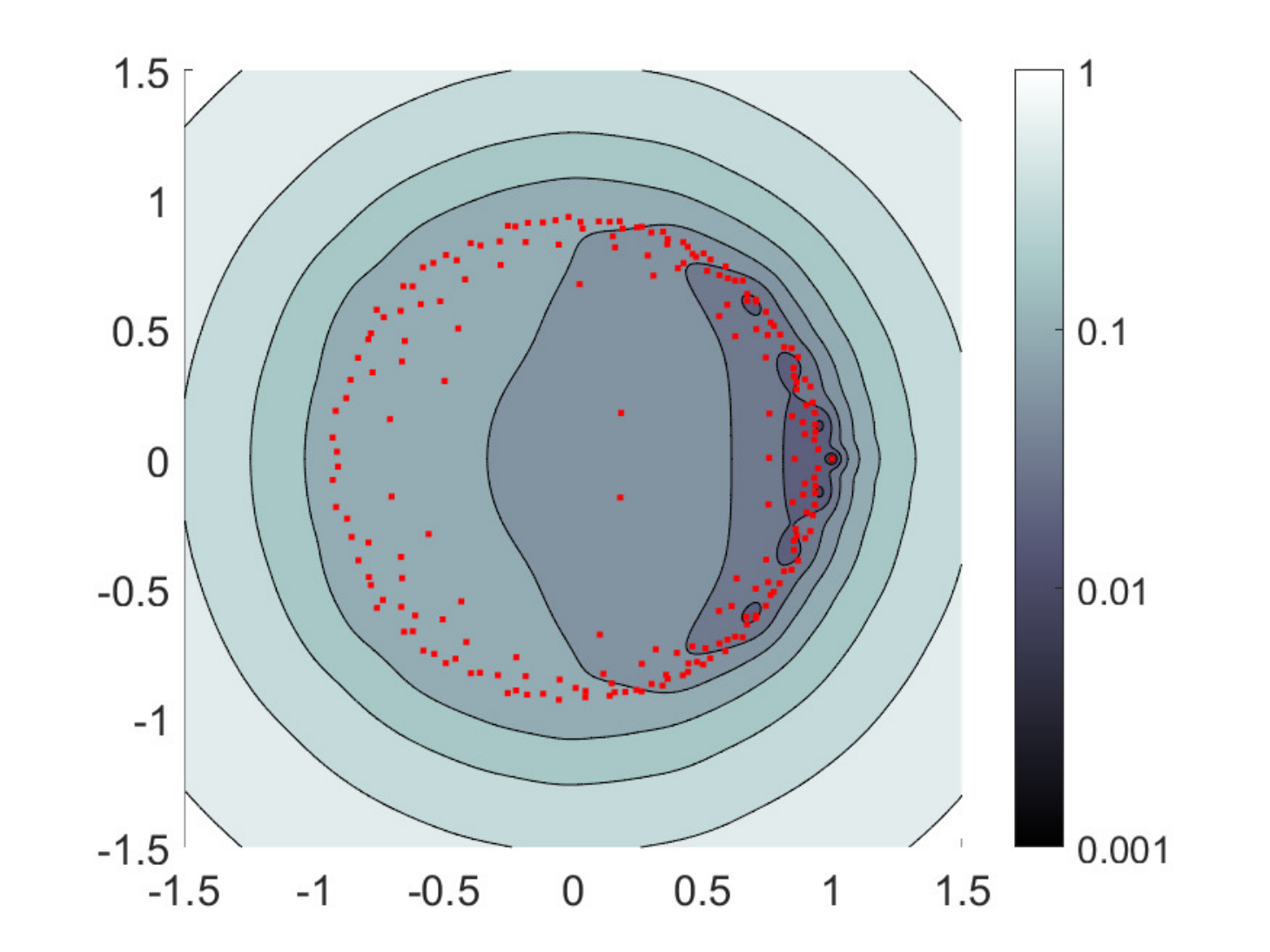}
			\put (40,73) {$\tau_{N}(\lambda)$}
   \put (40,-2) {$\mathrm{Re}(\lambda)$}
		\put (2,33) {\rotatebox{90}{$\mathrm{Im}(\lambda)$}}
   \end{overpic}
 \end{minipage}
 \begin{minipage}[b]{0.49\textwidth}
  \begin{overpic}[width=\textwidth,trim={0mm 0mm 0mm 0mm},clip]{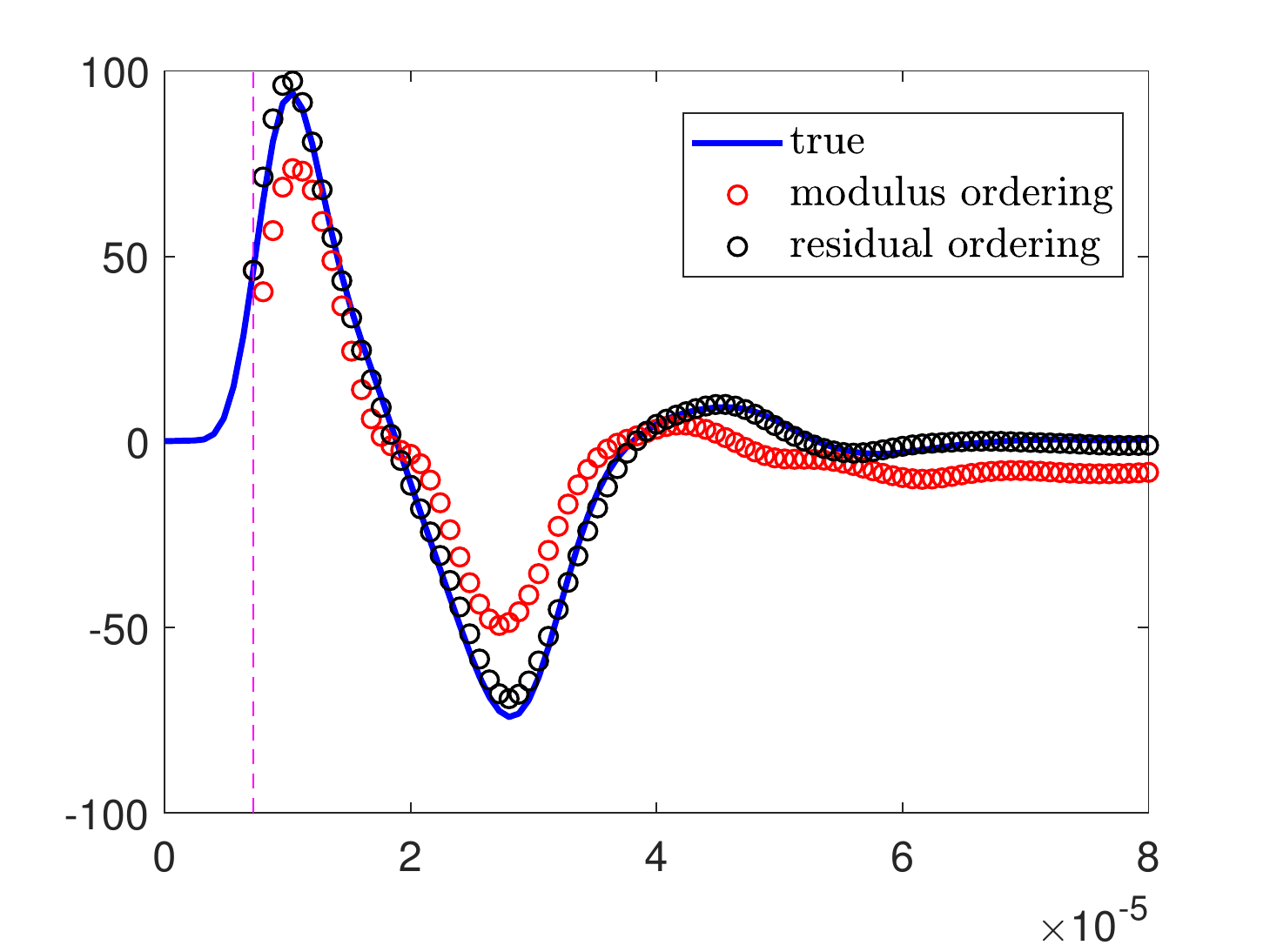}
	\put (41,-2) {time (s)}
   \end{overpic}
 \end{minipage}
\vspace{1mm}
	  \caption{Left: Pseudospectral contours, computed using \cref{alg:res_EDMD} with the non-linear dictionary and $d=10$. The eigenvalues of $\mathbb{K}$ are shown as red dots. Right: Prediction using \eqref{koop_mode_estimate2} on the first experiment in the test set. The values of $p$ to the left of the vertical dashed line correspond to $\pmb{x}_0$. For each type of ordering, we use $40$ modes.}
\label{fig:LIP2}
\end{figure}

\begin{figure}
 \centering
\begin{minipage}[b]{0.9\textwidth}
  \begin{overpic}[width=\textwidth,trim={0mm 0mm 0mm 0mm},clip]{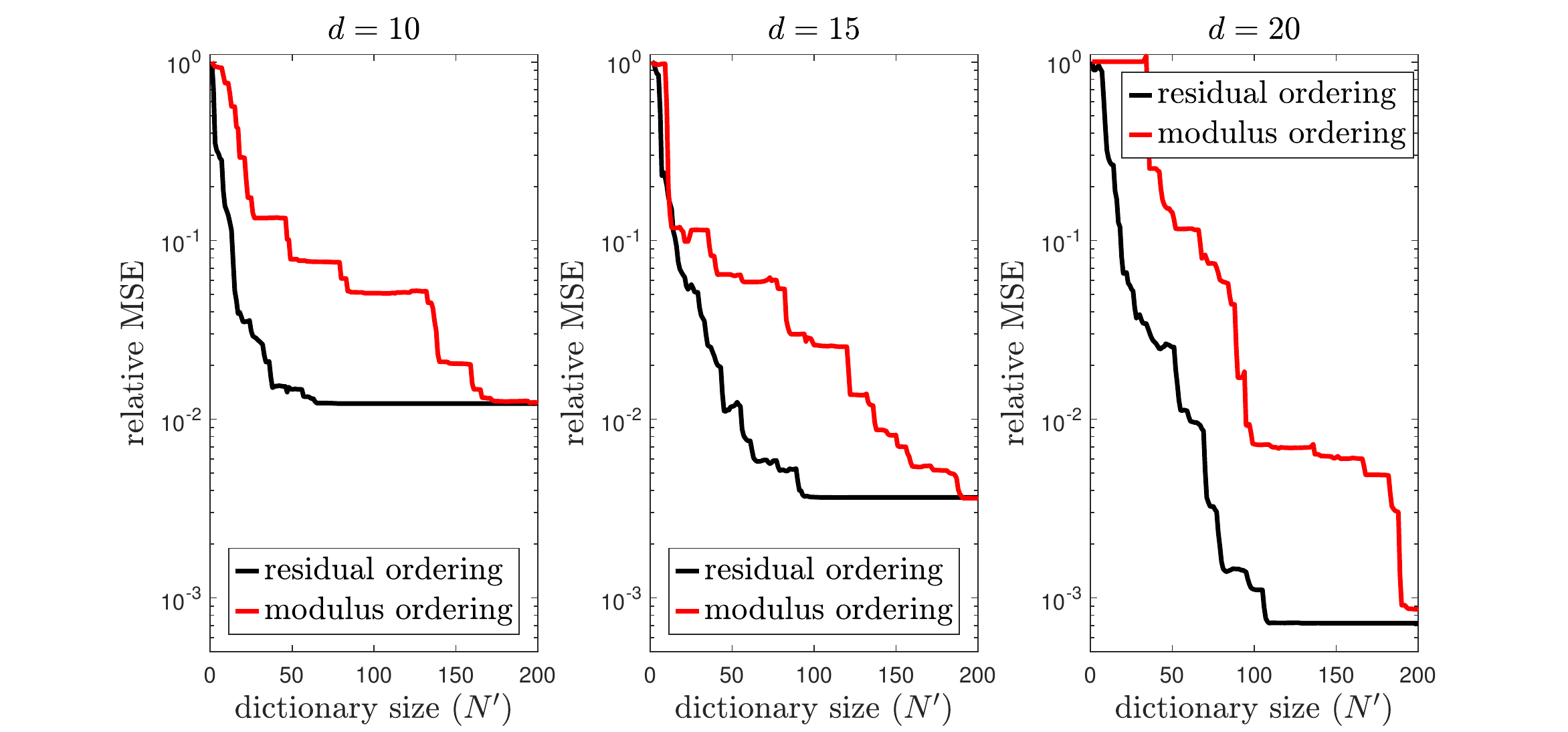}
   \end{overpic}
 \end{minipage}
\vspace{1mm}
	  \caption{RMSE averaged over the test set for two types of compression. `residual ordering' (black curves) corresponds to ordering approximate eigenvalues based on their residual. `modulus ordering' (red curves) corresponds to ordering approximate eigenvalues based on their modulus.}
\label{fig:LIP3}
\end{figure}

\cref{fig:LIP2} (left) shows the corresponding pseudospectral contours, computed using \cref{alg:res_EDMD} with $d=10$.
We can use ResDMD to compress the representation of the dynamics, by ordering the Koopman eigenvalues $\lambda_j$, eigenfunctions $g_j$, and modes according to their (relative) residual $\mathrm{res}(\lambda_j,g_j)$ (defined in \eqref{eq:abs_res}). For a prescribed value of $N'$, we can alter $\epsilon$ in \eqref{koop_mode_estimate2} to produce a Koopman mode decomposition of the $N'$ eigenfunctions with the smallest residual. In \cref{fig:LIP2} (right), we compare this to a compression based on the modulus of the eigenvalues using $40$ modes in each expansion. It is clear that ordering the eigenvalues by their residuals gives a much better compression of the dynamics. To investigate this further, \cref{fig:LIP3} shows the error curves of the two different compressions for various dictionary sizes and choices of $d$. This suggests ResDMD may be effective in the construction of reduced order models.

\section{Discussion and conclusions}
\label{sec:last_section_conc}

This paper has implemented a new form of Dynamic Mode Decomposition (DMD), namely ResDMD, which permits accurate calculation of residuals during the modal decomposition of general Koopman operators. ResDMD computes spectra and pseudospectra of general Koopman operators with error control, and computes smoothed approximations of spectral measures (including continuous spectra) with explicit high-order convergence theorems. We have illustrated the strength of this new method through four examples of turbulent flow. For the canonical example of flow past a circular cylinder, the residual allows accurate identification of spurious modes when using a linear dictionary. For turbulent boundary layer flow, the residual ensures we can accurately identify physical transient modes (and filter out spurious modes), which lead to greater understanding of the turbulent structures within the boundary layer. Further, by relating the Koopman spectral measure to the power spectral density (PSD) we show that processing experimental data via the new ResDMD algorithm reduces low-frequency broadening, which is commonplace in conventional PSD calculations, and provides convergence guarantees. Finally, through analysing the acoustic signature of laser induced plasma, we show that ordering modes via their residual enables greater data compression than ordering modes by their modulus. There are many potential extensions of these ideas. For example, ResDMD may open the door to computing more exotic features of spectra of Koopman operators~\citep{colbrook2019b,colbrook2019computation}, and physical questions such as studying and controlling the transition to turbulence. Other generalisations could include continuous-time dynamical systems, control, and stochastic systems. Since calculating the residual is as easy and computationally efficient as traditional DMD, these new methods provide a useful toolbox for robust and verified Koopmanism.


\backsection[Funding]{This work was supported by a FSMP Fellowship at École Normale Supérieure (M.J.C.); EPSRC Early-Career Fellowship EP/P015980/1 (L.J.A.); and NSF Grant CBET-1802915 (M. Sz.).}

\backsection[Declaration of interests]{The authors report no conflict of interest.}

\appendix

\section{Spectral measures of Koopman operators}\label{sec:unitarySpectralMeasure}
Any normal matrix $A\in \mathbb{C}^{n\times n}$, i.e., $A^*A = AA^*$, has an orthonormal basis of eigenvectors $v_1,\dots,v_n$ for $\mathbb{C}^n$ such that
\begin{equation}\label{eqn:disc_decomp}
v = \left(\sum_{k=1}^n v_kv_k^*\right)v, \quad v\in\mathbb{C}^n \qquad\text{and}\qquad Av = \left(\sum_{k=1}^n\lambda_k v_kv_k^*\right)v, \quad v\in\mathbb{C}^n,
\end{equation}
where $\lambda_1,\ldots,\lambda_n$ are eigenvalues of $A$, i.e., $Av_k = \lambda_kv_k$ for $1\leq k\leq n$. The projections $v_kv_k^*$ simultaneously decompose the space $\mathbb{C}^n$ and diagonalise the operator $A$. This idea carries over to our infinite-dimensional setting with two changes.
First, a Koopman operator that is an isometry does not necessarily commute with its adjoint~\citep{colbrook2021rigorous} and hence is not normal. Therefore, we must consider a unitary extension of $\mathcal{K}$. Namely, $\mathcal{K}'$ defined on an extended Hilbert space $\mathcal{H}'$ with $L^2(\Omega,\omega)\subset\mathcal{H}'$~\cite[Proposition I.2.3]{nagy2010harmonic}. Even though such an extension is not unique, it allows us to understand the spectral information of $\mathcal{K}$. The spectral measures discussed in this paper are canonical and do not depend on the choice of extension.

Second, if $\mathcal{K}'$ has non-empty continuous spectrum, then the eigenvectors of $\mathcal{K}'$ do not form a basis for $\mathcal{H}'$ or diagonalise $\mathcal{K}'$. Instead, the projections $v_kv_k^*$ in~\eqref{eqn:disc_decomp} can be replaced by a projection-valued measure $\mathcal{E}$ supported on the spectrum of $\mathcal{K}'$~\citep[Thm.~VIII.6]{reed1972methods}. $\mathcal{K}'$ is unitary and hence $\sigma(\mathcal{K}')$ lies inside the unit circle $\mathbb{T}$. The measure $\mathcal{E}$ assigns an orthogonal projector to each Borel measurable subset of $\mathbb{T}$ such that
\[
f=\left(\int_\mathbb{T} d\mathcal{E}(y)\right)f \qquad\text{and}\qquad \mathcal{K}'f=\left(\int_\mathbb{T} y\,d\mathcal{E}(y)\right)f, \qquad f\in\mathcal{H}'.
\]
Analogous to~\eqref{eqn:disc_decomp}, $\mathcal{E}$ decomposes $\mathcal{H}'$ and diagonalises the operator $\mathcal{K}'$. For example, if $U\subset\mathbb{T}$ contains only discrete eigenvalues of $\mathcal{K}'$ and no other types of spectra, then $\mathcal{E}(U)$ is simply the spectral projector onto the invariant subspace spanned by the corresponding eigenfunctions. More generally, $\mathcal{E}$ decomposes elements of $\mathcal{H}'$ along the discrete and continuous spectrum of $\mathcal{K}'$ (see \eqref{eqn:spec_meas}).

Given an observable $g\in L^2(\Omega,\omega)\subset\mathcal{H}'$ of interest, the spectral measure of $\mathcal{K}'$ with respect to $g$ is a scalar measure defined as $\mu_{g}(U):=\langle\mathcal{E}(U)g,g\rangle$, where $U\subset\mathbb{T}$ is a Borel measurable set. It is convenient to equivalently consider the corresponding measures $\nu_g$ defined on the periodic interval $[-\pi,\pi]_{\mathrm{per}}$ after a change of variables $\lambda=\exp(i\theta)$.

\bibliographystyle{jfm}
\bibliography{koopman}

\end{document}